\RequirePackage{fix-cm}
\documentclass[smallextended]{svjour3}
\smartqed
\usepackage{graphicx}
\usepackage{algorithm}
\usepackage{algpseudocode}
\usepackage{caption}
\usepackage{xcolor}
\usepackage{tabularx}
\usepackage{booktabs}
\usepackage{array}
\usepackage{multirow}
\usepackage{amsmath}
\usepackage{amsfonts}
\usepackage{rotating}
\usepackage{adjustbox}
\usepackage{caption}
\usepackage{mdframed}

\usepackage{dsfont}

\usepackage{enumitem}

\usepackage[colorlinks,
            citecolor={blue!50!black}, 
            linkcolor={blue!50!black}]{hyperref}

\usepackage[many]{tcolorbox}
\usetikzlibrary{calc}
\tcbuselibrary{skins}

\newtcolorbox{resp}[1][]{%
	enhanced jigsaw,%
	colback=blue!0.5!white,%
	colframe=blue!5!black,%
	size=small,%
	boxrule=1pt,%
	halign title=flush center,%
	coltitle=black,%
	breakable,%
	drop shadow=black!50!white,%
	attach boxed title to top left={xshift=1cm,yshift=-\tcboxedtitleheight/2,yshifttext=-\tcboxedtitleheight/2},%
	minipage boxed title=3cm,%
	boxed title style={%
		colback=white,%
		size=fbox,%
		boxrule=1pt,%
		boxsep=2pt,%
		underlay={%
			\coordinate (dotA) at ($(interior.west) + (-0.5pt,0)$);
			\coordinate (dotB) at ($(interior.east) + (0.5pt,0)$);
			\begin{scope}[gray!80!black]
				\fill (dotA) circle (2pt);
				\fill (dotB) circle (2pt);
			\end{scope}
		}%
	},%
	#1%
}
\DeclareRobustCommand\sampleline[1]{%
	\tikz\draw[#1] (0,0) (0,\the\dimexpr\fontdimen22\textfont2\relax)
	-- (2em,\the\dimexpr\fontdimen22\textfont2\relax);%
}

\newtheorem{assumption}{Assumption}

\usepackage[T1]{fontenc}
\newcommand{\R}{\mathbb{R}}

\usepackage{physics}
\usepackage{listings}
\usepackage{xcolor}

\definecolor{codebg}{RGB}{250,250,250}
\definecolor{keyword}{RGB}{0,102,204}
\definecolor{string}{RGB}{0,153,0}
\definecolor{comment}{RGB}{150,150,150}
\definecolor{number}{RGB}{180,50,50}
\definecolor{identifier}{RGB}{40,40,40}

\lstdefinestyle{mystyle}{
    language=Python,
    backgroundcolor=\color{codebg},
    basicstyle=\ttfamily\small\color{identifier},
    keywordstyle=\bfseries\color{keyword},
    stringstyle=\color{string},
    commentstyle=\itshape\color{comment},
    stepnumber=1,
    numbersep=8pt,
    showstringspaces=false,
    breaklines=true,
    frame=single,
    rulecolor=\color{black!30},
    frameround=tttt,
    tabsize=4,
    captionpos=b,
    breakatwhitespace=true,
columns=fullflexible
}

\lstdefinelanguage{json}{morestring=[b]"}
\lstdefinestyle{json}{style=mystyle, language=json, basicstyle=\ttfamily\footnotesize\color{identifier}, breaklines=false}

\newcommand{\toolname}{\textsf{Scen-Opt}}

\def\makeheadbox{\relax}

\usepackage{tikz}
\usepackage{subcaption}

\usepackage{optidef}

\begin{document}

\title{\toolname: A Scenario Optimization Toolbox for Data-Driven Convex Programming}

\titlerunning{\toolname: A Scenario Optimization Toolbox}
\authorrunning{Wooding et al.}

\author{Ben Wooding$^1$ \and Simone Garatti$^2$ \and Marco C. Campi$^3$ \and Abolfazl Lavaei$^4$}

\institute{$^1$ Institute for Convergent Software Integrated Systems, Vanderbilt University, USA \\
        \email{ben.wooding@vanderbilt.edu}\vspace{0.2cm}\\  
       $^2$ Department of Electronics, Information and Bioengineering, Politecnico di Milano, Italy \\
       \email{simone.garatti@polimi.it}\vspace{0.2cm}\\ 
       $^3$ Department of Information Engineering, University of Brescia, Italy \\
       \email{marco.campi@unibs.it}\vspace{0.2cm}\\ 
       $^4$ School of Computing, Newcastle University, UK \\
       \email{abolfazl.lavaei@newcastle.ac.uk}\\ 
}

\date{}
\maketitle

\begin{abstract}
    The \emph{scenario approach} is a well-established statistical framework for data-driven decision-making. In particular, in data-driven optimization, the scenario approach unveils how the problem structure governs \emph{out-of-sample} generalization, and offers a principled basis for assessing and certifying the reliability of the optimal solution as per constraint satisfaction. Despite its strong theoretical development and wide applicability, no software toolbox has been available to date that enables user-friendly, data-driven convex optimization within the scenario-approach framework. In this paper, we introduce \toolname, an \emph{open-source} software tool that integrates convex programming with data samples while providing statistical guarantees grounded in scenario theory. \toolname~is implemented in Python, supporting data-driven \emph{linear, quadratic, and semidefinite programming}, and offers a Python-based web application with an intuitive and reactive graphical user interface (GUI) built using modern web technologies. \toolname~can be used directly through its online interface or installed locally, accommodating both manual input and data-file uploads (\texttt{CSV}, \texttt{JSON}, \texttt{TXT}, \texttt{TSV}, \texttt{MAT}, Excel, \texttt{NPY}, \texttt{NPZ}, \texttt{Parquet}). Built on a Python backend with a modern JavaScript frontend, \toolname~offers a highly user-friendly experience and efficient usability across desktops, laptops, tablets, and mobile devices. In this paper, \toolname~is applied to a set of representative benchmarks, demonstrating its practical effectiveness for data-driven convex optimization with guaranteed performance.
    
\end{abstract}

\section{Introduction}

The increasing availability of data across scientific and engineering domains has fundamentally reshaped decision-making workflows: rather than depending solely on analytical models, which may be inaccurate, overly conservative, or impractical to obtain, there is growing interest in methods that leverage sampled observations directly to inform decision-making~\cite{hou2013model}. Data can reveal structure, expose model limitations, and support more adaptive strategies. Yet, most data-driven decision procedures offer limited assurances about how the resulting decisions will perform when confronted with \emph{new, unseen situations}. In many cases, solutions are derived from data but their reliability is assessed only through ad-hoc or post-hoc testing, leaving open the fundamental question of whether such solutions generalize with some robustness beyond the observed sample.

The \emph{scenario approach}~\cite{calafiore2006scenario,campi2009scenario,campi2018introduction} provides a principled answer to this challenge. Originating in convex optimization and now encompassing a broad class of data-driven decision-making schemes, scenario theory establishes rigorous probabilistic guarantees for decisions obtained from independent and identically distributed (i.i.d.) data. By characterizing the intersection between the decision and the observed data through the notion of \emph{complexity}, scenario theory offers explicit bounds on the probability that a newly observed uncertainty instance will render the decision inappropriate. This \emph{risk quantification} is entirely data-driven and does not rely on assumptions about the underlying probability distribution, making it suitable for a wide spectrum of applications~\cite{campi2021scenarioreview,garatti2025non}.

When applied to \emph{convex} optimization, scenario theory attains a particularly elegant form. Each data point is treated as a constraint, the complexity is the number of constraints strictly needed to reconstruct the solution (the so called \emph{support list}), and the risk corresponds to the probability that a new constraint, \emph{unseen at design time}, is violated by the computed solution. The structural properties of convex optimization ensure that solutions are globally optimal and unique (when the problem is strictly convex), so the connection between complexity, sample size, and generalization becomes especially transparent. These features have driven the widespread adoption of convex optimization techniques across control theory, robotics, machine learning, data science, signal processing, and many other fields.

\noindent\textbf{Original contributions.} Motivated by the central challenge of making reliable decisions when only sampled data are available but robust performance guarantees are still required, this paper introduces \toolname~ a \emph{first-of-its-kind tool} that leverages scenario theory to deliver user-friendly, data-driven convex optimization with rigorous probabilistic guarantees, even when only limited sampled data are available.

The main features of the proposed tool are the following:
\begin{enumerate}[label=(\roman*)]
	\item  \toolname~leverages a Python backend with a modern JavaScript frontend, powered by a suite of state-of-the-art solvers (\emph{$27$ solvers} supported at the time of writing) including CVX toolbox \cite{grant2009cvx} and MOSEK\footnote{https://www.mosek.com/documentation/}, and supports three classes of scenario optimization programs (i) \emph{linear programs (LP)}, (ii) \emph{quadratic programs (QP)} and (iii) \emph{semidefinite programs (SDP)}.
     \item For each of the mentioned scenario programs, \toolname~supports \emph{four problem configurations}: (i) the robust problem, (ii) the relaxed problem, (iii) the regularized problem, and (iv) the general formulation combining relaxation, and regularization.
	\item Implemented as a responsive and reactive Python Flask\footnote{https://flask.palletsprojects.com/en/stable/} web application, \toolname~provides an intuitive, user-friendly interface that enables efficient interaction with all scenario-optimization functionalities. Users can access and use the application directly through the web across desktops, laptops, tablets, and mobile devices without any download or installation\footnote{The server has $24$GB RAM and $4$ vCPUs, enabling a predicted $20+$ simultaneous users with a $50$MB maximum uploaded file size and $2$ hour timeout for any single program. Heavier computations can be run by installing \toolname~to their local machine.}.
	\item In addition, \toolname~provides a \emph{Docker-based version}\footnote{\texttt{docker pull ghcr.io/kiguli/scen-opt:latest}\\
\texttt{docker run -p 5000:5000 ghcr.io/kiguli/scen-opt:latest}} that can be downloaded and executed locally, enabling users to achieve higher performance when leveraging their own computing resources.
	\item The tool supports both manual data entry and data-file uploads for sampled trajectories, providing a user-friendly interface for specifying all required inputs across the different scenario optimization programs. The full program can be uploaded in one-shot via \texttt{.mat} or \texttt{.json}, and results downloaded in the same formats. Scenario-dependent constraints can be provided either \emph{symbolically}, as expressions in the uncertainty, or \emph{numerically}, as pre-evaluated per-sample matrices.
\end{enumerate}
Besides describing \toolname, this paper demonstrates its effectiveness through a series of physical benchmarks, spanning all three supported classes of scenario optimization programs and showcasing their corresponding four sub-classes for each program type.

The web-based version of \toolname~is accessible at:
\begin{center}
	\href{https://scen-opt.woodingben.com}{https://scen-opt.woodingben.com}
\end{center}
The \toolname~source code, accompanied by comprehensive installation and usage instructions for the Docker-based version, can be accessed at\footnote{The version described in this paper (v1.0) is permanently archived at:
	\href{https://doi.org/10.5281/zenodo.23177690}{https://doi.org/10.5281/zenodo.23177690}}:
\begin{center}
	\href{https://github.com/Kiguli/Scen-Opt}{https://github.com/Kiguli/Scen-Opt}
\end{center}

\noindent\textbf{Organization.} The remainder of the paper is structured as follows. Section~\ref{sec:underpinning-theory} reviews the theoretical foundations of the scenario approach in a convex optimization setting that underpin the methods and guarantees used throughout this paper. Section~\ref{sec:scenario-convex-optimization} presents the convex optimization programs supported by \toolname, outlining the three program classes and their four scenario-based variants implemented for each. Implementation details of the toolbox, including the underlying algorithms, are provided in Section~\ref{sec:implementation}, while the web-based graphical interface is described in Section~\ref{sec:gui}. Comprehensive benchmarking results across all supported problem classes are reported in Section \ref{sec:benchmarking}, and the paper concludes, together with directions for future development, in Section \ref{sec:futurework}.

\noindent\textbf{Related studies on scenario approach.} Over the past two decades, the scenario approach has played a central role in advancing data-driven methods for decision-making under uncertainty. The seminal work in \cite{calafiore2006scenario} introduced scenario-based convex programming for robust control, showing that semi-infinite constraints can be handled through randomized sampling. This line of work was further strengthened in \cite{campi2008exact,campi2011sampling}, which established explicit probabilistic bounds on violation and extended the framework to settings with discarded or a-posteriori-violated constraints. These developments form the foundation for a broad class of randomized and scenario-based optimization techniques that have since been generalized and applied across numerous domains.

Numerous contributions to the field have been made by many authors. Methods for convex scenario programs with multiple chance constraints are investigated in \cite{schildbach2013randomized}. The sample complexity of random convex programs with structured uncertainty is characterized in \cite{zhang2015sample}, leading to less conservative bounds when exploiting problem structure. A probabilistic connection between the optimal value of scenario convex programs and that of robust and chance-constrained formulations is established in \cite{esfahani2014performance}, with extensions to certain mixed-integer nonconvex settings; the results are further applied to infinite-dimensional linear programs in \cite{mohajerin2018infinite}. The relationship between compression learning and the original results in scenario optimization is analyzed in \cite{margellos2015connection}, while \cite{margellos2014road} proposes an intermediate approach to chance-constrained optimization that sits between robust optimization and fully randomized scenario methods. Distributed scenario optimization in uncertain multi-agent framework is studied in \cite{margellos2017distributed}. Probabilistic guarantees for convex scenario programs where validation tests are exploited to derive tighter confidence bounds on the violation probability are developed in \cite{shang2020posteriori}. Finally, \cite{romao2022exact,romao2023probabilistic} study exact feasibility properties of convex scenario programs with discarded constraints, providing conditions under which the tightest feasibility guarantees remain valid despite constraint removal.

A major turning point in the theory of the scenario approach is given by the works~\cite{campi2018general,garatti2025non,campi2018wait,garatti2022risk} which introduce the wait-and-judge philosophy and permit assessing generalization after a solution is obtained, using the size of a suitably defined support subsample as a measure of complexity. This enables tight, instance-specific guarantees for a quite general class of decision problems, broadening the applicability of scenario methods far beyond traditional convex optimization settings.

A further shift to enhance the broadest applicability is the new compression-based theory in~\cite{campi2023compression} establishes rigorous control over the probability that newly observed data, which alter a compressed representation, providing tight finite-sample bounds under minimal assumptions. By showing that the compressed-set cardinality consistently estimates this probability in a fully agnostic setting, this abstract framework offers strong support for general observation-driven decision-making and constitutes a foundational tool for modern learning tasks involving uncertainty quantification.

\section{Foundational Concepts in Scenario (Convex) Optimization}\label{sec:underpinning-theory}

This section reviews the foundational concepts underpinning scenario-based decision making, focusing on the theoretical principles that support data-driven convex optimization with probabilistic guarantees. We summarize the key definitions, assumptions, and central results from the scenario approach literature~\cite{garatti2022risk,garatti2025non} that form the basis of the methods implemented in \toolname. 

We begin by formalizing the class of convex optimization problems considered in this work.

\begin{definition}[Convexity]
Let $\R^d$ denote a Euclidean space with $d$ components. A set $C \subseteq \R^d$ is convex if the line segment between any two points in $C$ lies entirely within $C$; that is, for all $x_1, x_2 \in C$ and any $0 \leq a \leq 1$,
 \[ax_1 + (1-a)x_2 \in C.\] 
 A function $f : \R^d \to \R$ is convex if its domain is a convex set and, for all $x, y$ in its domain and any $0 \leq a \leq 1$,
  \[f(ax+(1-a)y) \leq af(x) + (1-a)f(y).\] 
  Geometrically, this means that the line segment connecting $(x, f(x))$ and $(y, f(y))$ lies above the graph of $f$. \hfill $\square$
\end{definition}

Let $x \in \mathcal{X}$ be a vector of design variables, where $\mathcal{X} \subset \mathbb{R}^d$. Let $\delta\in\mathbb{R}^q$, with $q$ components, be a parameter that describes the environment to which a decision is applied. In this setting, one can formulate the following scenario program:

\begin{subequations}\label{eq:robust-scenario-optimization}
\begin{align} 
\min_{x\in\mathcal{X},\zeta_i\geq 0}& \quad c(x)~+~\tau\Vert x-\overline{x}\Vert~+~ \rho\sum_{i=1}^{N}\zeta_i\label{eq:RSO-objective} \\
    \text{subject to:}\quad& f(x,\delta_i)\leq\zeta_i, \quad i=1,\ldots,N.\label{eq:RSO-constraints}
\end{align}
\end{subequations}

The constraints~\eqref{eq:RSO-constraints} are expressed as $f(x,\delta_i)\leq 0$, for each data point $\delta_i$, and we allow both regularization of the solution and relaxation of the constraints. The values $\zeta_i$ are per-constraint scalar slack optimization variables used to relax the problem. By selecting $\zeta_i>0$, a scenario constraint $f(x,\delta_i)\leq 0$ may be violated, thereby enlarging the feasibility domain for $x$ and creating the possibility of finding a solution with lower cost $c(x)$. To control the regret associated with constraint violations, however, the sum of the relaxation parameters, $\sum_i\zeta_i$, is also penalized in the objective function~\eqref{eq:RSO-objective}. A fixed weight parameter $\rho\geq0$ determines the relative importance assigned to the cost $c(x)$ and the constraint violations. Regularization is likewise imposed in the objective to encourage the solution to remain near some $\overline{x}$, with the fixed weight parameter $\tau\geq0$ controlling the strength of this regularization. There are special cases of~\eqref{eq:robust-scenario-optimization} that typically arise as limiting cases, but which we explicitly consider in this work. \\

\noindent\textbf{Only Constraint Relaxation ($\tau =0$)}\begin{align*} 
\min_{x\in\mathcal{X},\zeta_i\geq 0}& \quad c(x)~+~ \rho\sum_{i=1}^{N}\zeta_i \\
    \text{subject to:}\quad& f(x,\delta_i)\leq\zeta_i, \quad i=1,\ldots,N.
\end{align*}
\noindent\textbf{Only Regularization ($\rho = 0$ and $\zeta_i = 0$, $\forall i$)}\begin{align*} 
\min_{x\in\mathcal{X}}& \quad c(x)+~\tau\Vert x-\overline{x}\Vert \\
    \text{subject to:}\quad& f(x,\delta_i)\leq0, \quad i=1,\ldots,N.
\end{align*}
\noindent\textbf{Robust problem ($\tau =0$, $\rho = 0$ and $\zeta_i = 0$, $\forall i$)}\begin{align*} 
\min_{x\in\mathcal{X}}& \quad c(x) \\
    \text{subject to:}\quad& f(x,\delta_i)\leq0, \quad i=1,\ldots,N.
\end{align*}

The user can choose the formulation which is most appropriate for the problem at hand among these four alternatives. Once a formulation has been selected, let $x^*_N$ denote the $x$-component of the corresponding optimal solution (in what follows, there is no need to explicitly consider the optimal $\zeta^*_i$ components, if any). When problem~\eqref{eq:robust-scenario-optimization} is solved, it yields $x^\ast_N$, for which the cost $c(x^\ast_N)$ can be directly computed.

\begin{assumption}
    We assume that a solution to~\eqref{eq:robust-scenario-optimization} exists and is unique for every $N$ and for every realization of the $\delta_i$'s. \hfill $\square$
\end{assumption}

Existence is always guaranteed when there is constraint relaxation (for a compact $\mathcal{X}$; otherwise a small $\rho$ can make the relaxed problem unbounded), while uniqueness is guaranteed whenever there is a strictly convex regularization term with $\tau> 0$, provided existence holds. The results below apply to $x^*_N$ regardless of the formulation adopted. The variables $x$ and $\delta$ are linked through the concept of \emph{appropriateness} as in the following definition.

\begin{definition}[Appropriateness]
The variable $x$ is said to be \emph{appropriate} for $\delta$ if it satisfies the corresponding requirement induced by $\delta$; that is, the satisfaction condition associated with $\delta$ holds when evaluated at $x$. \hfill $\square$
\end{definition}

In robust optimization, $x$ is appropriate for $\delta$ if it lies within the region of feasible decisions induced by $\delta$. The set of all values of $x$ that are appropriate for $\delta$ is denoted by $\mathcal{X}_{\delta}$. We assume access to a list $(\delta_{1},\ldots,\delta_{N})$ of observations of $\delta$, where each $\delta_{i}$ is referred to as a \emph{scenario} and is used to inform the decision variable $x$. By abuse of notation, the same expression is used even when $N=0$, in which case the list is interpreted as empty. Formally, $(\delta_{1},\ldots,\delta_{N})$ is an i.i.d. sample from a probability space $(\Delta,\mathcal{D},\mathds{P})$, where $\Delta$ is the set of all possible uncertainty realizations, $\mathcal{D}$ is the associated $\sigma$-algebra of measurable subsets of $\Delta$, and $\mathds{P}$ is the underlying probability measure governing the sampling process. Although the scenario approach does not require knowledge of the underlying probability space, a \emph{distribution-free} paradigm, the resulting generalization guarantees still yield insight into the reliability of the decision.

However, the appropriateness of $x^\ast_N$ for a \emph{new, unseen} scenario $\delta$ remains uncertain. This motivates the introduction of the notion of \emph{risk}, which quantifies the probability that $x^\ast_N$ fails to satisfy the requirement induced by a future scenario.

\begin{definition}[Risk]
	The \emph{risk} $V(x)$ quantifies the probability that a newly drawn scenario $\delta$ renders $x$ inappropriate. For any $x \in \mathcal{X}$, $V(x) = \mathds{P}\big\{\delta : f(x,\delta)> 0\big\}$. \hfill $\square$
\end{definition}

The fundamental problem is to estimate $V(x^*_N)$, which cannot be computed directly since the probability measure $\mathds{P}$ is unknown. Scenario optimization provides a powerful \emph{generalization theory} that enables one to bound or estimate $V(x^\ast_{N})$ \emph{without requiring any additional samples} or any knowledge of the underlying probability distribution.

To analyze how the solution evolves as additional scenarios are incorporated, we consider~\eqref{eq:robust-scenario-optimization} for some general $N$. This formulation allows us to track how the feasible set and its optimizer change as new scenarios are added, which is central to the generalization properties of scenario optimization. The following concept of \emph{complexity} plays a central role in scenario-based theory, as it characterizes the minimal subset of scenarios that fully determine the solution.

\begin{definition}[Support List and Complexity]
\label{def:support-list-and-complexity}
    Given a list of scenarios $(\delta_1,\ldots,\delta_N)$, a \emph{support list} is the sublist of scenarios that include all those for which $f(x^*_N,\delta_i)> 0$ and some of those for which $f(x^*_N,\delta_i)=0$, such that:
    \begin{enumerate}[label=(\roman*)]
        \item it allows reconstruction of the solution $x^*_N$ of the considered problem;
\item the sublist is \emph{irreducible}: no element can be further removed from the sublist without the decision changing.
\end{enumerate}
For a given list $(\delta_1,\ldots,\delta_N)$, multiple support lists may exist with different cardinalities $k$. The minimal such cardinality among all support lists is called the \emph{complexity} and is denoted by $s^\ast_N$. \hfill $\square$
\end{definition}

Note that both $x^\ast_N$ and $s^\ast_N$ depend on the list $(\delta_1,\ldots,\delta_N)$ and are therefore random elements defined on the product probability space $(\Delta^N,\mathcal{D}^N,\mathds{P}^N)$, where the product measure arises from the independence of scenario draws. Finally, we can state the theorems where $x^*_N$ is the solution of the considered problem for a known $N$.

\begin{theorem}[Risk Certificate]
\label{thm:riskcert}
Let $\epsilon(k)$ be defined as
    \begin{equation*}
        \epsilon(k) := 1-t(k), \quad k=0,1,\ldots,N-1, ~\text{and}~\epsilon(N) = 1,
    \end{equation*}
    where, for a prescribed confidence level $1-\beta$, the scalar $t(k)\in(0,1)$ is the unique solution of
    \begin{equation*}
        \frac{\beta}{N}\sum_{i=k}^{N-1}\binom{i}{k}t^{i-k}-\binom{N}{k}t^{N-k}=0, \qquad k=0,1,\ldots,N-1.
    \end{equation*}
Then, for any probability measure $\mathds{P}$, the following distribution-free risk bound holds:
    \begin{equation*}
        \mathds{P}^N\Big\{V(x^*_N) > \epsilon(s^*_N)\Big\} \leq \beta.
    \end{equation*}\hfill $\square$
\end{theorem}

In Theorem~\ref{thm:riskcert}, the parameter $\beta$ is referred to as the \emph{confidence level} and is typically chosen to be very small, for example $10^{-5}$ or $10^{-6}$. The result asserts that $V(x^\ast_N)$, the \emph{risk} that the solution $x^\ast_N$ fails to be appropriate for a new out-of-sample scenario, is upper-bounded with confidence $1-\beta$, by $\epsilon(s^\ast_N)$. Here, $\epsilon(s^\ast_N)$ is a known deterministic function evaluated at the \emph{complexity} $s^\ast_N$ of the scenario optimization problem.

\begin{remark}
	The upper bound on the \emph{risk} provided in Theorem~\ref{thm:riskcert} holds for any solution computed by \toolname. In the sequel, we also present a result that yields a corresponding \emph{lower bound}; however, this requires additional assumptions. While \toolname~attempts to detect violations of these assumptions, such checks cannot be performed for out-of-sample scenarios. Consequently, it is ultimately the user’s responsibility to assess whether the lower bound remains applicable in cases where \toolname~has not flagged a violation.\hfill $\square$
\end{remark}

We now recall a structural property of scenario programs that will play a central role in the forthcoming theorem.

\begin{property}[Non-degeneracy]
\label{def:non-degeneracy}
For any $N$ and with probability one with respect to the generation of the scenarios, there exists a unique support list for $(\delta_1,\ldots,\delta_{N})$. \hfill $\square$
\end{property}

It is often instructive to illustrate degeneracy concretely, in order to clarify the meaning of non-degeneracy. In any optimization problem, one may identify subsets of support constraints whose removal does not alter the optimal solution. Degeneracy arises when multiple distinct subsets of support constraints can be removed while still preserving the solution. A simple example occurs when a support constraint appears multiple times in the list: each repeated instance may be removed independently without affecting the optimizer. As a result, there exist multiple support lists, each omitting a different repetition, despite all having the same complexity.

\begin{theorem}[Two-sided Risk Certificate]
\label{thm:riskcert-nondeg}
Given a confidence parameter $\beta \in (0,1)$, for any $k = 0,1,\ldots,N-1$ consider the polynomial equation in the variable $t$
\begin{subequations}
    \begin{equation}
    \label{eq:nondeg1_con}
        \binom{N}{k}t^{N-k} - \frac{\beta}{2N}\sum_{i=k}^{N-1}\binom{i}{k}t^{i-k} - \frac{\beta}{6N}\sum_{i=N+1}^{4N}\binom{i}{k}t^{i-k}=0,
    \end{equation}
and for $k = N$, consider instead the polynomial equation
    \begin{equation}
    \label{eq:nondeg2_con}
        1 - \frac{\beta}{6N}\sum_{i=N+1}^{4N}\binom{i}{k}t^{i-N}=0.
    \end{equation}
    \end{subequations}
    For any $k = 0,1,\ldots,N-1$, ~\eqref{eq:nondeg1_con} admits exactly two solutions in $[0,+\infty)$, denoted $\underline{t}(k)$ and $\overline{t}(k)$ with $\underline{t}(k) \leq \overline{t}(k)$.
    For $k = N$,~\eqref{eq:nondeg2_con} admits exactly one solution in $[0,+\infty)$, denoted $\overline{t}(N)$, which we define with $\underline{t}(N) = 0$. Let $\underline\epsilon(k):= \max\{0,1-\overline{t}(k)\}$ and $\overline{\epsilon}(k):= 1 - \underline{t}(k),$ for all $k=0,1,\ldots,N.$ Under the assumptions of consistency and non-degeneracy, it holds for any uncertainty space $\Delta$ and any probability measure $\mathds{P}$ that:
    \begin{equation*}
        \mathds{P}^N\{\underline{\epsilon}(s^*_N) \leq V(x^*_N)\leq\overline{\epsilon}(s^*_N)\} \geq 1-\beta.
    \end{equation*} \hfill $\square$
\end{theorem}

With this theorem, one can bound the risk associated with scenario optimization for any chosen confidence level and any list of scenarios $(\delta_{1},\ldots,\delta_{N})$. In our tool, we implement this result using the same numerical procedure described in \cite[Appendix B]{campi2023compression}.

\section{Scenario Optimization Schemes Covered by the Toolbox}\label{sec:scenario-convex-optimization}

In this section, we present the data-driven convex optimization schemes supported by our toolbox \toolname. The problem formulations are intentionally aligned with the classical exposition in \cite{boyd2004convex} to maintain familiarity, and the underlying implementation of \toolname~draws inspiration from the design principles established in the CVX toolbox \cite{grant2009cvx}.

We present below the implementation of scenario convex optimization for LP, QP, and SDP programs. As outlined in Section~\ref{sec:underpinning-theory}, \toolname~supports four subclasses for each of these optimization types, : \emph{robust}, \emph{relaxation}, \emph{regularization}, and the combined \emph{relaxation-regularization} formulation. For consistency, we introduce each class using the relaxation-regularization formulation, as it is the \emph{most general case}, and then indicate how the other subclasses are obtained through appropriate adaptations. 

Throughout, we denote the decision variable by $x \in \mathbb{R}^d$. Each optimization program incorporates the constraints induced by every scenario $\delta_i\in\mathbb{R}^q$ in the list $(\delta_1,\ldots,\delta_N)$, together with any additional hard constraints required by the problem structure.

\subsection{Linear Programming (LP)}\label{sec:lp}

We consider the most general case of \emph{relaxed-regularized} LP as 

\begin{resp}\vspace{-0.3cm}
\begin{subequations}\label{eq:LP-full}
\begin{align} 
\min_{x,\zeta_i \geq 0}& \quad c^\top x~+~\tau\Vert  x-\bar{ x}\Vert_p~+~\rho\sum_{i=1}^{N} \zeta_i  \label{eq:LP-objective} \\
    \text{subject to:}& \quad A(\delta_i) x~+~ b(\delta_i) \leq \zeta_i, \quad i = 1,\ldots, N\label{eq:LP-constraint1}
    \\
    & \quad G x~+~ h \leq 0, \label{eq:LP-constraint2}
\end{align}
\end{subequations}
where $\tau \geq 0$ and $\rho\geq0$.
\end{resp}

LPs are so named because the objective function $c^\top x$, where $c \in \mathbb{R}^d$, and all constraint functions in \eqref{eq:LP-constraint1} and \eqref{eq:LP-constraint2} are required to be linear. For arbitrary integers $\mathfrak{n} > 0$ and $\mathfrak{m} > 0$, the matrix $G \in \mathbb{R}^{\mathfrak{n} \times d}$ and vector $h \in \mathbb{R}^{\mathfrak{n}}$ define the hard linear constraints, while the matrix-valued function $A(\cdot) \in \mathbb{R}^{\mathfrak{m} \times d}$ and vector-valued function $b(\cdot) \in \mathbb{R}^{\mathfrak{m}}$ specify the linear constraints associated with each scenario $\delta_i$.

In what follows, we discuss how the \emph{relaxed–regularized} LP in~\eqref{eq:LP-full} can be specialized to obtain each of the other LP subclasses.

\noindent\textbf{Robust LP.} The \emph{robust LP} is obtained by setting $\tau = 0$, $\rho = 0$, and $\zeta_i = 0$ for all $i$, and is the strictest LP form available in \toolname. Under these restrictions, the cost function reduces to \[\min_ x \quad c^\top x\] and constraint~\eqref{eq:LP-constraint1} collapses to the \emph{hard} linear constraints $A(\delta_i) x~+~ b(\delta_i) \leq 0$.

\noindent\textbf{Relaxed LP.} The \emph{relaxed LP} is obtained by setting $\tau=0$, thereby reducing the objective function to:
\[\min_{ x,\zeta_i} \quad c^\top x~+~\rho\sum_{i=1}^{N} \zeta_i,\] while leaving the constraints \eqref{eq:LP-constraint1},\eqref{eq:LP-constraint2} unchanged.

\noindent\textbf{Regularized LP.} The \emph{regularized LP} corresponds to set $\rho = 0$ and $\zeta_i = 0$ for all $i$. The cost function is given as \[\min_ x \quad c^\top x~+~\tau\Vert  x-\bar{ x}\Vert_p,\] while scenario constraints \eqref{eq:LP-constraint1} become $A(\delta_i) x~+~ b(\delta_i) \leq 0$.

\noindent When regularization is present, different choices of norm $p$ are supported by \toolname; the field accepts any order $p\geq 1$ (\emph{e.g.}, $p=1$ or $p=2$), \texttt{inf} for the $\infty$-norm, or \texttt{fro} for the Frobenius norm, which for the vector $x-\bar{x}$ coincides with $p=2$.

To illustrate the use of scenario optimization for linear programs, we present a simple yet instructive example.

\begin{figure}
\begin{center}     
	\begin{tikzpicture}
		\draw[<->] (-5,0) -- (5,0) node[right] {};
		
		\foreach \x in {-4,-2,-1,1,3} {
			\draw[red, thick] (\x,0) node[below] {} -- ++(0,0.2);
			\node[red] at (\x,0.4) {\Huge $\times$};
		}
		
		\def\center{-0.5}
		\def\halfwidth{3.5}
		
		\draw[thick] (\center-\halfwidth,1) -- (\center+\halfwidth,1);
		\draw[thick] (\center-\halfwidth,0.9) -- (\center-\halfwidth,1.1);
		\draw[thick] (\center+\halfwidth,0.9) -- (\center+\halfwidth,1.1);
		
		\node[above] at (\center+0.3,0.3) {$o^*$};
		\node[above] at (\center+\halfwidth/2,1) {$w^*$};
		
		\draw[dashed] (\center,0) -- (\center,1);
	\end{tikzpicture}
\end{center} 
 \caption{Robust LP: Smallest enclosing interval example.}
\label{fig:Robust LP}
\end{figure}
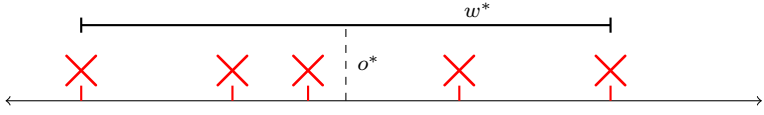

\begin{example}[Smallest Enclosing Interval.]
\label{ex:LP}
   \noindent Consider the problem of determining the smallest interval that contains a collection of sampled points $\delta_i$ from a one-dimensional space. This task can be naturally formulated as a \emph{robust} LP, where the decision variable consists of the center $o^\ast$ and half-width $w^\ast$ of the interval (cf. Fig.~\ref{fig:Robust LP}), such that $x^\ast_N = \begin{bmatrix} o^\ast & w^\ast \end{bmatrix}^\top$\!\!.
   
  \noindent In this formulation, the robust LP of the form in $($\ref{eq:LP-full}$)$ is specified by the matrices
    \begin{align}\label{new}
        c = \begin{bmatrix}
            0 \\ 1
        \end{bmatrix}\!\!, \quad
        A(\delta_i) = \begin{bmatrix}
            -1 &&& -1 \\
            1 &&& -1
        \end{bmatrix}\!\!, \quad
        b(\delta_i) = \begin{bmatrix}
            \delta_i \\ -\delta_i
        \end{bmatrix}\!\!,
    \end{align}
    and the constraint $A(\delta_i)x + b(\delta_i) \leq 0$ is imposed for each sampled point (scenario) $\delta_i$. Since this is a robust LP, we set $\tau = 0$, $\rho = 0$, and $\zeta_i = 0$ for all $i$. The corresponding JSON program file is given in Listing~\ref{lst:json-lp} in Appendix~\ref{appendix:json}. \hfill $\square$
\end{example} 

\toolname~supports LP scenario optimization through the following function in Listing 1, with its parameters described in detail in Appendix \ref{appendix:imp-solve_lp}. This function solves the LP defined in \eqref{eq:LP-full} and returns the optimal solution together with the information required to analyze the associated risk, including the complexity and the number of scenarios $N$. For diagnostic and analysis purposes, the full set of constraints is returned, along with an indicator specifying whether any degeneracy was detected during the solution process (the full solution can also be downloaded).\footnote{Each Listing shows python code, the python syntax preassigns values to the optional function variables equivalent to omitting those parameters), but all variables can be assigned by a user in practice. See Appendix~\ref{appendix:imp} for further details.}

\begin{lstlisting}[language=Python, caption=Function to solve LP.]
def solve_lp(deltas, A_d, b_d, G, h, c, tau=0.0, x_ref=np.array([0.0]), rho=0.0, norm_type=2, solver=None, include_slack=None):
return x_out, zeta_out, cost_out, N, complexity, constraints, degeneracy
\end{lstlisting}

\subsection{Quadratic Programming (QP)}\label{sec:qp}

We consider the most general case of \emph{relaxed-regularized} QP as 
\newpage

\begin{resp}\vspace{-0.3cm}
\begin{subequations}\label{eq:QP-full}
\begin{align} 
\min_{ x,\zeta_i \geq 0}& \quad c^\top x~+~\frac{1}{2} x^\top Q x~+~\tau\Vert  x-\bar{ x}\Vert_p~+~\rho\sum_{i=1}^{N} \zeta_i  \label{eq:QP-objective} \\
    \text{subject to:}& \quad A(\delta_i) x~+~ b(\delta_i) \leq \zeta_i, \quad i = 1,\ldots, N \label{eq:QP-constraint1}
    \\
    & \quad G x~+~ h \leq 0, \label{eq:QP-constraint2}
\end{align}
\end{subequations}
where $\tau \geq 0$ and $\rho\geq0$. 
\end{resp}

QPs are so named as their objective includes the quadratic term $\tfrac{1}{2} x^\top Q x$, where $Q \in \mathbb{R}^{d \times d}$ is a symmetric positive-semidefinite matrix, besides the linear term $ c^\top x$, where $c \in \mathbb{R}^d$. A QP therefore minimizes a convex quadratic function over a polyhedral feasible set. For arbitrary numbers of hard constraints $\mathfrak{n} > 0$ and scenario-dependent constraints $\mathfrak{m} > 0$, the matrix $G \in \mathbb{R}^{\mathfrak{n} \times d}$ and vector $h \in \mathbb{R}^{\mathfrak{n}}$ encode the hard linear constraints, while the matrix-valued function $A(\cdot) \in \mathbb{R}^{\mathfrak{m} \times d}$ and vector-valued function $b(\cdot) \in \mathbb{R}^{\mathfrak{m}}$ encode the linear constraints associated with each scenario $\delta_i$.

In the following, we describe how the robust–relaxed–regularized QP in \eqref{eq:QP-full} can be specialized to recover each of the QP subclasses.

\noindent\textbf{Robust QP.} The \emph{robust} QP in \toolname~is obtained by setting $\tau = 0$, $\rho = 0$, and all slack variables $\zeta_i = 0$. Under these restrictions, the cost function becomes \[\min_ x \quad c^\top x~+~\frac{1}{2} x^\top Q x,\] and \eqref{eq:QP-constraint1} reduces to the hard constraints $A(\delta_i) x~+~ b(\delta_i) \leq 0$.

\noindent\textbf{Relaxed QP.} The \emph{relaxed} QP is obtained by setting $\tau=0$, so resulting in the objective function \[\min_{ x,\zeta_i} \quad c^\top x~+~\frac{1}{2} x^\top Q x~+~\rho\sum_{i=1}^{N} \zeta_i,\] while leaving the constraints \eqref{eq:QP-constraint1},\eqref{eq:QP-constraint2} unchanged.

\noindent\textbf{Regularized QP.} The \emph{regularized} QP is obtained by setting $\rho=0$ and $\zeta_i=0$ for all $i$. The objective function becomes \[\min_ x \quad c^\top x~+~\frac{1}{2} x^\top Q x~+~\tau\Vert  x-\bar{ x}\Vert_p,\] while constraints \eqref{eq:QP-constraint1} reduce to the hard constraints $A(\delta_i) x~+~ b(\delta_i) \leq 0$.

As in the LP case, different norms $\Vert \cdot\Vert_p$ are supported by \toolname when regularization is present.

We motivate the use of scenario optimization for quadratic programs with a common machine-learning classification problem, as illustrated in the following example.

\begin{figure}
	\begin{center}
		\includegraphics[width=0.5\textwidth]{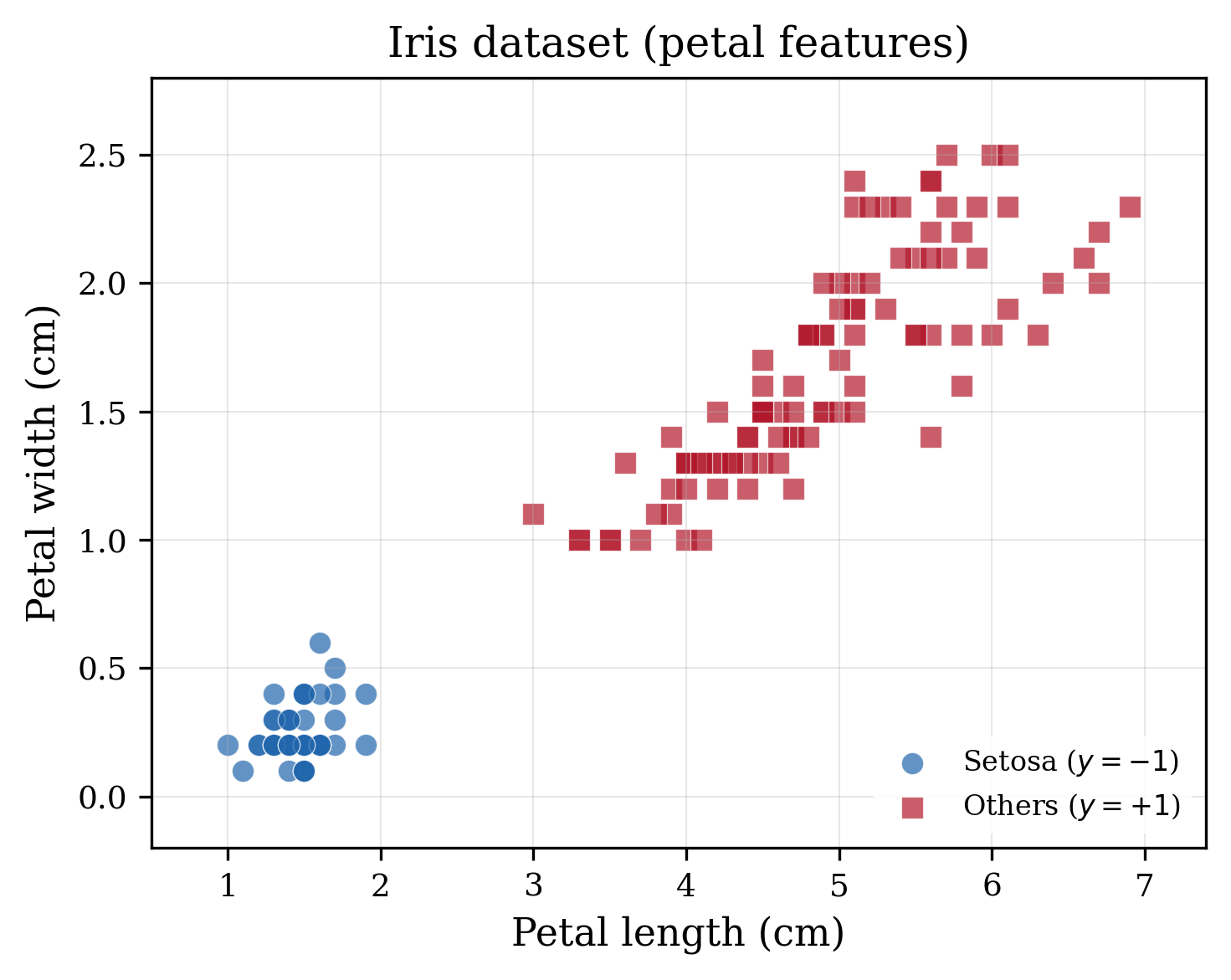}
	\end{center}
	\caption{Robust QP: Support vector machine example. \vspace{-0.3cm}}
	\label{fig:Robust QP}
\end{figure}

\begin{example}[Support Vector Machine]\label{ex:QP}
Consider the problem of binary classification, where one seeks a hyperplane that maximizes the separation margin between two labeled classes of data points. We sample points from a $2$-dimensional space, where each point is assigned either a blue or a red label (cf. Fig.~\ref{fig:Robust QP}), and the aim is to seek the optimal separating hyperplane that maximizes the margin between the two classes. This can be formulated as a robust QP, where the decision variable encodes the hyperplane $\mathbf{w}^\top x + h = 0$, with gradient $\mathbf{w} = \begin{bmatrix} \mathbf{w}_1 && \mathbf{w}_2 \end{bmatrix}^\top$ and intercept $h$. The scenario optimization solution can therefore be written as	 $x^\ast_N = \begin{bmatrix} \mathbf{w}^\ast_1 && \mathbf{w}^\ast_2 && h^\ast \end{bmatrix}^\top\!\!.$

\noindent A scenario $\delta_i\in\mathbb{R}^3$ consists of the triple $(\delta_i(1), \delta_i(2), \delta_i(3))$, where $(\delta_i(1), \delta_i(2))$ is the coordinate of the sample and $\delta_i(3)$ is its class label. Then the robust QP in~\eqref{eq:QP-full} is described by the following matrices:
    \begin{align}\label{new2}
        c = \begin{bmatrix}
            0 \\ 0\\ 0
        \end{bmatrix}\!\!, \quad
        Q = \begin{bmatrix}
            1 && 0 && 0 \\
            0 && 1 && 0 \\
            0 && 0 && 0
        \end{bmatrix}\!\!, \quad
        A(\delta_i) = \begin{bmatrix}
            -\delta_i(1)\delta_i(3) \\ -\delta_i(2)\delta_i(3) \\ -\delta_i(3)
        \end{bmatrix}^\top\!\!\!\!\!\!, \quad
        b(\delta_i) = 1,
    \end{align}
where the constraint $A(\delta_i)x + b(\delta_i) \leq 0$ is enforced for each scenario $\delta_i$.

    \noindent\textbf{Relaxed QP.} If the data are not linearly separable, the relaxed QP allows one to find a hyperplane that best accommodates the samples by introducing slack variables to handle misclassified points. The JSON program file of the robust QP is given in Listing~\ref{lst:json-qp} in Appendix~\ref{appendix:json}. \hfill $\square$
\end{example}

\toolname~supports QP scenario optimization through the following function  in Listing 2, with its parameters described in detail in Appendix~\ref{appendix:imp-solve_qp}. The function solves the QP in~\eqref{eq:QP-full} and returns the solution together with all information required for subsequent \emph{risk} analysis, including the complexity and the number of scenarios $N$. For completeness, the full set of constraints is also returned, along with a flag indicating whether any degeneracy was detected.

\begin{lstlisting}[caption=Function to solve QP.]
def solve_qp(deltas, A_d, b_d, G, h, c, Q, tau=0.0, x_ref=np.array([0.0]), rho=0.0, norm_type=2, solver=None, include_slack=None):
return x_out, zeta_out, cost_out, N, complexity, constraints, degeneracy
\end{lstlisting}

\subsection{Semidefinite Programming (SDP)}\label{sec:sdp}

Conic optimization is a structured subclass of convex optimization, and \emph{semidefinite programming} constitutes a fundamental instance of conic optimization, since the set of semidefinite matrices forms a convex cone. \toolname~supports SDPs in the \emph{inequality form SDP}, which naturally generalizes the inequality-based formulations used for LPs and QPs.

Several convex optimization problems can be reformulated or relaxed as an SDP, including LPs, QPs, quadratically-constrained QPs (QCQPs), second-order cone programs (SOCPs), linear matrix inequalities (LMIs), and least-squares problems. This makes \toolname~applicable to a wide range of problem classes.

We first consider the \emph{inequality form SDP}, in which no equality constraints are present, and both hard and soft constraints are expressed as LMIs. Let $x = [x_{1}, x_{2}, \ldots, x_{d}] \in \mathbb{R}^{d}$ denote the vector of decision variables. The \emph{robust–relaxed–regularized} SDP in \emph{inequality} form can then be written as 
\begin{resp}\vspace{-0.3cm}
\begin{subequations}\label{eq:SDP-full-ineq}
\begin{align} 
\min_{ x,\zeta_i \geq 0}& \quad c^\top x~+~\frac{1}{2} x^\top Q x~+~\tau\Vert  x-\bar{ x}\Vert_p~+~\rho\sum_{i=1}^{N} \zeta_i  \label{eq:SDP-objective-ineq} \\
    \text{subject to:}& \quad F_0(\delta_i)~+~ \sum_{j=1}^{d}  x_jF_j(\delta_i) \ \preceq \ \zeta_i \mathbb{I}_{\mathfrak{m}\times \mathfrak{m}}, \quad i = 1,\ldots, N \label{eq:SDP-constraint1-ineq}
    \\
    & \quad E_0~+~ \sum_{j=1}^{d}  x_jE_j \ \preceq \ 0,  \label{eq:SDP-constraint2-ineq}
\end{align}
\end{subequations}
where $\tau \geq 0$, $\rho\geq0$, and $\mathbb{I}_{\mathfrak{m}\times \mathfrak{m}}$ denotes the $\mathfrak{m}\times \mathfrak{m}$ identity matrix. 
\end{resp}
In~\eqref{eq:SDP-full-ineq}, $Q$ is a \emph{positive semidefinite symmetric matrix} in $\mathbb{R}^{d\times d}$ and $c\in\mathbb{R}^{d}$. $E_{0},~\ldots,~E_{d}$ are \emph{symmetric matrices} in $\mathbb{R}^{\mathfrak{n}\times \mathfrak{n}}$, these objects define the hard LMI constraint in~\eqref{eq:SDP-constraint2-ineq}. Likewise, the symmetric matrix-valued functions $F_{j}(\cdot)\in\mathbb{R}^{\mathfrak{m}\times \mathfrak{m}}$, for $j=0,\ldots,d$, generate the LMI constraints associated with each scenario $\delta_{i}\in\mathbb{R}^q$.

\noindent\textbf{Robust SDP.} The \emph{robust} SDP is supported by \toolname~and obtained by setting $\tau = 0$, $\rho = 0$, and all $\zeta_i = 0$, which yields the objective \[\min_ x \quad c^\top x~+~\frac{1}{2} x^\top Q x\] while \eqref{eq:SDP-constraint1-ineq} reduces to the hard LMI constraints $F_0(\delta_i)~+~ \sum_{j=1}^{d}  x_jF_j(\delta_i) \ \preceq \ 0$, applied for each scenario $\delta_i$.

\noindent\textbf{Relaxed SDP.} The \emph{relaxed SDP} corresponds to setting $\tau=0$ and results in the objective \[\min_{ x,\zeta_i} \quad c^\top x~+~\frac{1}{2} x^\top Q x~+~\rho\sum_{i=1}^{N} \zeta_i,\] while constraints \eqref{eq:SDP-constraint1-ineq},\eqref{eq:SDP-constraint2-ineq} remain the same.

\noindent\textbf{Regularized SDP.} The \emph{regularized SDP} is supported by \toolname, with its cost function taking the form \[\min_ x \quad c^\top x~+~\frac{1}{2} x^\top Q x~+~\tau\Vert  x-\bar{ x}\Vert_p,\] and constraints \eqref{eq:SDP-constraint1-ineq} becoming the hard LMI constraints $F_0(\delta_i)~+~ \sum_{j=1}^{d}  x_jF_j(\delta_i) \ \preceq \ 0$.

Similarly to the LP and QP cases, different norms $\Vert \cdot\Vert_p$ for SDP are supported by the tool when regularization is present.

To demonstrate the use of scenario optimization for SDP inequality form, we present the following illustrative example.

\begin{example}[SDP Inequality Form]
\label{ex:SDP1}

 Consider a linear parameter-varying (LPV) system 
$\dot{\xi} = A(\delta)\,\xi$ with
\begin{equation*}
A(\delta) = \underbrace{\begin{bmatrix} 0 & 1 \\ -2 & -1 \end{bmatrix}}_{A_0}
+ \;\delta\;\underbrace{\begin{bmatrix} 0 & 0 \\ 0.3 & 0.1 \end{bmatrix}}_{A_1},
\qquad \delta \in [-0.22,\; 1.0],
\end{equation*}
where $A_0$ is a stable damped oscillator and $A_1$ captures
parameter-varying uncertainty in the restoring and damping terms (see Figure~\ref{fig:LPV} for the state trajectories achieved for various fixed values of $\delta$ and initializations).
The goal is to find a common quadratic Lyapunov function
$V(x) = x^\top P\,x$ with $P \succ 0$ certifying stability across
all parameter values, i.e.\ $A(\delta)^\top P + P\,A(\delta) \preceq 0$.

This maps to the SDP inequality form~\eqref{eq:SDP-full-ineq} with
decision variable $x = [p_{11},\; p_{12},\; p_{22}]^\top \in \mathbb{R}^3$,
the upper-triangular entries of the $2\times 2$ symmetric matrix~$P$.
Let $\Pi_1 = \left[\begin{smallmatrix}1&0\\0&0\end{smallmatrix}\right]$,
$\Pi_2 = \left[\begin{smallmatrix}0&1\\1&0\end{smallmatrix}\right]$,
$\Pi_3 = \left[\begin{smallmatrix}0&0\\0&1\end{smallmatrix}\right]$
be the basis matrices for the symmetric $2\times 2$ space, so that
$P = \sum_{j=1}^{3} x_j \Pi_j$.
The scenario-dependent constraint
matrices in~\eqref{eq:SDP-constraint1-ineq} are
\begin{equation*}
F_0(\delta_i) = 0, \qquad
F_j(\delta_i) = A(\delta_i)^\top \Pi_j + \Pi_j\, A(\delta_i),
\quad j = 1,2,3,
\end{equation*}
\begin{figure}
    \centering
    \includegraphics[width=0.5\linewidth]{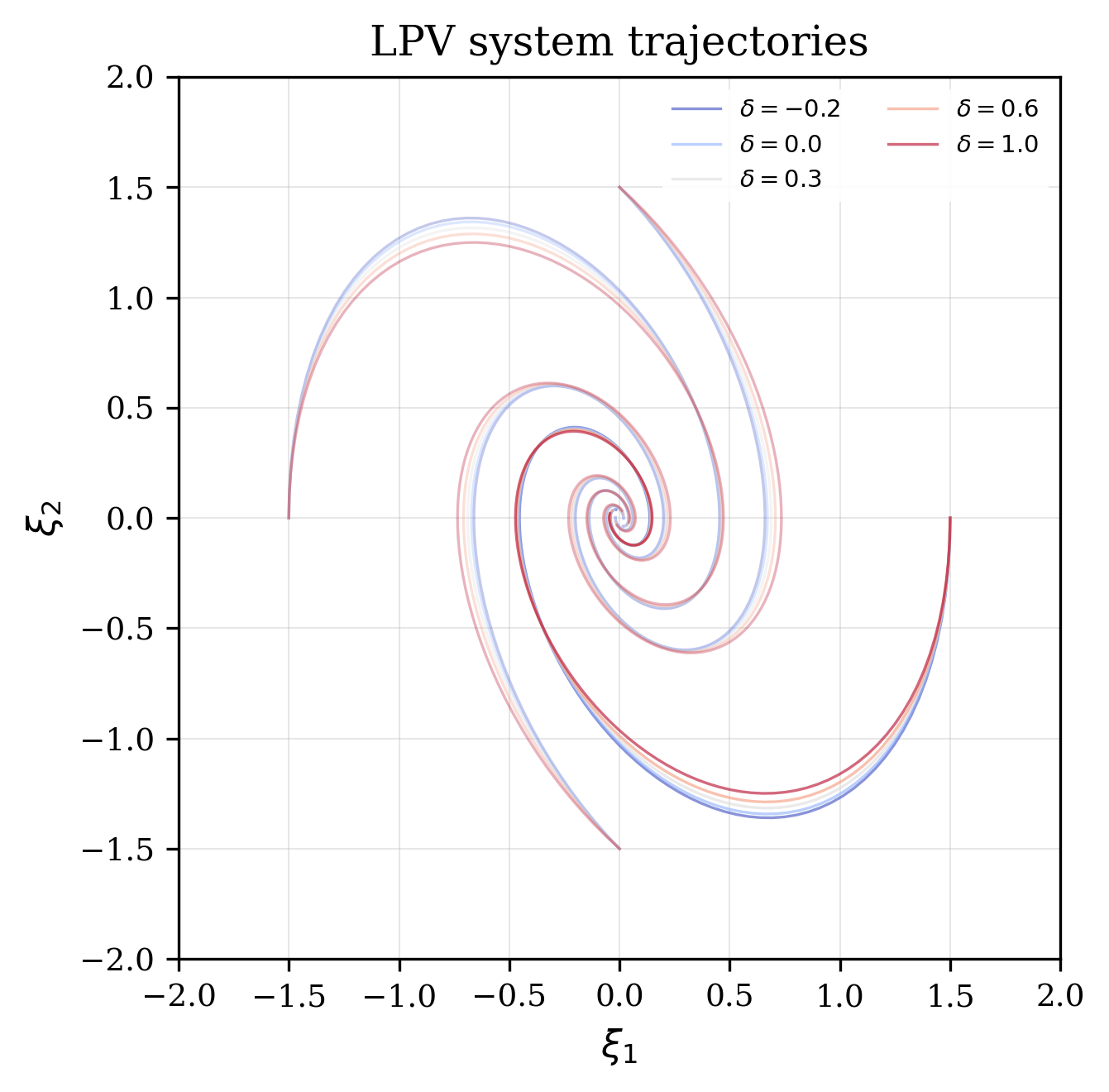}
    \caption{Robust SDP: LPV System Example.}
    \label{fig:LPV}
\end{figure}
encoding the Lyapunov inequality
$A(\delta_i)^\top P + P\,A(\delta_i) \preceq \zeta_i I$.
The hard constraint in~\eqref{eq:SDP-constraint2-ineq} enforces
$P \succ 0$ via
\begin{equation*}
E_0 = 0, \qquad E_j = -\Pi_j, \quad j = 1,2,3.
\end{equation*}
The objective uses $c = [-1,\; 0,\; -1]^\top$ and
$Q = 0.1\,\mathbb{I}_{3\times3}$, while we set $\tau = 0$ and $\rho =1$. This amounts to maximizing $\mathrm{trace}(P)$
with light quadratic regularization and constraint relaxation. The corresponding JSON program file is given in Listing~\ref{lst:json-sdp} in Appendix~\ref{appendix:json}.
 \hfill $\square$
\end{example}

\toolname~provides SDP scenario optimization via a dedicated function in Listing 3, with parameters described in Appendix~\ref{appendix:imp-solve_sdp1}. This function solves the SDP specified in~\eqref{eq:SDP-full-ineq} and returns the optimal solution together with the quantities needed for \emph{risk} assessment, including the complexity and the number of scenarios $N$. It also outputs the complete set of constraints and a diagnostic flag indicating whether degeneracy was detected.

\begin{lstlisting}[caption=Function to solve SDP (Inequality Form)]
def solve_sdp(deltas, F_d, E, c, Q, tau=0.0, x_ref=np.array([0.0]), rho=0.0, norm_type=2, solver=None, include_slack=None):
return x_out, zeta_out, cost_out, N, complexity, constraints, degeneracy
\end{lstlisting}
\vspace{-0.3cm}

\section{\toolname: Implementation Details}\label{sec:implementation}

In this section, we outline the algorithms used to evaluate the \emph{risk} and \emph{complexity} of a scenario-based convex optimization problem, in accordance with the theory presented in Section~\ref{sec:underpinning-theory}. As established by Theorem~\ref{thm:riskcert} and Theorem~\ref{thm:riskcert-nondeg}, the risk associated with a scenario convex optimization solution depends on three quantities: the number of scenarios $N$, the complexity $s^*_N$, and the user-selected confidence level $\beta$. Among these, the computation of $s^*_N$ is the most nontrivial task, as it requires identifying a support list, which comprises all constraints violated at the optimum together with an irreducible subset of the active constraints that, jointly, allow reconstruction of the solution (cf.\ Definition~\ref{def:support-list-and-complexity}). We begin by describing the computation of the risk, followed by the procedures used to determine the complexity and other auxiliary quantities. While $s^*_N$ is the true complexity, in the implementation we use $k$ for the calculated minimal cardinality, where $k = s^*_N$ in non-degenerate cases but $k\geq s^*_N$ for degeneracy (cf. Remark~\ref{rem-degen}).

\vspace{-0.2cm}
\subsection{Risk}

\begin{lstlisting}[label=lst:risk, caption=Function to compute risk bounds.]
def quantify_risk(k,N,beta):
return epsL, epsU
\end{lstlisting}

The ultimate aim of \toolname~is to compute the risk bounds associated with a convex optimization problem solved under the scenario approach. Algorithm~\ref{alg:risk} summarizes this procedure, assuming that the number of scenarios $N$, the complexity $k$, and the user-defined confidence level $\beta$ have already been obtained. The pseudocode follows the implementation strategy of~\cite[Appendix B]{campi2023compression}, originally presented in \textsf{MATLAB}. The algorithm evaluates the bounds stated in Theorem~\ref{thm:riskcert-nondeg}, which hold under the non-degeneracy assumption and rely on the incomplete beta function \textsf{betainc}. The corresponding Python implementation used by \toolname~is provided in Listing~\ref{lst:risk}.

\begin{algorithm}[t!]
\caption{Quantifying risk bounds $\underline{\epsilon}~\text{and}~\overline{\epsilon}$}
\label{alg:risk}
\begin{algorithmic}[1]
\Require Complexity $k$ and number of scenarios $N$ with \(0 \le k \le N\), confidence \(\beta\in(0,1)\)
\Ensure Lower bound \(\underline{\epsilon}\) and upper bound \(\overline{\epsilon}\)
\State \(\;t_{\min}\gets 0.0\) \Comment{Compute lower bound \(\underline{\epsilon}\)}
\State \(\;t_{\max}\gets k/N\)
\State \(\;\text{threshold}\gets 10^{-10}\)
\While{\((t_{\max}-t_{\min}) > \text{threshold}\)}
  \State \(t \gets (t_{\min}+t_{\max})/2\)
  \State \(L \gets \dfrac{\beta}{3}\,\textsf{betainc}(k+1,\;N-k,\;t)\;+\;\dfrac{\beta}{6}\,\textsf{betainc}(k+1,\;4N+1-k,\;t)\)
  \State \(R \gets \bigl(1+\dfrac{\beta}{6N}\bigr)\,t\,N\!\bigl(\textsf{betainc}(k,\;N-k+1,\;t)-\textsf{betainc}(k+1,\;N-k,\;t)\bigr)\)
  \If{\(L > R\)}
    \State \(t_{\min}\gets t\)
  \Else
    \State \(t_{\max}\gets t\)
  \EndIf
\EndWhile
\State \(\underline{\epsilon} \gets t_{\min}\) \Comment{when \(k=0\) the loop is skipped and \(\underline{\epsilon}=0\)}
\If{\(k = N\)} \Comment{Compute upper bound \(\overline{\epsilon}\)}
  \State \(\overline{\epsilon} \gets 1.0\)
\Else
  \State \(t_{\min}\gets k/N\)
  \State \(t_{\max}\gets 1.0\)
  \While{\((t_{\max}-t_{\min}) > \text{threshold}\)}
    \State \(t \gets (t_{\min}+t_{\max})/2\)
    \State \(L \gets \bigl(\tfrac{\beta}{2}-\tfrac{\beta}{6}\bigr)\,\textsf{betainc}(k+1,\;N-k,\;t)\;+\;\dfrac{\beta}{6}\,\textsf{betainc}(k+1,\;4N+1-k,\;t)\)
    \State \(R \gets \bigl(1+\tfrac{\beta}{6N}\bigr)\,t\,N\!\bigl(\textsf{betainc}(k,\;N-k+1,\;t)-\textsf{betainc}(k+1,\;N-k,\;t)\bigr)\)
    \If{\(L > R\)}
      \State \(t_{\max}\gets t\)
    \Else
      \State \(t_{\min}\gets t\)
    \EndIf
  \EndWhile
  \State \(\overline{\epsilon} \gets t_{\max}\)
\EndIf

\State \Return \([\underline{\epsilon},\;\overline{\epsilon}]\)
\end{algorithmic}
\end{algorithm}

\begin{remark}
When the assumption of non-degeneracy, as defined in Definition~\ref{def:non-degeneracy}, is violated, the validity of the two-sided bound in Theorem~\ref{thm:riskcert-nondeg} can no longer be guaranteed. In this case, only the \emph{upper} bound on the risk remains theoretically sound, whereas the lower bound may fail to hold (see Theorem~\ref{thm:riskcert}). Consequently, the risk certificate becomes one-sided, reflecting an additional risk reduction introduced by degeneracy. We recall that, although the toolbox issues a warning when the specified optimization problem instance is detected to be degenerate, the use of the lower bound remains at the discretion of the user, who is responsible for verifying the validity of the non-degeneracy property.\hfill $\square$
\end{remark}

\subsection{Support List Validation}

In Definition \ref{def:support-list-and-complexity}, the support list is characterized by two conditions: (i) the solution obtained when only the candidate constraints are enforced coincides with the solution obtained using the full set of constraints, and (ii) the candidate list is irreducible, meaning that removing any of its elements alters the solution. Given any candidate support list, that is any subset of the constraints, we can verify property (i) using Algorithm \ref{alg:support-list}. This algorithm summarizes the procedure implemented in Listing \ref{lst:test_support}, and its operation is independent of the particular solver employed.

\begin{lstlisting}[label=lst:test_support, caption=Function to validate candidate support lists.]
def test_support(objective, support, non_risk, ref, solver=None):
return bool
\end{lstlisting}

Algorithm~\ref{alg:support-list} takes a solved problem \textsf{prob} and its objective \textsf{obj}. The set $A$ is a subset of the scenario constraints, i.e.\ a \emph{candidate} support list. \textsf{TEST\_SUPPORT} re-solves the problem for candidate constraints $A$ together with the non-scenario constraints $R$\footnote{$R$ are the hard constraints; enforced always and not counted in the complexity.}, and returns \texttt{True} if and only if the optimal value is reproduced. Deciding support list membership using the optimal value keeps the test well posed when the optimizer is non-unique: removing a genuinely-supporting scenario strictly changes the optimum, whereas a redundant one would leave it unchanged. A \texttt{True} result establishes the first requirement of Definition~\ref{def:support-list-and-complexity}; irreducibility is checked separately.

\begin{algorithm}[t!]
	\caption{Validating candidate support list}
	\label{alg:support-list}
	\begin{algorithmic}[1]
		\Require Candidate constraints $A$, solved optimization problem $\textsf{prob}$,  objective $\textsf{obj}$, non-scenario constraints $R$

		\Function{TEST\_SUPPORT}{$\textsf{prob},\ \textsf{obj},\ A,\ R$}
		\State Solve problem $\textsf{prob}' \gets \textsf{Problem}(\textsf{obj},\ A\,\cup\, R)$
		\State \Return True iff the optimal values match: $\textsf{prob.value} \approx \textsf{prob}'.\textsf{value}$
		\EndFunction
	\end{algorithmic}
\end{algorithm}

\newpage 

\subsection{Complexity Computation}
\begin{lstlisting}[label=lst:complexity, caption=Function to determine complexity.]
def get_support(constraints, non_risk_constraints, prob, objective, solver=None, threshold=1e-8, sol_tol=1e-6):
return complexity, support, degeneracy
\end{lstlisting}

The computation of complexity is the most intricate component of \toolname’s implementation. Although the full procedure is realized as a single function in Listing~\ref{lst:complexity}, for clarity we describe its constituent steps separately.

Algorithm~\ref{alg:complexity} computes the support list. Lines 3--7 screen the scenario constraints by dual value, adding each constraint with a nonzero dual to the candidate list $A$. Such a constraint is either \emph{active} (binding) at the optimum or \emph{violated}: under relaxation, stationarity of the Lagrangian in the slack $\zeta_i$ forces the dual of a violated constraint to sum to $\rho>0$, so the check captures both. Line 8 tests, through \textsc{test\_support}, whether $A$ alone reproduces the optimum. If it does, the screened constraints $A$ are the \emph{support scenarios}, and line 9 greedily prunes them via \textsc{prune} (Algorithm~\ref{alg:resolve-degeneracy}). Non-degeneracy is equivalent to the support scenarios forming a support list (Property~\ref{def:non-degeneracy}), so this prune doubles as a degeneracy test: the flag is raised exactly when \textsc{prune} removes a constraint that is \emph{active or violated} at the optimum, since the support scenarios are then reducible and more than one support list exists. The removal of a \emph{spuriously screened} constraint --- one that is strictly slack at the optimum, yet carries a small nonzero dual --- is instead ignored (see Remark~\ref{rem-spurious}). If only spurious screens are removed, the remaining $A$ is an irreducible support list and the instance is certified non-degenerate. In the degenerate case the pruned list is irreducible but not guaranteed to be of minimum cardinality. If \textsc{test\_support} is \texttt{False}, the dual screening was inaccurate (\emph{i.e.} from numerical errors in the solver). To be conservative the degeneracy flag is raised and \textsc{prune} restarts from the full scenario list $C$, which always recovers a valid support list. The complexity is the cardinality of the returned support list, $k = |A|$.

\begin{algorithm}[t!]
\caption{Computing the support list and complexity}
\label{alg:complexity}
\begin{algorithmic}[1]
\Require Scenario constraints $C$, non-scenario constraints $R$, problem $\textsf{prob}$, objective $\textsf{obj}$
\Ensure Complexity $k$, support list $A$, flag for degeneracy $d$
  \State $d \gets \text{False}$
  \State $A \gets [\,]$ \Comment{candidate support list}
    \ForAll{constraint $c$ in $C$}
    \If{$\max(|c.\text{dual\_value}|) > \text{threshold}$} \Comment{default threshold $10^{-8}$}
      \State append $c$ to $A$ \Comment{$c$ is violated or active}
    \EndIf
  \EndFor
  \If{\Call{TEST\_SUPPORT}{$\textsf{prob},\ \textsf{obj},\ A,\ R$}} \Comment{screened set reproduces the optimum}
    \State $A_{\mathrm{s}} \gets A$ \Comment{the screened candidates}
    \State $A \gets$ \Call{PRUNE}{$A,\ \textsf{prob},\ \textsf{obj},\ R$}
    \If{any $c \in A_{\mathrm{s}} \setminus A$ is active or violated at the optimum}
      \State $d \gets \text{True}$ \Comment{support scenarios reducible $\Rightarrow$ degenerate}
    \EndIf
  \Else \Comment{inaccurate duals: conservative}
    \State $d \gets \text{True}$
    \State $A \gets$ \Call{PRUNE}{$C,\ \textsf{prob},\ \textsf{obj},\ R$} \Comment{restart from the full list}
 \EndIf
  \State $k \gets |A|$ \Comment{cardinality of the support list}
  \State \Return $(k,\ A,\ d)$
\end{algorithmic}
\end{algorithm}

The greedy reduction \textsc{prune} (Algorithm~\ref{alg:resolve-degeneracy}) turns a candidate list into an irreducible support list. It is used in both branches of Algorithm~\ref{alg:complexity}: on the small screened set in the normal case, and on the full scenario list when the screening fails to reproduce the optimum. It checks each constraint in turn and removes it (via \lstinline|TEST_SUPPORT|) whenever doing so leaves the optimum unchanged. The returned list is irreducible but not guaranteed to be of minimum cardinality. The pass is $\mathcal{O}(n)$ in the number of constraints: convexity guarantees that a constraint which is not of support cannot become of support as others are removed, so no constraint need be rechecked.

\begin{remark}
\label{rem-degen}
In cases of degeneracy, it is possible the complexity value calculated is suboptimal, \emph{i.e.}, for multiple support lists, the smallest support list is not necessarily identified. This sub-optimality is acceptable in practice, as a larger complexity leads to a more conservative (higher) upper bound on the risk. Consequently, the true complexity would yield a strictly tighter, and therefore less risky certificate than the one obtained from a suboptimal result.\hfill $\square$
\end{remark}

\begin{remark}
\label{rem-spurious}
Some solvers (\emph{e.g.} interior point) may return small nonzero duals on constraints that are neither active nor violated, so the dual screen of lines 3--7 may retain such \emph{spuriously screened} constraints. Left unaddressed, their removal by \textsc{prune} would raise the degeneracy flag even in non-degenerate instances, which is overly conservative. \toolname~resolves this by recording every screened constraint's activity margin at the optimum (before any re-solve overwrites the solution). A removed constraint whose margin clearly exceeds the tolerance is recognized as a spurious screen and discounted by the degeneracy test. When the margin is ambiguous, the constraint is conservatively treated as active, so the test still leans towards flagging degeneracy.\hfill $\square$
\end{remark}

\begin{algorithm}[t!]
\caption{\textsc{prune}: greedy support-list reduction}
\label{alg:resolve-degeneracy}
    \begin{algorithmic}[1]
        \Require Candidate constraints $A$, problem $\textsf{prob}$, objective $\textsf{obj}$, non-scenario constraints $R$
        \Ensure Irreducible support list $A$
        \State i = 0
        \Repeat
          \State $A_{\text{temp}} \gets A[:,i] + A[i+1, :]$ \Comment{Temporary list without constraint in position $i$}
          \If{\Call{TEST\_SUPPORT}{$\textsf{prob},\ \textsf{obj},\ A_{\text{temp}},\ R$}}
            \State $A.\textsf{pop}(i)$ \Comment{Remove constraint in position $i$}
        \Else
            \State $i = i+1$ \Comment{increment $i$}
          \EndIf
      \Until{i == |A|}
      \State \Return $A$
    \end{algorithmic}
\end{algorithm}
\vspace{-0.8cm}
\subsection{Solvers}

\toolname~builds upon CVXPY~\cite{diamond2016cvxpy}, a well-established convex optimization framework that interfaces with a wide range of solvers\footnote{\url{https://www.cvxpy.org/tutorial/solvers/index.html}}
. By leveraging CVXPY for scenario optimization, \toolname~can access \emph{$27$ solvers} (at the time of writing), including both \emph{open-source} options (\emph{e.g.}, \textsf{CLARABEL}~\cite{clarabel}, \textsf{CVXOPT}~\cite{cvxopt}, \textsf{SCS}~\cite{scs}) and \emph{commercial} solvers (\emph{e.g.}, \textsf{MOSEK}~\cite{mosek}, \textsf{GUROBI}~\cite{gurobi}). The function \lstinline|get_solvers|, shown in Listing~\ref{lst:solver}, automatically detects and returns all compatible solvers available on the host machine\footnote{The server contains the following solvers \texttt{CLARABEL,CVXOPT,DAQP,ECOS,ECOS\_BB,GLPK,}\\\texttt{GLPK\_MI,HIGHS,MOSEK,OSQP,PIQP,PROXQP,SCIPY,SCS}. Further solvers can be used by installing \toolname~locally.}. This list is then used to populate the dynamic solver-selection menu in the web interface, as discussed in Section~\ref{sec:gui}.

\begin{lstlisting}[label=lst:solver, caption=Function to get installed solvers.]
def get_solvers():
return solver_dict
\end{lstlisting}

The returned \lstinline|solver_dict| is a mapping from each installed solver's name to the list of program types it supports (`\lstinline{LP}', `\lstinline{QP}', and `\lstinline{SDP}'), which is used to filter the dropdown in the web interface based on the active program tab. Commercial solvers requiring a license are handled by a dedicated subprocess entry point (\texttt{mosek\_solve.py}) that sets the license environment variable before importing CVXPY, allowing concurrent users to each upload their own license without conflicts; the license is held only in RAM and never persists on disk.

\subsection{Parallelism}
With modern computational resources, one might expect significant gains from distributing computations across multiple processors or GPUs. However, for convex optimization it is well established that parallelization, particularly GPU-based acceleration, offers limited practical benefit~\cite{balkanski2018parallelization,nemirovski1994parallel}. The current version of \toolname~therefore runs sequentially. In Algorithm~\ref{alg:complexity}, the screening loop in lines 3--7 only reads dual values and is inexpensive, while \textsc{prune} (Algorithm~\ref{alg:resolve-degeneracy}) admits no meaningful parallelism: although it contains a loop, each iteration depends on the outcome of the previous one, since the decision to remove a constraint affects all subsequent checks. The clearest opportunity for future versions is to solve the programs for different values of $\rho$ and $\tau$ in parallel, since they are independent. Such an implementation must give every parallel solve its own copy of the optimization problem: if parallel solves share optimization variables, the values written by one solve can overwrite those of another, the support list may then be identified incorrectly, and the complexity $k$ can be under-counted, which would make the risk certificate too optimistic.

\section{Web Interface}\label{sec:gui}

\begin{figure}[t!]
	\centering
	\includegraphics[width=\textwidth]{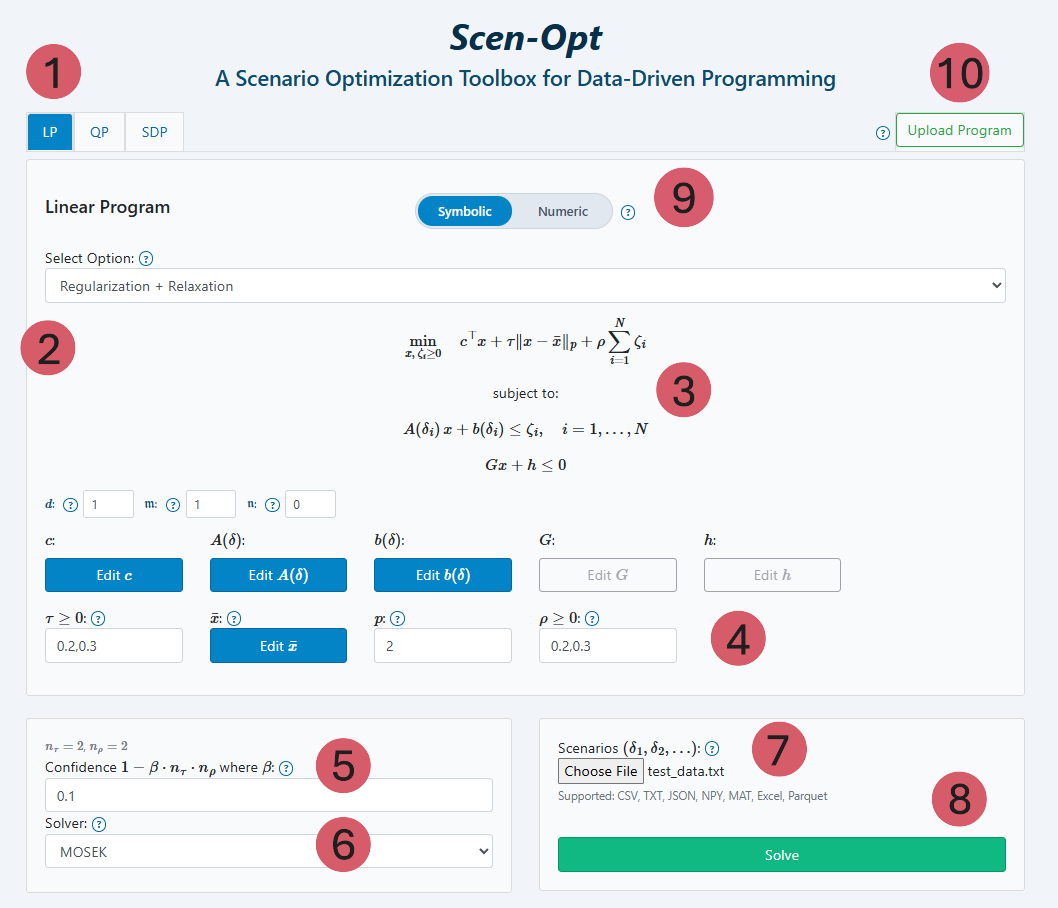}
	\caption{Web interface of \toolname~illustrating the relaxed and regularized LP.}
	\label{fig:web-interface}
\end{figure}

\toolname~is accompanied by an interactive web interface designed to make the toolbox highly accessible and user-friendly. Figure \ref{fig:web-interface} illustrates the interface for a relaxed and regularized LP, with annotated labels <1>–<10>. In the interface, <1> denotes the tab system that allows navigation between the different convex optimization programs supported by \toolname. Each tab displays the corresponding optimization scheme, where <2> provides a dropdown menu for selecting whether regularization and/or relaxation are included or whether the basic robust formulation is used. The mathematical description of the chosen program is shown in <3>, with variable names aligned to the input fields and buttons in <4>. A row of dimension fields ($d$, $\mathfrak{m}$, $\mathfrak{n}$, and, for SDP, the LMI size) sets the problem sizes; changing any of them automatically reshapes the matrices already entered, zero-padding or truncating as needed.

\noindent Pressing any blue button opens a modal dialog for entering or uploading matrices or vectors.
When relaxation or regularization is not applicable, the corresponding buttons and fields automatically disappear, ensuring a clean and adaptive interface. A help icon ($?$) beside each field opens a short tooltip explaining the corresponding input. Each matrix modal provides both a cell-by-cell grid editor and a JSON-paste box, with live validation.  The scenario-dependent data $A(\delta)$, $b(\delta)$, and $F_j(\delta)$ can be entered either symbolically or numerically --- both input modes, and their corresponding scenario file format, are detailed in the \emph{Symbolic and numeric input modes} paragraph below. The confidence parameter can be specified in <5>; when several values of $\rho$ or $\tau$ are swept (see below), this field instead reports the overall confidence $1-\beta\,n_\tau n_\rho$ obtained from a union bound over the $n_\tau n_\rho$ solved instances, with $\beta$ the user-entered per-instance value, while the scenario data file\footnote{Supported formats include \lstinline|.csv|,\lstinline|.json|, \lstinline|.mat|, \lstinline|.txt|, \lstinline|.tsv|, \lstinline|.xlsx|, \lstinline|.xls|, \lstinline|.npy|,\lstinline|.npz|, and \lstinline|.parquet|} is uploaded via <7>, whose row-wise format depends on the active input mode (see below). As \toolname~supports up to 27 solvers, the dynamically populated solver dropdown <6> enables selection among the solvers available on the host machine, automatically filtered to those compatible with the active program type (LP, QP, or SDP). Once all parameters are set, the optimization problem can be executed by clicking Solve at <8>. Label <9> denotes the input-mode toggle, which switches between the \emph{Symbolic} and \emph{Numeric} ways of entering the scenario-dependent constraints (detailed below). Label <10> shows the button used to do a full upload of a program in \lstinline|.json| or \lstinline|.mat| formats, with \toolname~automatically detecting the suitable program choice. The program can be supplied either as a file or by pasting its \lstinline|.json| text, and the dialog provides a format reference listing the required fields for each program type; in \lstinline|.mat| files the symbolic matrices are passed as cell arrays of strings.

\noindent\textbf{Symbolic and numeric input modes.} Selected via the toggle <9>,
\toolname~offers two complementary ways of providing the scenario-dependent
constraints, in both cases uploaded through <7>. Throughout, the uncertainty
$\delta$ is a $q$-dimensional vector whose components are addressed as
\lstinline|delta[0]|, \ldots, \lstinline|delta[q-1]|. In \emph{symbolic} mode,
each entry of $A(\delta)$ and $b(\delta)$ (LP/QP), or of $F_j(\delta)$ (SDP), is
entered in the matrix editors of <4> as a closed-form expression in
\lstinline|delta[0]|, \ldots, \lstinline|delta[q-1]|, \emph{e.g.},
\lstinline|2*delta[0]-delta[3]|. The file uploaded via <7> then contains the
scenarios themselves, as an $N \times q$ table whose $i$-th row lists the
components $\delta_i[0], \ldots, \delta_i[q-1]$ of scenario $\delta_i$; each row
is substituted into the symbolic expressions to generate the $i$-th scenario
constraint. This mode is convenient whenever the dependence on $\delta$ is
available in closed form. In \emph{numeric} mode, the scenario constraints are
instead supplied already evaluated: the scenarios are given directly as the
instances $A_i$, $b_i$ (or $F_{j,i}$), so nothing is specified in the matrix
editors (they are replaced by simple dimension fields), and the $i$-th row of
the file uploaded via <7> lists the row-wise flattened numerical blocks
$[\,A_i \mid b_i\,]$ for LP/QP or $[\,F_{0,i} \mid F_{1,i} \mid \cdots \mid F_{d,i}\,]$
for SDP. Numeric mode is useful when the constraint data are produced externally (\emph{e.g.}, precomputed from raw measurements or another processing pipeline) and do not admit a convenient symbolic form. Both modes solve the same underlying scenario program and return identical results. In numeric mode, \toolname~verifies that each uploaded row has the length implied by the declared dimensions, reporting a descriptive error otherwise.

\begin{remark}
    The Mosek~\cite{mosek} solver requires a license, which is not provided for the user when using the server. Instead, when trying to use Mosek, a user can upload their valid license to be checked and used. The license is only temporarily stored. For convenience, an uploaded license is cached in the browser's session storage so that subsequent solves skip the upload prompt; a cache indicator with a \emph{Clear} button appears beneath the solver menu, while users with a locally installed Mosek can instead skip the upload and use their system license. Similarly, no scenario or program files uploaded persist on the disk. If privacy concerns remain, the tool can be downloaded/installed locally.
\end{remark}

\begin{figure}[t!]
\centering
\includegraphics[width=\textwidth]{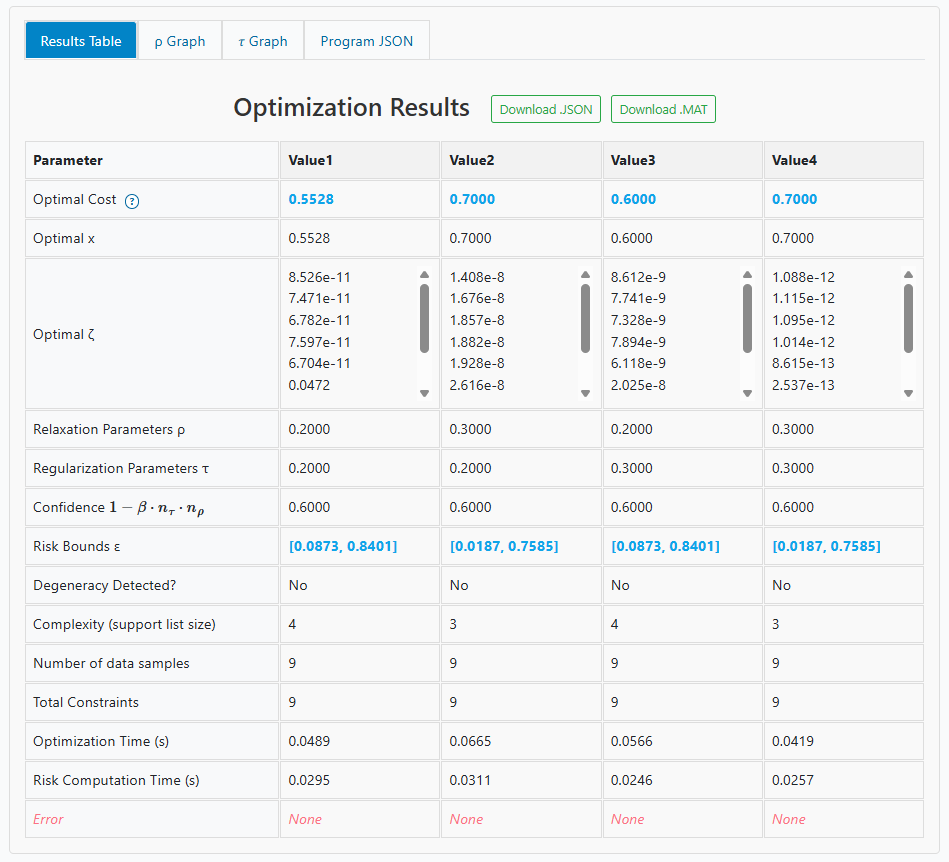}
        \caption{\toolname~web interface displaying the results table for the relaxed and regularized LP set up in Fig.~\ref{fig:web-interface} ($d=1$, $N=9$ scenarios, $\beta=0.1$), with one column for each pair $(\rho,\tau)$, $\rho,\tau\in\{0.2,0.3\}$.}
        \label{fig:table}
\end{figure}

As shown in Fig.~\ref{fig:table}, pressing \emph{Solve} triggers the optimization, after which the results are dynamically displayed. The interface provides up to four analysis tabs (the $\rho$ and $\tau$ graphs appear only when relaxation or regularization is selected); the first tab presents a results table containing key information such as the optimal cost, the solution values, the associated upper and (when applicable) \emph{lower} risk bounds, and an indicator flag identifying whether degeneracy was detected. The table also reports the computed complexity, the total number of constraints, the timings for acquiring the results (reported separately as the optimization solve time and the risk-bound computation time), and any numerical issues encountered during the solution process. When degeneracy is flagged, the lower risk bound is withheld and only the certified upper bound is shown, in line with Theorem~\ref{thm:riskcert-nondeg}.

\begin{figure*}[t!]
    \centering
    \begin{subfigure}[t]{0.5\textwidth}
        \centering
        \includegraphics[height=2in]{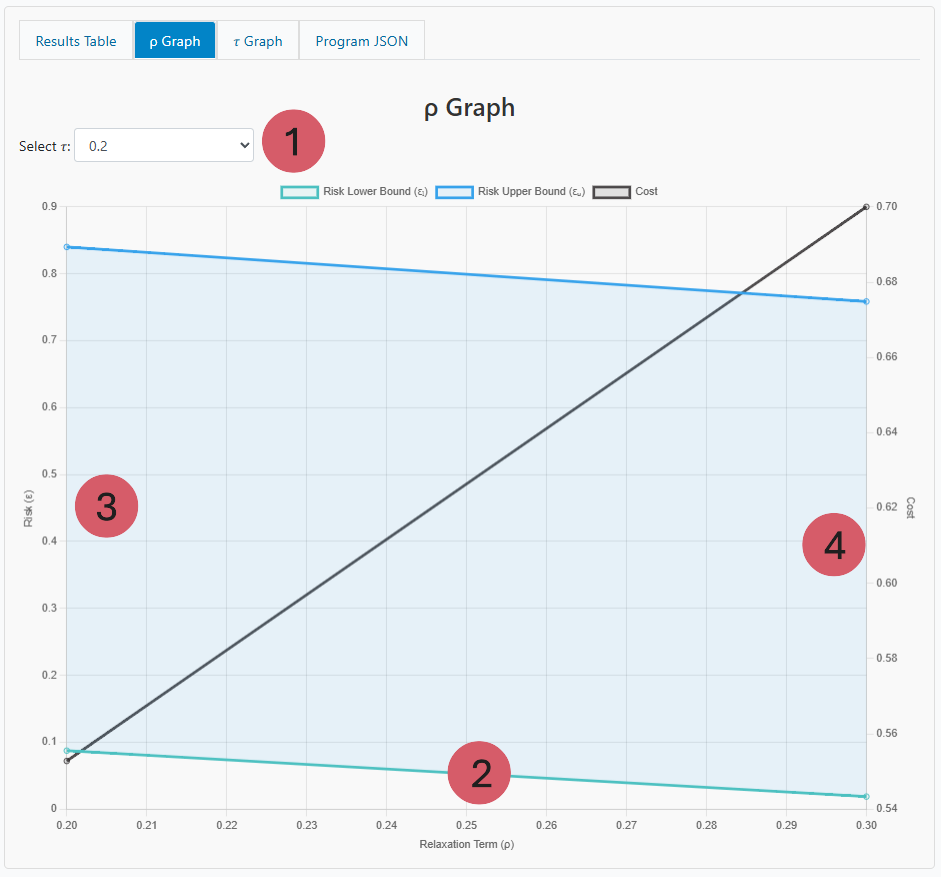}
        \caption{Comparing $\rho$ for different selections of $\tau$}
        \label{fig:rho-graph}
    \end{subfigure}%
    ~ 
    \begin{subfigure}[t]{0.5\textwidth}
        \centering
        \includegraphics[height=2in]{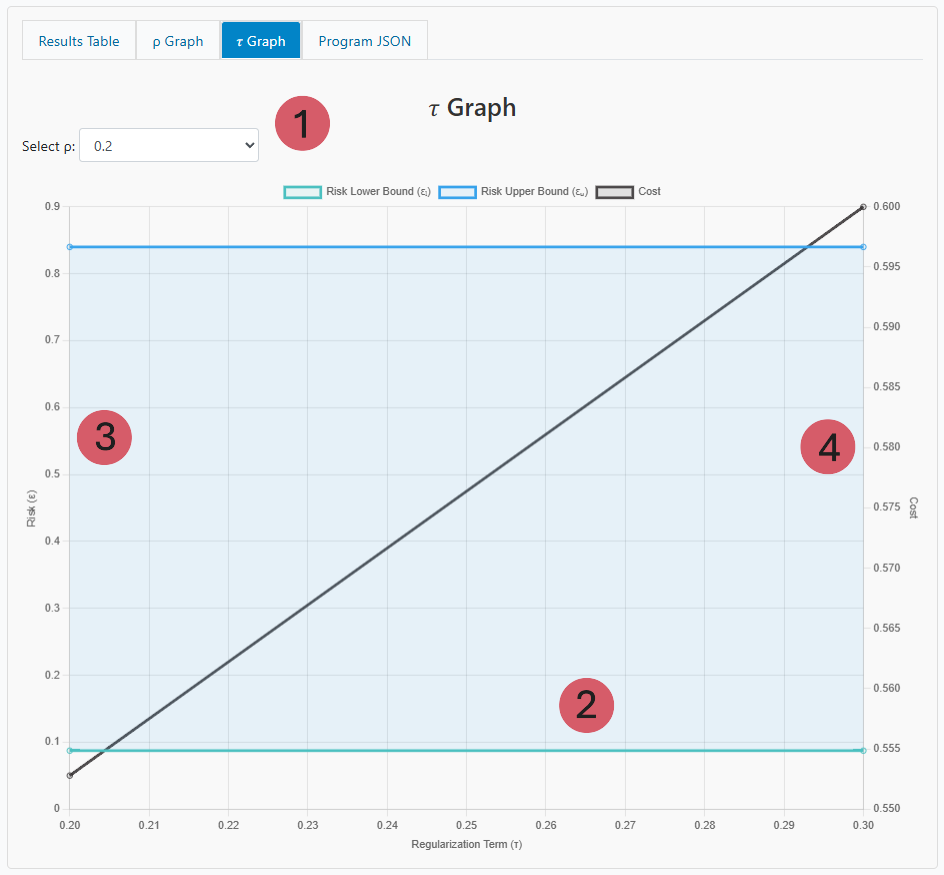}
        \caption{Comparing $\tau$ for different selections of $\rho$}
        \label{fig:tau-graph}
    \end{subfigure}
    \caption{Example graphical output from \toolname~for cases involving relaxation or regularization, for the run of Figs.~\ref{fig:web-interface} and~\ref{fig:table}.}
    \label{fig:graphs}
\end{figure*}

\toolname~allows users to specify multiple scalar values for $\rho$ and $\tau$, separated by commas. When multiple values are provided, each resulting optimization problem is solved independently with the entered confidence parameter $\beta$, and the table reports the overall confidence $1-\beta\,n_\tau n_\rho$, which holds simultaneously for all $n_\tau n_\rho$ cases. When multiple cases are present, the table automatically expands to include additional columns for each trial, and a horizontal scroll bar enables convenient navigation across all displayed results. Similarly, when $d$ or $N$ is large, the optimal $x$ and optimal $\zeta_i$ windows become scrollable. All the results can be subsequently downloaded in \texttt{.json} or \texttt{.mat} formats using the respective green download buttons.

To provide deeper insight, \toolname~also offers graphical summaries of the solutions when multiple values of $\rho$ and $\tau$ are tested. Figure~\ref{fig:graphs} illustrates this functionality for the run of Figs.~\ref{fig:web-interface} and~\ref{fig:table}, with two choices of $\rho$ and two choices of $\tau$. Since visualizing all $(\rho,\tau)$ combinations simultaneously would lead to cluttered plots, the interface instead presents Pareto front slices: one parameter (either $\tau$ or $\rho$) is fixed <1> via an interactive dropdown, while the other <2> is varied and plotted against the resulting risk <3> and cost <4>. The risk plot displays both the upper (blue) and lower (green) bounds, whereas the cost curve is shown in black. These visualizations allow users to conveniently explore and interpret the trade-off between risk and objective value across different regularization and relaxation settings in the spirit of \cite{garatti2022risk,garatti2025non}.

The final tab in the results displays the form data, which allows users to verify that the optimization problem has been specified correctly. The \texttt{.json} representation shown in this tab can also be saved, via a dedicated \emph{Download Program} button, and later re-uploaded through the \emph{Upload Program} dialog, providing a convenient workflow for experimenting with multiple related optimization problems or repeating similar analyses.

\section{Benchmarking}\label{sec:benchmarking}

Table~\ref{tab:benchmarks} summarizes the twelve case studies used to evaluate \toolname. The benchmarks span all three supported program types (LP, QP, SDP) and cover decision-variable dimensions from $d=2$ to $d=120$, with scenario uncertainty dimensions up to $q=48$. Solve times are wall-clock measurements on a desktop with an Intel Core Ultra 9 285K (24 cores) and 64\,GB RAM running Windows 11 using MOSEK~\cite{mosek} as the underlying conic solver. For two benchmarks (Growth Bound and Radiation Therapy) \toolname~detects degeneracy --- the constraints active at the optimum do not form a support list --- so their lower risk bounds are not certified (see Theorem~\ref{thm:riskcert-nondeg}) and are marked N/A in the table. Full problem descriptions are provided in Appendix~\ref{appendix}.

\begin{table}
\footnotesize
\centering
\caption{Summary of all twelve benchmark case studies solved with \toolname, grouped by program type. $d$~=~decision variables, $q$~=~uncertainty dimension, $m_s$~=~scenario constraints (rows of $A(\delta)$ for LP/QP, LMI dimension for SDP), $m_h$~=~hard constraints (rows of $G$ for LP/QP, LMI dimension for SDP), $N$~=~number of sampled scenarios, $k$~=~complexity (number of support constraints). All benchmarks use the confidence parameter $\beta=10^{-6}$. Solve times are wall-clock seconds on a desktop with an Intel Core Ultra 9 285K (24 cores) and 64\,GB RAM running Windows 11. $^{\dagger}$Power Dispatch, Inventory and Portfolio CVaR use an augmented formulation in which the slack penalty is embedded in the objective coefficient $c$; the value of \lstinline|rho| passed to the solver is therefore $0$, while the displayed value reflects the user-facing penalty weight. N/A~=~degeneracy detected, so the lower risk bound is not certified. Scenario data come from Yahoo Finance (Portfolio CVaR) and the UCI repository (Iris SVM~\cite{iris_53}, Covariance Estimation~\cite{wine_109}, Min.\ Encl.\ Ellipsoid~\cite{breast_cancer_wisconsin_diagnostic_17}); Radiation Therapy uses synthetic data inspired by TROTS~\cite{breedveld2019trots}, and the scenarios of all other benchmarks are generated as described in Appendix~\ref{appendix}.}
\label{tab:benchmarks}
\setlength{\tabcolsep}{3pt}
\begin{tabular}{@{}lrrrrrrrrrrr@{}}
\toprule
Benchmark & $d$ & $q$ & $m_s$ & $m_h$ & $\rho$ & $\tau$ & $N$ & $k$ & $\underline{\varepsilon}$ & $\bar{\varepsilon}$ & Time\,(s) \\
\midrule
\multicolumn{12}{@{}l}{\textit{Linear Programs (LP)}} \\[2pt]
Half Width               &   2 &  1 &  2 &   0 & $0$    & $0$   &  100 &  2 & $0$      & $0.2085$ &   1.2 \\
Growth Bound             &  12 &  6 &  3 &   6 & $0$    & $0$   & 3127 &  6 & N/A      & $0.0103$ & 1276.8 \\
Inventory                &  11 & 15 &  6 &  17 & $100^{\dagger}$ & $0$ &  500 &  7 & $0$      & $0.0666$ &   4.0 \\
Portfolio CVaR           &  13 & 12 &  4 &  27 & $0.016^{\dagger}$ & $0$  & 1255 &  3 & $0$      & $0.0203$ &  22.0 \\
Power Dispatch           & 120 & 48 & 48 & 330 & $100^{\dagger}$ & $0$ &  150 & 34 & $0.0791$ & $0.4496$ &   2.9 \\[4pt]
\multicolumn{12}{@{}l}{\textit{Quadratic Programs (QP)}} \\[2pt]
Iris SVM                 &   3 &  3 &  1 &   0 & $0$    & $0$   &  150 &  2 & $0$      & $0.1441$ &   1.2 \\
Robot Navigation         & 118 &  1 &  3 & 320 & $0$    & $0$   &  500 &  1 & $0$      & $0.0403$ &   1.6 \\
Radiation Therapy        &  50 &  3 & 80 & 100 & $0$    & $0.1$ &  200 &  5 & N/A      & $0.1414$ & 112.3 \\[4pt]
\multicolumn{12}{@{}l}{\textit{Semidefinite Programs (SDP)}} \\[2pt]
LPV Stability            &   3 &  1 &  2 &   2 & $1$    & $0$   &  100 &  1 & $0$      & $0.1858$ &   1.7 \\
Quadratic Stability      &   6 &  2 &  3 &   3 & $0$    & $0$   &  500 &  1 & $0$      & $0.0403$ &   2.2 \\
Covariance Estimation    &  15 &  5 &  5 &   5 & $0$    & $0$   &  178 &  7 & $0$      & $0.1780$ &   2.1 \\
Min.\ Encl.\ Ellipsoid   &  15 &  5 &  1 &   5 & $0$    & $0$   &  569 &  5 & $0$      & $0.0518$ & 235.6 \\
\bottomrule
\end{tabular}
\end{table}

\section{Conclusions and Future Work}\label{sec:futurework}

This work presented \toolname, the \emph{first open-source} software tool designed to bring the scenario approach into practical, user-friendly data-driven convex optimization. Building on the rigorous theoretical foundations of scenario theory, which links sampled constraints, their support list, and the generalization properties of the resulting decision, \toolname~enabled researchers and practitioners to solve linear, quadratic, and semidefinite programs using only sampled uncertainty realizations while obtaining certified probabilistic guarantees on performance. The tool was implemented in Python and complemented with a modern JavaScript-based web interface, providing an intuitive and highly accessible user experience across devices. It supported both \emph{web-based usage} and \emph{local installation}, allowing flexible data entry through manual input or file uploads. Through a set of representative benchmarks, we demonstrated the practical effectiveness of \toolname~for reliable, data-driven decision making. Looking ahead, the framework established in this work opens the door to extending \toolname~to accommodate emerging advances in \emph{nonconvex} scenario optimization.

\bibliographystyle{spmpsci}  

\bibliography{biblio.bib} 

@book{boyd2004convex,
  title={Convex optimization},
  author={Boyd, Stephen P and Vandenberghe, Lieven},
  year={2004},
  publisher={Cambridge university press}
}

@article{campi2023compression,
  title={Compression, generalization and learning},
  author={Campi, Marco C and Garatti, Simone},
  journal={Journal of Machine Learning Research},
  volume={24},
  number={339},
  pages={1--74},
  year={2023}
}

@article{campi2021scenarioreview,
  title={The scenario approach: A tool at the service of data-driven decision making},
  author={Campi, Marco C and Car{\`e}, Algo and Garatti, Simone},
  journal={Annual Reviews in Control},
  volume={52},
  pages={1--17},
  year={2021},
  publisher={Elsevier}
}

@article{campi2011sampling,
  title={A sampling-and-discarding approach to chance-constrained optimization: feasibility and optimality},
  author={Campi, Marco C and Garatti, Simone},
  journal={Journal of optimization theory and applications},
  volume={148},
  number={2},
  pages={257--280},
  year={2011},
  publisher={Springer}
}

@article{campi2008exact,
  title={The exact feasibility of randomized solutions of uncertain convex programs},
  author={Campi, Marco C and Garatti, Simone},
  journal={SIAM Journal on Optimization},
  volume={19},
  number={3},
  pages={1211--1230},
  year={2008},
  publisher={SIAM}
}

@article{hou2013model,
	title={From model-based control to data-driven control: Survey, classification and perspective},
	author={Hou, Z. S. and Wang, Z.},
	journal={Information Sciences},
	volume={235},
	pages={3--35},
	year={2013}
}

@article{calafiore2006scenario,
	title={The scenario approach to robust control design},
	author={Calafiore, G. C. and Campi, M. C.},
	journal={IEEE Transactions on automatic control},
	volume={51},
	number={5},
	pages={742--753},
	year={2006},
	publisher={IEEE}
}

@article{campi2009scenario,
	title={The scenario approach for systems and control design},
	author={Campi, Marco C and Garatti, Simone and Prandini, Maria},
	journal={Annual Reviews in Control},
	volume={33},
	number={2},
	pages={149--157},
	year={2009}
}

@article{esfahani2014performance,
	title={Performance bounds for the scenario approach and an extension to a class of non-convex programs},
	author={Mohajerin Esfahani, P.  and Sutter, T. and Lygeros, J.},
	journal={IEEE Transactions on Automatic Control},
	volume={60},
	number={1},
	pages={46--58},
	year={2014},
	publisher={IEEE}
}

@article{mohajerin2018infinite,
	title={From infinite to finite programs: explicit error bounds with applications to approximate dynamic programming},
	author={Mohajerin Esfahani, P. and Sutter, T. and Kuhn, D. and Lygeros, J.},
	journal={SIAM journal on optimization},
	volume={28},
	number={3},
	pages={1968--1998},
	year={2018},
	publisher={SIAM}
}

@article{campi2018general,
	title={A general scenario theory for nonconvex optimization and decision making},
	author={Campi, M. C. and Garatti, S. and Ramponi, F. A.},
	journal={IEEE Transactions on Automatic Control},
	volume={63},
	number={12},
	pages={4067--4078},
	year={2018}
}

@article{garatti2025non,
	title={Non-convex scenario optimization},
	author={Garatti, S. and Campi, M. C.},
	journal={Mathematical Programming},
	volume={209},
	number={1},
	pages={557--608},
	year={2025}
}

@article{margellos2014road,
	title={On the road between robust optimization and the scenario approach for chance constrained optimization problems},
	author={Margellos, Kostas and Goulart, Paul and Lygeros, John},
	journal={IEEE Transactions on Automatic Control},
	volume={59},
	number={8},
	pages={2258--2263},
	year={2014}
}

@article{garatti2022risk,
  title={Risk and complexity in scenario optimization},
  author={Garatti, Simone and Campi, Marco C},
  journal={Mathematical Programming},
  volume={191},
  number={1},
  pages={243--279},
  year={2022},
  publisher={Springer}
}

@misc{grant2009cvx,
  title={{CVX} users’ guide},
  author={Grant, Michael and Boyd, Stephen and Ye, Yinyu},
  year={2009}
}

@article{diamond2016cvxpy,
  title={CVXPY: A Python-embedded modeling language for convex optimization},
  author={Diamond, Steven and Boyd, Stephen},
  journal={Journal of Machine Learning Research},
  volume={17},
  number={83},
  pages={1--5},
  year={2016}
}

@article{mosek,
  title={Mosek optimizer API for python},
  author={ApS, Mosek},
  journal={Version},
  volume={9},
  number={17},
  pages={6--4},
  year={2022}
}

@misc{gurobi,
  author = {{Gurobi Optimization, LLC}},
  title = {{Gurobi Optimizer Reference Manual}},
  year = 2024,
  url = "https://www.gurobi.com"
}

@misc{iris_53,
  author       = {Fisher, R. A.},
  title        = {{Iris}},
  year         = {1936},
  howpublished = {UCI Machine Learning Repository},
  note         = {{DOI}: https://doi.org/10.24432/C56C76}
}

@article{kazemi2024data,
  title={Data-driven abstraction-based control synthesis},
  author={Kazemi, Milad and Majumdar, Rupak and Salamati, Mahmoud and Soudjani, Sadegh and Wooding, Ben},
  journal={Nonlinear Analysis: Hybrid Systems},
  volume={52},
  year={2024},
  publisher={Elsevier}
}

@article{nemirovski1994parallel,
  title={On parallel complexity of nonsmooth convex optimization},
  author={Nemirovski, Arkadi},
  journal={Journal of Complexity},
  volume={10},
  number={4},
  pages={451--463},
  year={1994},
  publisher={Elsevier}
}

@article{balkanski2018parallelization,
  title={Parallelization does not accelerate convex optimization: Adaptivity lower bounds for non-smooth convex minimization},
  author={Balkanski, Eric and Singer, Yaron},
  journal={arXiv:1808.03880},
  year={2018}
}

@article{scs,
    author       = {Brendan O'Donoghue and Eric Chu and Neal Parikh and Stephen Boyd},
    title        = {Conic Optimization via Operator Splitting and Homogeneous Self-Dual Embedding},
    journal      = {Journal of Optimization Theory and Applications},
    month        = {June},
    year         = {2016},
    volume       = {169},
    number       = {3},
    pages        = {1042-1068},
    url          = {http://stanford.edu/~boyd/papers/scs.html},
}

@article{cvxopt,
  title={{CVXOPT}: Convex optimization},
  author={Andersen, Martin and Dahl, Joachim and Vandenberghe, Lieven},
  journal={Astrophysics Source Code Library},
  year={2020}
}

@article{clarabel,
  title={Clarabel: An interior-point solver for conic programs with quadratic objectives},
  author={Goulart, Paul J and Chen, Yuwen},
  journal={arXiv:2405.12762},
  year={2024}
}

@book{campi2018introduction,
  title={Introduction to the scenario approach},
  author={Campi, Marco C and Garatti, Simone},
  year={2018},
  publisher={SIAM}
}

@article{buzby2014estimated,
  title={The estimated amount, value, and calories of postharvest food losses at the retail and consumer levels in the United States},
  author={Buzby, Jean C and Farah-Wells, Hodan and Hyman, Jeffrey},
  journal={USDA-ERS Economic Information Bulletin},
  number={121},
  year={2014}
}

@article{perez2017fruit,
  title={Fruit and tree nuts outlook},
  author={Perez, Agnes and Ferreira, Gustavo and Minor, Travis},
  year={2017}
}

@article{draxl2015wind,
  title={The wind integration national dataset (wind) toolkit},
  author={Draxl, Caroline and Clifton, Andrew and Hodge, Bri-Mathias and McCaa, Jim},
  journal={Applied Energy},
  volume={151},
  pages={355--366},
  year={2015},
  publisher={Elsevier}
}

@article{grantham2017generating,
  title={Generating synthetic five-minute solar irradiance values from hourly observations},
  author={Grantham, AP and Pudney, PJ and Ward, LA and Belusko, M and Boland, JW},
  journal={Solar Energy},
  volume={147},
  pages={209--221},
  year={2017},
  publisher={Elsevier}
}

@article{rockafellar2000optimization,
  title={Optimization of conditional value-at-risk},
  author={Rockafellar, R Tyrrell and Uryasev, Stanislav and others},
  journal={Journal of risk},
  volume={2},
  pages={21--42},
  year={2000}
}

@article{deasy2003cerr,
  title={CERR: a computational environment for radiotherapy research},
  author={Deasy, Joseph O and Blanco, Angel I and Clark, Vanessa H},
  journal={Medical physics},
  volume={30},
  number={5},
  pages={979--985},
  year={2003},
  publisher={Wiley Online Library}
}

@article{breedveld2019trots,
  title={TROTS-The Radiotherapy Optimisation Test Set},
  author={Breedveld, Sebastiaan and Heijmen, Ben},
  year={2019}
}

@misc{wine_109,
  author       = {Aeberhard, Stefan and Forina, M.},
  title        = {{Wine}},
  year         = {1992},
  howpublished = {UCI Machine Learning Repository},
  note         = {{DOI}: https://doi.org/10.24432/C5PC7J}
}

@misc{breast_cancer_wisconsin_diagnostic_17,
  author       = {Wolberg, William and Mangasarian, Olvi and Street, Nick and Street, W.},
  title        = {{Breast Cancer Wisconsin (Diagnostic)}},
  year         = {1993},
  howpublished = {UCI Machine Learning Repository},
  note         = {{DOI}: https://doi.org/10.24432/C5DW2B}
}

@article{schildbach2013randomized,
  title={Randomized solutions to convex programs with multiple chance constraints},
  author={Schildbach, Georg and Fagiano, Lorenzo and Morari, Manfred},
  journal={SIAM Journal on Optimization},
  volume={23},
  number={4},
  pages={2479--2501},
  year={2013},
  publisher={SIAM}
}

@article{margellos2015connection,
  title={On the connection between compression learning and scenario based single-stage and cascading optimization problems},
  author={Margellos, Kostas and Prandini, Maria and Lygeros, John},
  journal={IEEE Transactions on Automatic Control},
  volume={60},
  number={10},
  pages={2716--2721},
  year={2015},
  publisher={IEEE}
}

@article{zhang2015sample,
  title={On the sample size of random convex programs with structured dependence on the uncertainty},
  author={Zhang, Xiaojing and Grammatico, Sergio and Schildbach, Georg and Goulart, Paul and Lygeros, John},
  journal={Automatica},
  volume={60},
  pages={182--188},
  year={2015},
  publisher={Elsevier}
}

@article{margellos2017distributed,
  title={Distributed constrained optimization and consensus in uncertain networks via proximal minimization},
  author={Margellos, Kostas and Falsone, Alessandro and Garatti, Simone and Prandini, Maria},
  journal={IEEE Transactions on Automatic Control},
  volume={63},
  number={5},
  pages={1372--1387},
  year={2017},
  publisher={IEEE}
}

@article{shang2020posteriori,
  title={A posteriori probabilistic bounds of convex scenario programs with validation tests},
  author={Shang, Chao and You, Fengqi},
  journal={IEEE Transactions on Automatic Control},
  volume={66},
  number={9},
  pages={4015--4028},
  year={2020},
  publisher={IEEE}
}

@article{romao2022exact,
  title={On the exact feasibility of convex scenario programs with discarded constraints},
  author={Romao, Licio and Papachristodoulou, Antonis and Margellos, Kostas},
  journal={IEEE Transactions on Automatic Control},
  volume={68},
  number={4},
  pages={1986--2001},
  year={2022},
  publisher={IEEE}
}

@article{romao2023probabilistic,
  title={Probabilistic feasibility guarantees for convex scenario programs with an arbitrary number of discarded constraints},
  author={Romao, Licio and Margellos, Kostas and Papachristodoulou, Antonis},
  journal={Automatica},
  volume={149},
  pages={110601},
  year={2023},
  publisher={Elsevier}
}

@article{campi2018wait,
  title={Wait-and-judge scenario optimization},
  author={Campi, Marco C and Garatti, Simone},
  journal={Mathematical Programming},
  volume={167},
  number={1},
  pages={155--189},
  year={2018},
  publisher={Springer}
}

\appendix
\suppressfloats[t]

\section{Case Studies}\label{appendix}

All case study benchmarks can be found in the repository at \url{https://github.com/Kiguli/Scen-Opt/tree/v1.0/benchmarks}, and in the archived release at \url{https://doi.org/10.5281/zenodo.23177690}. The results presented here were run on a desktop with an Intel Core Ultra 9 285K (24 cores) and 64\,GB RAM running Windows 11.

\subsection{Linear Programming Cases}\label{appendix:LP}

\subsubsection{Inventory Management}

\begin{figure}[t]
    \centering
    \includegraphics[width=\linewidth]{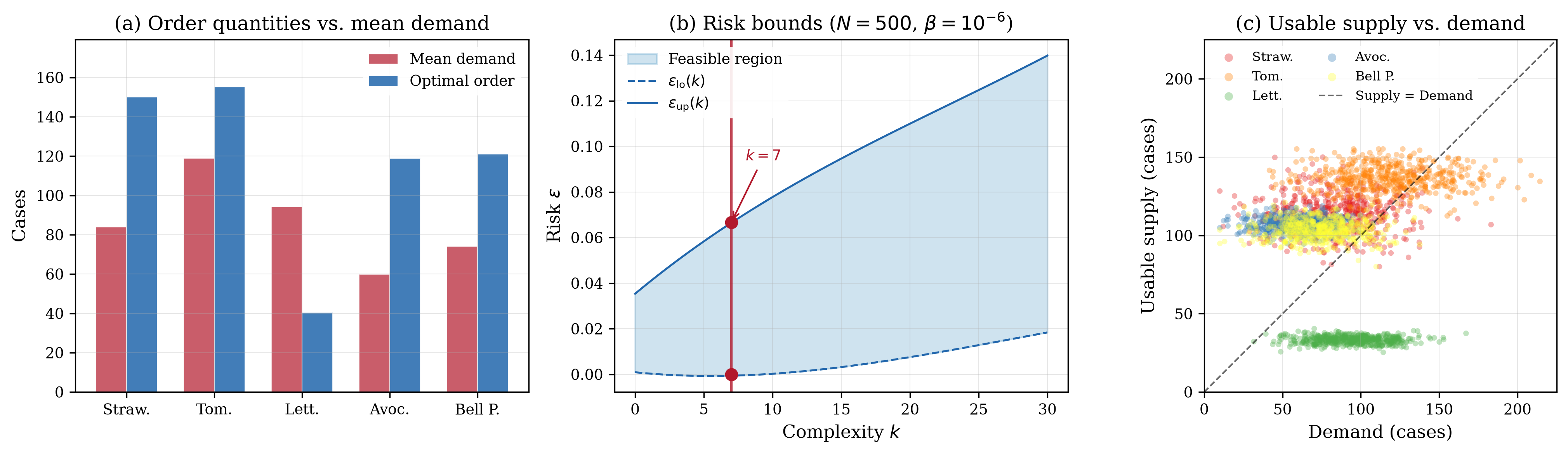}
    \caption{\textbf{Left.} Comparison of optimal order against mean demand. \textbf{Center.} Quantified risk bounds for $N=500$ and confidence $99\%$. \textbf{Right.} Graphic of scenario samples for different produce.}
    \label{fig:inventory}
\end{figure}

A purchasing manager must decide how many cases of fresh produce to order daily from the Oakland Terminal Market. Fresh produce is highly perishable, \textit{e.g.} strawberries might have $25\%$ perished by the time they reach stores. Demand fluctuates based on weather, local events and seasonal patterns. The purchasing manager has a fixed budget and limited warehouse space. Too little orders means empty shelves and lost sales. Too many and produce spoils on the shelf, a wasted purchase.

Determine the optimal daily order quantities to minimize wastage for $5$ perishable produce, as labeled in Table~\ref{tab:produce}. The daily budget is $\$12,000$ and warehouse has $800$ cubic feet of refrigerator space. Products have a maximum order limit of $150-200$ cases, item dependent.

\begin{table}[]
    \centering
    \begin{tabular}{c|c|c|c|c}
    \textbf{Product} & \textbf{Cost} (\$) & \textbf{Mean Demand} (cases) & \textbf{Yield} ($\%$) & \textbf{Space} (cu. ft) \\
    \hline
    Strawberries &	18.50	&85&	75 (25 loss) &	1.2\\
Tomatoes	&14.25	&120 &	88 (12 loss)&	1.5\\
Lettuce	&12.00	&95&	82 (18 loss)&	2.0 \\
Avocados	&32.00	&60 &	90 (10 loss)	&0.8\\
Bell Peppers&	22.50	&75&	85 (15 loss)	&1.3
    \end{tabular}
    \caption{Produce purchasing details. The data is synthetic but calibrated to real data~\cite{buzby2014estimated,perez2017fruit}, along with USDA Terminal Market Prices for highest grade products.}
    \label{tab:produce}
\end{table}

The decision variable is $x \in \mathbb{R}^5$, representing daily order quantities for five perishable produce items (strawberries, tomatoes, lettuce, avocados, bell peppers), with per-case ordering cost $c = [18.50,\; 14.25,\; 12.00,\; 32.00,\; 22.50]^\top$ USD, slack penalty $\rho = 100$, and no regularization ($\tau = 0$). Each scenario ${\delta}_i \in \mathbb{R}^{15}$ encodes uncertain yields $\delta_i^{(1:5)}$, demands $\delta_i^{(6:10)}$, and warehouse packing efficiencies $\delta_i^{(11:15)}$, drawn from correlated distributions calibrated to USDA wholesale market data. The scenario-dependent constraint matrices in~\eqref{eq:LP-constraint1} are
\begin{equation*}\label{eq:inv-Ad}
A({\delta}_i) =
\begin{bmatrix}
-\delta_i^{(1)} & & & & \\
 & -\delta_i^{(2)} & & & \\
 & & \ddots & & \\
 & & & & -\delta_i^{(5)} \\[4pt]
s_1\,\delta_i^{(11)} & s_2\,\delta_i^{(12)} & \cdots & \cdots & s_5\,\delta_i^{(15)}
\end{bmatrix}\!, \quad
b({\delta}_i) =
\begin{bmatrix}
\delta_i^{(6)} \\ \delta_i^{(7)} \\ \vdots \\ \delta_i^{(10)} \\[4pt] -C_{\mathrm{wh}}
\end{bmatrix}\!,
\end{equation*}
where $s_j$ is the warehouse space per case of product~$j$ and $C_{\mathrm{wh}} = 800\;\text{ft}^3$. The first five rows enforce that usable supply meets demand; the last row enforces refrigerated warehouse capacity. The hard constraints in~\eqref{eq:LP-constraint2} are
\begin{equation*}\label{eq:inv-Gh}
G =
\begin{bmatrix}
-\mathbb I_5 \\ \mathbb I_5 \\ c^\top
\end{bmatrix}\!, \quad
h =
\begin{bmatrix}
0_5 \\ -\bar{q} \\ -B
\end{bmatrix}\!,
\end{equation*}
encoding non-negativity, supplier limits $\bar{q} = [150,\; 200,\; 180,\; 120,\; 140]^\top$ cases, and a daily budget $B = \$12{,}000$. $\mathbb I_n$ is the $n\times n$ identity matrix.

\toolname~solves the LP in $4.0$ seconds considering $N=500$ scenarios. The optimal cost is $\$45,099.67$ with all $\$12,000$ of budget used. The penalty $\rho=100$ implies a cost per case of lost sales for an item, therefore the expected daily loss from stockouts is only $\$66.20$ or about $0.5\%$ of the ordering budget. Lettuce is bulky (2.0 cu.ft/case), cheap ($\$12$/case), and has high spoilage ($18\%$). Under a $\$12,000$ budget, it is more efficient to spend budget on avocados ($\$32$/case but $90\%$ yield) and accept lettuce shortfalls. With confidence $10^{-6}$, the LP problem had complexity $k=7$, no degeneracy, and risk bounds $[0.0000, 0.0666]$. This gives a maximum $6.7\%$ chance of the optimal cost being nullified if another sample was acquired.

\begin{table}[]
    \centering
    \begin{tabular}{c|c|c|c|c}
    \textbf{Product} & \textbf{Order (cases)} & \textbf{Mean Demand} & \textbf{Service Level} ($\%$) \\
    \hline
    Strawberries &	150	&83.9&	 85.8\\
Tomatoes	&155.2	&118.9 &	72.4\\
Lettuce	&40.4	&94.3&	0.2 \\
Avocados	&118.9	&59.8 &	99.8\\
Bell Peppers&	121.0	&74.1&	90.0
    \end{tabular}
    \caption{Optimal Purchasing Order}
    \label{tab:produce-optimal}
\end{table}

\subsubsection{Power Dispatch}

A grid operator must schedule three thermal generators over a $24$-hour horizon to meet electricity demand while integrating uncertain wind and solar generation. Demand follows a typical residential-commercial daily profile peaking at $1000$~MW. The system includes a $200$~MW wind farm and $150$~MW solar farm whose output cannot be known in advance.

Determine the optimal hourly generation schedule $P_{m,t}$ for each thermal unit~$m$ at each hour~$t$. The decision variable is $x = P \in \mathbb{R}^{72}$ ($3$~generators $\times$ $24$~hours), with per-MWh generation cost $c = [\underbrace{40, \ldots, 40}_{24},\; \underbrace{50, \ldots, 50}_{24},\; \underbrace{30, \ldots, 30}_{24}]^\top$, slack penalty $\rho = 100$, and no regularization ($\tau = 0$). Each scenario $\delta_i \in \mathbb{R}^{48}$ encodes uncertain hourly wind generation $\delta_i^{(1:24)}$ and solar generation $\delta_i^{(25:48)}$, drawn from a Weibull distribution (shape $2.0$, hour-varying scale) for wind~\cite{draxl2015wind} and a Beta$(2,5)$ cloudiness factor applied to a clear-sky envelope for solar~\cite{grantham2017generating}. The scenario-dependent constraint matrices in~\eqref{eq:LP-constraint1} enforce power balance at each hour:
\begin{equation*}
A(\delta_i) =
\begin{bmatrix}
\phantom{-}e_t^\top \\[2pt]
-e_t^\top
\end{bmatrix}_{t=0}^{23}\!, \quad
b(\delta_i) =
\begin{bmatrix}
\delta_i^{(t)} + \delta_i^{(24+t)} - D_t \\[2pt]
-\delta_i^{(t)} - \delta_i^{(24+t)} + D_t
\end{bmatrix}_{t=0}^{23}\!,
\end{equation*}
where $e_t \in \mathbb{R}^{72}$ selects all generators at hour~$t$ and $D_t$ is the demand at hour~$t$. Each pair of rows bounds the power imbalance from above and below. The hard constraints in~\eqref{eq:LP-constraint2} are
\begin{equation*}
G x + h \leq 0,
\end{equation*}
encoding generator capacity limits $P_{\min} \leq P_{m,t} \leq P_{\max}$ and ramp-rate constraints $|P_{m,t+1} - P_{m,t}| \leq R_m$ ($282$ rows total).

\begin{table}[t]
    \centering
    \begin{tabular}{c|c|c|c|c}
    \textbf{Generator} & \textbf{Capacity} & \textbf{Min Output} & \textbf{Cost} & \textbf{Ramp Rate} \\
    \hline
    Gas 1  & 400 MW & 100 MW & 40 \$/MW & 200 MW/h \\
    Gas 2  & 300 MW & 75 MW & 50 \$/MW & 150 MW/h \\
    Coal   & 500 MW & 200 MW & 30 \$/MW & 100 MW/h
    \end{tabular}
    \caption{Thermal generator parameters. Costs and capacities are synthetic but calibrated against standard values.}
    \label{tab:generators}
\end{table}

\begin{figure}[t]
    \centering
    \includegraphics[width=\linewidth]{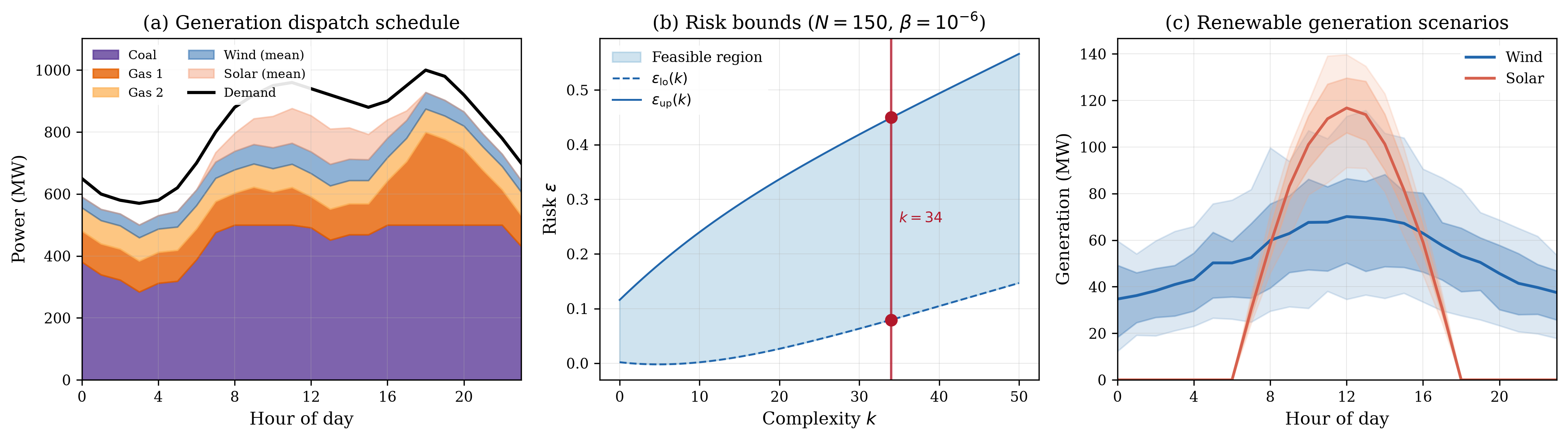}
    \caption{\textbf{Left.} Stacked generation dispatch over $24$ hours: coal provides baseload, gas ramps for the evening peak. \textbf{Center.} Quantified risk bounds for $N=150$ and confidence $99\%$. \textbf{Right.} Wind and solar scenario variability (median with $25-75$th and $10-90$th percentile bands).}
    \label{fig:power_dispatch}
\end{figure}

\toolname~solves the LP considering $N=150$ renewable generation scenarios in $2.9$ seconds. The total cost is $\$831{,}357$, comprising $\$537{,}670$ in fuel costs and $\$293{,}687$ in power imbalance penalty ($\rho=100$ per MW of mismatch). Coal, the cheapest unit at $\$30$/MWh, provides baseload at $284-500$~MW. Gas~1 ($\$40$/MWh) ramps from its $100$~MW minimum to $300$~MW during the evening peak (hour~18). Gas~2 ($\$50$/MWh) remains at its $75$~MW minimum throughout, dispatching it further is never cost-effective. The LP problem has confidence $10^{-6}$, complexity $k=34$, no degeneracy, and risk bounds $[0.0791, 0.4496]$. This gives a $\approx 55\%$ likelihood of having the correct result for unseen data, therefore, more data is likely required to increase the confidence in this power dispatch solution.

\subsubsection{Portfolio Management - Conditional Value at Risk (CVaR)}

A pension fund must allocate capital across $8$ ETF asset classes to minimize tail risk under market uncertainty. The fund uses Conditional Value-at-Risk (CVaR) at the $95\%$ level (the average loss in the worst $5\%$ of scenarios) as its primary risk measure, following the linear formulation of~\cite{rockafellar2000optimization}.

The decision variable is $x = [w,\,\alpha] \in \mathbb{R}^{9}$, where $w \in \mathbb{R}^8$ are portfolio weights and $\alpha$ is the VaR threshold. The objective $c = [0,\ldots,0,\,1]^\top$ minimizes~$\alpha$, with slack penalty $\rho = 1/((1-0.95) \cdot N) \approx 0.016$ so that $\alpha + \rho\,\sum \zeta_i$ approximates CVaR at the $95\%$ level. There is no regularization ($\tau = 0$).

Each scenario ${\delta}_i \in \mathbb{R}^{12}$ encodes $8$~asset returns $\delta_i^{(1:8)}$ (bootstrapped from historical data), a market stress indicator $\delta_i^{(9)} \in [0,1]$, credit spread shock $\delta_i^{(10)}$, interest rate shock $\delta_i^{(11)}$, and volatility scaling $\delta_i^{(12)}$. The scenario-dependent constraint matrix in~\eqref{eq:LP-constraint1} has $4$ rows:
\begin{equation*}
A(\delta_i) = \begin{bmatrix}\gamma\,\delta_i^{(1)} & \gamma\,\delta_i^{(2)} & \gamma\,\delta_i^{(3)} & \gamma\,\delta_i^{(4)} & \gamma\,\delta_i^{(5)} & \gamma\,\delta_i^{(6)} & \gamma\,\delta_i^{(7)} &\gamma\,\delta_i^{(8)} & -1 \\
0 & -0.5\,\delta_i^{(10)} & 0 & 0 & 0 & -0.3\,\delta_i^{(10)} & 0 &-\delta_i^{(10)} & 0 \\
0 & -5\,\delta_i^{(11)} & 0 & 0 & 0 & -18\,\delta_i^{(11)} & 0 & -8\,\delta_i^{(11)} & 0 \\
-\tfrac{\delta_i^{(9)}}{2}\delta_i^{(1)} & 0 & 0 & 0 & -\tfrac{\delta_i^{(9)}}{2}\delta_i^{(5)} & 0 & -\tfrac{\delta_i^{(9)}}{2}\delta_i^{(7)} & 0 & 0
\end{bmatrix}\!,
\end{equation*}
with $\gamma = -(1{+}0.3\,\delta_i^{(9)})$ and $b({\delta}_i) = 0_4$. The first row is the main CVaR loss constraint with stress amplification; rows $2$ and $3$ capture credit spread and interest rate duration exposure on bond assets (AGG, TLT, LQD); row~$4$ models equity correlation breakdown during crises (SPY, EFA, VWO). The hard constraints in~\eqref{eq:LP-constraint2} encode position limits ($5-30\%$ per asset), full investment ($\sum w_j = 1$), equity allocation range ($30-60\%$), fixed income minimum ($25\%$), and VaR threshold bounds ($23$ rows total).

\begin{table}[t]
    \centering
    \begin{tabular}{c|c|c|c}
    \textbf{Ticker} & \textbf{Asset} & \textbf{Class} & \textbf{Role} \\
    \hline
    SPY & S\&P 500 ETF     & US Equity    & Core Growth \\
    AGG & US Aggregate Bond & Fixed Income & Stability \\
    VNQ & Real Estate REIT  & Real Assets  & Inflation Hedge \\
    GLD & Gold              & Commodities  & Crisis Hedge \\
    EFA & Intl Developed    & Intl Equity  & Diversification \\
    TLT & 20+ Year Treasury & Long Bonds   & Duration \\
    VWO & Emerging Markets  & Intl Equity  & Growth \\
    LQD & Inv.\ Grade Corp  & Credit       & Yield
    \end{tabular}
    \caption{ETF asset universe. Historical returns are downloaded from Yahoo Finance via \texttt{yfinance} (5 years daily data).}
    \label{tab:etfs}
\end{table}

\begin{figure}[t]
    \centering
    \includegraphics[width=\linewidth]{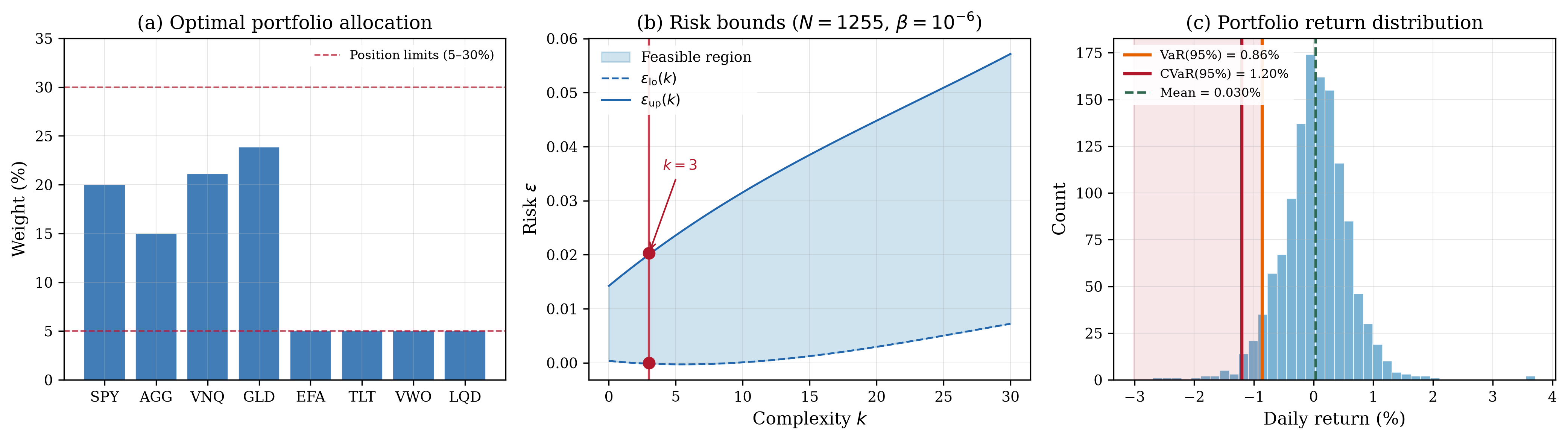}
    \caption{\textbf{Left.} Optimal portfolio weights with position limit constraints. \textbf{Center.} Quantified risk bounds for $N=1255$ and $\beta=10^{-6}$. \textbf{Right.} Portfolio return distribution with VaR$(95\%)$ and CVaR$(95\%)$ marked.}
    \label{fig:portfolio_cvar}
\end{figure}

\toolname~solves the LP considering $N=1255$ market scenarios (one per historical trading day) with confidence $\beta=10^{-6}$ in $22.0$ seconds. The optimizer tilts heavily toward alternatives: GLD ($23.9\%$) and VNQ ($21.1\%$) absorb the allocation not consumed by regulatory minimums. Equities sit at the $30\%$ floor and fixed income at $25\%$, both at their binding lower limits. This minimizes tail risk, the CVaR$(95\%)$ is only $1.20\%$ daily loss. The LP problem has complexity $k=3$, no degeneracy, and risk bounds $[0.0000, 0.0203]$.

\subsubsection{Smallest Enclosing Interval}\label{appendix:LP-interval}
We sample $100$ points from the $1$-dimensional space $[0,1]$ and seek the smallest interval that encloses all samples, following the setup in~\cite{campi2018introduction}. This is formulated as a \emph{robust LP}, where the solution consists of the interval center $o^\ast$ and half-width $w^\ast$, \emph{i.e.}, $x^\ast_N = \begin{bmatrix} o^\ast & w^\ast \end{bmatrix}^\top$\!\!. The corresponding matrices defining the robust LP in~\eqref{eq:LP-full} are given in~\eqref{new}, where the constraint $A(\delta_i)x + b(\delta_i) \leq 0$ is enforced for each sampled point (scenario) $\delta_i$. Using the dataset with $N=100$, $\beta = 10^{-6}$, and complexity $k=2$, the resulting risk bounds are computed as $\underline{\epsilon}=0$ and $\bar{\epsilon}=0.209$. \toolname~solved the LP in $1.2$ seconds. The corresponding optimal solution is computed as $x^\ast_{100} = \begin{bmatrix}
        0.501 & 0.498
    \end{bmatrix}^\top$\!.

\begin{figure}
    \centering
    \includegraphics[width=\linewidth]{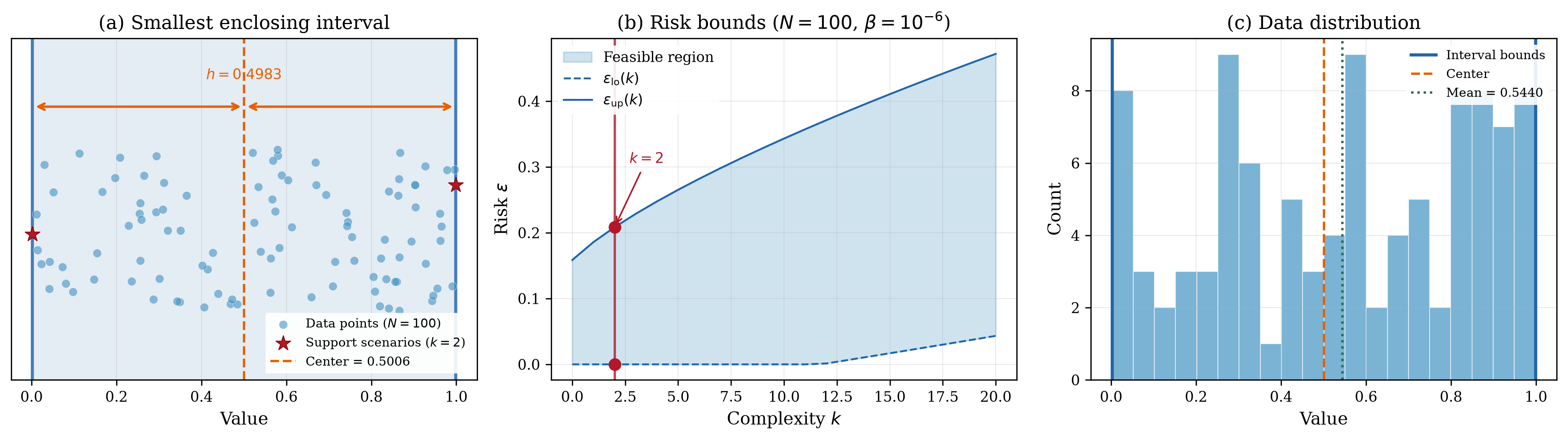}
    \caption{\textbf{Left.} Optimal half-width given sampled points. \textbf{Center.} Quantified risk bounds for $N=100$ and confidence $99.9999\%$. \textbf{Right.} Density of sampled points across the region.}
    \label{fig:half-width}
\end{figure}

    \subsubsection{Data-Driven Growth Bound for Finite Abstraction}

In the formal methods community, mathematical guarantees on system behavior are often obtained through hybridization, whereby a continuous-space system is approximated via a finite abstraction and the abstraction is analyzed to infer guarantees for the original continuous-space dynamics. The work~\cite{kazemi2024data} employs scenario optimization to construct a data-driven growth bound, formulated as a linear program and derived directly from sampled system trajectories, which in turn is used in building the finite abstraction of an unknown deterministic system. As a case study, we illustrate the computation of such a growth bound using \toolname, referring the reader to the original work for the full abstraction framework and controller synthesis procedure.

Finite abstractions group a set of continuous states into grid cells, each represented by a single point (typically the cell center). We consider a $3$-dimensional hypercube of side length $l = 1.6$, centered at $(0, 1.2, 0)$, together with vehicle dynamics of the form
    \begin{align*}
    \dot{x}(1) =  u(1)\frac{cos(\alpha + x(3))}{cos(\alpha)},
    \quad \dot{x}(2) = u(1) \frac{sin(\alpha + x(3))}{cos(\alpha)},
    \quad \dot{x}(3) = u(1)tan(u(2)),
    \end{align*}
where $\alpha = arctan(\frac{tan(u(2))}{2})$.

A total of $3127$ trajectories $(x_k, x_k')$ are sampled from the hypercube under the fixed input $u = (0.3, 0.3)$, where $(c, c')$ denotes the trajectory originating from the center. The growth bound
    \[\begin{bmatrix}
        m_{11} & m_{12} & m_{13} \\
        m_{21} & m_{22} & m_{23} \\
        m_{31} & m_{32} & m_{33} \\
    \end{bmatrix} \!\!\begin{bmatrix}0.5l\\ 0.5l\\ 0.5l\end{bmatrix}+\begin{bmatrix}
        \gamma\\ \gamma \\\gamma
    \end{bmatrix}\]
    with $\gamma$ being a bias term, is then computed using the following LP:

     \begin{align*} 
\min_{m_{ij},b_i}& \quad \sum_{i,j=1,i\neq j}^3 m_{ij} + \sum_{i=1}^3 b_i\\
    \text{subject to:}& \quad |m_{ii}| \leq b_i,\\
                    & \quad m_{i1} |x_k(1)-c(1)| + m_{i2} |x_k(2)-c(2)| + m_{i3} |x_k(3)-c(3)| \leq |x_k'(i)-c'(i)|\\
                    & ~~~\text{where}~i,j = 1,2,3~\text{and}~m_{ij} \geq 0~\text{when}~i\neq j.
\end{align*}

We consider a single scenario $\delta_k$ formed by the 6-dimensional concatenation $|x_k- c|, |x'_k - c'|$. The matrices and vectors defining the corresponding robust LP constraints are

\begin{align*}
    &A(\delta_i)=\begin{bmatrix}
        -\delta_i(1) && -\delta_i(2) && -\delta_i(3) && 0_6 && 0_3 \\
        0_3 && -\delta_i(1) && -\delta_i(2) && -\delta_i(3) && 0_6 \\
        0_6 && -\delta_i(1) && -\delta_i(2) && -\delta_i(3) && 0_3 \\
    \end{bmatrix}\!\!, \quad b(\delta_i) = \begin{bmatrix}
        \delta_i(4) \\
        \delta_i(5) \\
        \delta_i(6)
    \end{bmatrix}\!\!, \\
    &G = \begin{bmatrix}
        1 && 0_{2} && 0_6 && -1 && 0_2\\
        -1 && 0_{2} && 0_6 && -1 && 0_2\\
        0_4 && 1 && 0_{5} && -1 && 0\\
        0_4 && -1 && 0_{5} && -1 && 0\\
        0_4 && 0_4 && 1 && 0_2 && -1\\
        0_4 && 0_4 && -1 && 0_2 && -1\\
    \end{bmatrix}\!\!, \quad h = \begin{bmatrix}
        0\\0\\0\\0\\0\\0
    \end{bmatrix}\!\!.
\end{align*}

Solving the optimization yields the following rounded solution:
\begin{align*}
    M^\ast = \begin{bmatrix}
        1.000 &
0.000 &
0.005 \\
       0.000 &
1.000 &
0.009 \\
        0.000 &
0.000 &
1.000
    \end{bmatrix}\!\!, \quad b^\ast = \begin{bmatrix}
        1.000 &
1.000&
1.000
    \end{bmatrix}\!\!.
\end{align*}

Using the pre-selected bias $\gamma = 0.067$ and confidence parameter $\beta = 10^{-6}$, the resulting growth bound is computed as
$\begin{bmatrix}
    0.871 &
    0.874 &
    0.867
\end{bmatrix}^\top$.
The complexity is $k=6$; here \toolname~detects degeneracy: essentially all $3127$ scenario constraints are (numerically) active at the optimum, while only six are needed to reproduce it, so the active constraints do not form a support list and multiple support lists exist. The lower risk bound is therefore not certified (cf.\ Theorem~\ref{thm:riskcert-nondeg}) and only the upper bound is reported: $\overline{\epsilon} = 0.0103$. \toolname~computes this in $1276.8$ seconds.

\subsection{Quadratic Programming Cases}\label{appendix:QP}

\subsubsection{Radiation Therapy - Prostate Brachytherapy}

A medical physicist must design a brachytherapy treatment plan that delivers a prescribed dose of 145\,Gy to the prostate while minimizing dose to surrounding organs at risk (OARs). Catheter placement uncertainty of $\pm 5$\,mm occurs due to tissue deformation, needle deflection, and inter-fraction anatomy changes. 

The treatment uses dose-influence data~\cite{deasy2003cerr} from the Treatment and Outcome Tracking System (TROTS) Prostate\_BT\_01 patient~\cite{breedveld2019trots}. The original dataset has $55$ catheters/dwell positions and $7$ anatomical structures totaling approximately $35,000$ voxels. We subsample to $d = 50$ dwell positions (selected by total dose contribution) and $100$ representative voxels (selected by dose variance): $20$ prostate (PTV), $30$ rectum, $30$ bladder, and $20$ normal tissue. The nominal dose-influence matrix $D \in \mathbb{R}^{100 \times 50}$ relates dwell-position intensities to voxel doses, with entry $D_{v,b}$ representing the dose deposited in voxel $v$ by unit intensity of dwell position~$b$.

The TROTS dataset contains only $N=25$ scenarios, which is too small for proper analysis. Therefore, we design synthetic data inspired by this dataset. Each scenario $\delta_i \in \mathbb{R}^3$ encodes a catheter placement shift in three spatial directions (anterior--posterior, lateral, superior--inferior). A 5-level grid ($5^3 = 125$ scenarios) combined with 75 random samples from a truncated Gaussian ($\sigma = 2$\,mm) yields $N = 200$ scenarios. The perturbed dose-influence matrix follows an affine model:
\begin{equation*}
D(\delta) = D_0 + \nabla_x D \cdot \delta_1 + \nabla_y D \cdot \delta_2
+ \nabla_z D \cdot \delta_3,
\end{equation*}
where the gradient matrices are computed via finite differences on the nominal dose-influence data.

This is a Quadratic Program in the scenario approach framework. The scenario-dependent constraint matrices enforce dose bounds under each catheter shift:
\begin{equation*}
A(\delta_i)\,x + b(\delta_i) \leq 0, \quad i = 1,\ldots,N,
\end{equation*}
where $A(\delta_i) \in \mathbb{R}^{80 \times 50}$ encodes $20$ minimum tumor dose constraints ($D_v(\delta_i)\,x \geq 137.75$ Gy, i.e.\ $95\%$ of the $145$ Gy prescription), $30$ maximum rectum dose constraints ($D_v(\delta_i)\,x \leq 100$\,Gy), and 30 maximum bladder dose constraints ($D_v(\delta_i)\,x \leq 120$\,Gy). The 100 hard constraints enforce non-negativity and upper bounds on each dwell-position intensity ($0 \leq x_b \leq 150$ Gy$\cdot$s).

The objective minimizes negative tumor dose (maximizing target coverage) with $\ell_2$ regularization ($\tau = 0.1$) to encourage physically smooth dwell-time profiles:
\begin{equation*}
\min_{x} \quad c^\top x + \tfrac{1}{2}\,x^\top Q\,x + \tau\,\|x\|_2,
\end{equation*}
where $Q = D_{\mathrm{PTV}}^\top D_{\mathrm{PTV}}$ penalizes dose inhomogeneity across tumor voxels and $c$ encodes the negative mean tumor dose contribution of each dwell position. The problem is solved robustly ($\rho = 0$): all dose constraints must hold for every scenario without slack relaxation.

\begin{figure}[t]
\centering
\includegraphics[width=\textwidth]{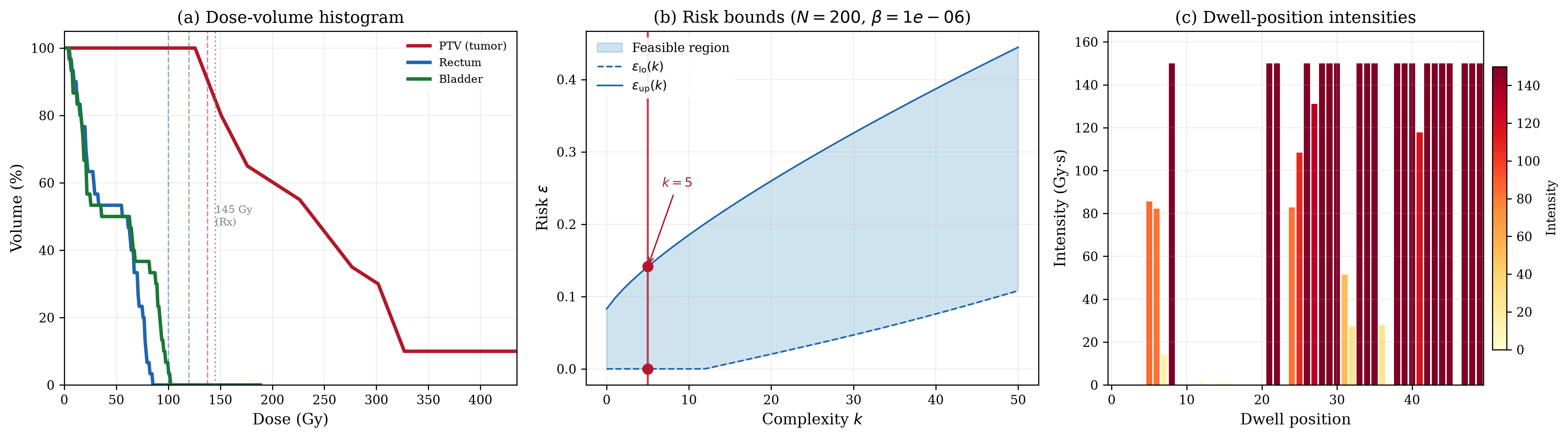}
\caption{
\textbf{Left.}~Dose-volume histogram showing tumor coverage above the 137.75\,Gy minimum with OAR doses within limits (rectum $< 100$\,Gy, bladder $< 120$\,Gy); the $x$-axis is capped at $3 \times$ prescription as brachytherapy produces extreme hot spots near dwell positions. \textbf{Center.}~Risk bounds with $k = 5$ support constraints. \textbf{Right.}~Dwell-position intensity profile showing most channels near maximum with selective suppression near OARs.}
\label{fig:brachytherapy}
\end{figure}

Figure~\ref{fig:brachytherapy} presents the solution. The optimal plan achieves a tumor $D_{95} = 145.12$ Gy (exceeding the $137.75$ Gy minimum), with maximum rectum dose of $84.6$ Gy (below the $100$ Gy limit) and maximum bladder dose of $101.6$ Gy (below the $120$ Gy limit). 

The left panel shows the dose-volume histogram (DVH). The tumor curve drops steeply beyond the prescription dose, confirming adequate target coverage, while the long tail reflects the hot spots inherent to brachytherapy (high doses near dwell positions are physically unavoidable and clinically acceptable). Both OAR curves fall well within their respective dose constraints.

The center panel shows the risk bounds. The complexity is $k = 5$; here \toolname~detects degeneracy: $20$ scenario constraints are active at the optimum while only five form the support list, so the lower risk bound is not certified (cf.\ Theorem~\ref{thm:riskcert-nondeg}) and only the upper bound is reported, $\varepsilon_{u} = 0.1414$ at $99.9999\%$ confidence ($\beta = 10^{-6}$). This guarantees that the dose constraints will be satisfied for a new random catheter placement with probability at least $85.9\%$. The low complexity relative to $d = 50$ decision variables indicates that only a small number of catheter-shift scenarios are critical to the solution geometry.

The right panel displays the dwell-position intensity profile. Most channels carry high intensity ($\approx150$\,Gy$\cdot$s, the upper bound), with a few positions at low or zero intensity. This reflects the geometry of the prostate target relative to the catheter array: dwell positions near the center of the PTV deliver maximum intensity, while those far from the target or near OARs are suppressed.

\subsubsection{Robot Motion Planning}

We consider a point robot navigating from a start position $(0.5, 2.5)$ to a goal $(9.5, 2.5)$ in a $10 \times 8$\,m workspace containing a wide rectangular wall obstacle spanning $x \in [3, 7]$, $y \in [0, 3]$. The robot has double-integrator dynamics discretized over $T = 20$ timesteps with $\Delta t = 0.4$\,s (total horizon $8$\,s). The state vector at each timestep is $(p_x, p_y, v_x, v_y)$ and the control input is $(a_x, a_y)$, yielding $d = 118$ decision variables (80 states and 38 controls).

The direct path at $y = 2.5$ is blocked by the wall, so the robot must arc above the wall face at $y = 3.0$ with a safety clearance of $d_{\mathrm{safe}} = R_{\mathrm{robot}} + d_{\mathrm{margin}} = 0.2 + 0.1 = 0.3$\,m. The wall face position is uncertain: each scenario $\delta_i \sim \mathcal{N}(0, \sigma^2)$ with $\sigma = 0.05$\,m perturbs the effective face location, generating $3$ scenario constraints per sample (one per constrained timestep in the wall's $x$-range), for a total of $N \times 3 = 1{,}500$ scenario constraints. In addition, $320$ hard constraints encode the dynamics ($4 \times 19 = 76$ equality-as-inequality pairs for state transitions), initial conditions ($8$ constraints fixing the start state), control limits ($2 \times 19 = 38$ upper and $38$ lower bounds on acceleration at $|a| \leq 2.5$\,m/s$^2$), workspace bounds ($2 \times 20 = 40$ upper and $40$ lower bounds on $y$-position), and goal proximity ($4$ constraints enforcing $\|p_T - p_{\mathrm{goal}}\|_\infty \leq 0.1$\,m). The total problem has $1,820$ inequality constraints.

The objective minimizes a weighted combination of control effort ($R = 1$) and position tracking error ($Q_{\mathrm{pos}} = 10$) relative to a straight-line reference at $y = 2.5$:
\begin{equation*}
    \min_{x} \;\; c^\top x + \frac{1}{2} x^\top Q x
    \quad \text{s.t.} \quad A(\delta_i) x \leq b(\delta_i), \;\; i = 1, \ldots, N, \quad G x \leq h.
\end{equation*}

With $N = 500$ scenarios, the solver returns a smooth arc trajectory with path length $9.79$\,m and minimum wall clearance $0.19$\,m from the safety boundary. The complexity is $k = 1$ (one support constraint). Calculating the risk bounds with $\beta = 10^{-6}$ yields $[0.0000,0.040]$, guaranteeing that with at least $99.9999\%$ confidence, the probability of constraint violation under a new scenario draw is at most $4\%$. No degeneracy is detected, and there is no regularization or relaxation ($\tau=0$ and $\rho=0$). The QP was solved in $1.6$ seconds.

\begin{figure}[t]
    \centering
    \includegraphics[width=\textwidth]{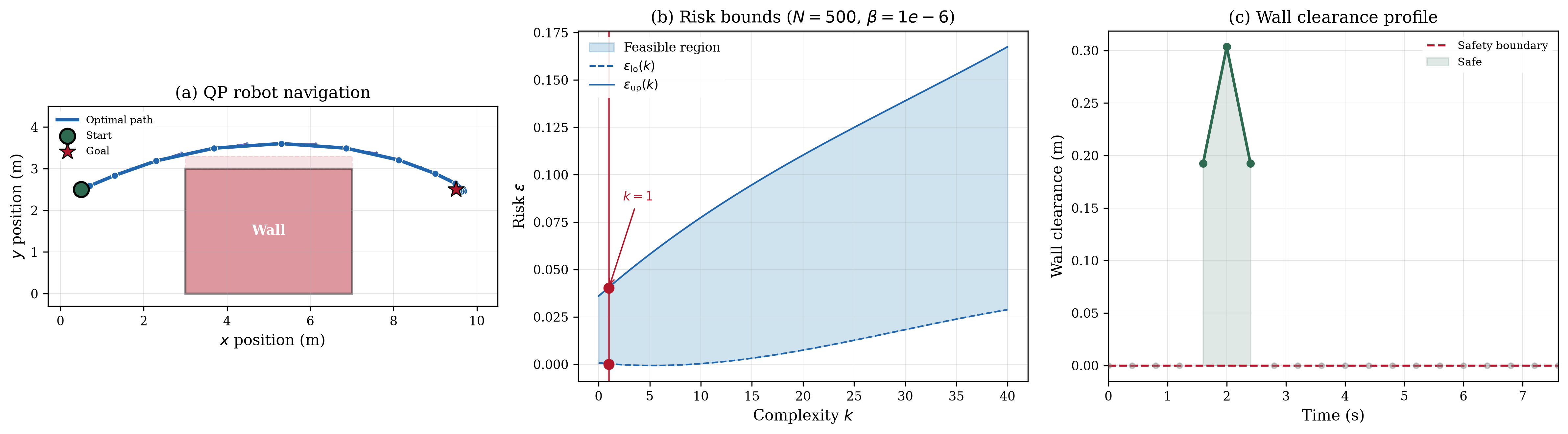}
    \caption{\textbf{Left.} Optimal trajectory arcing above the wall obstacle. \textbf{Center.} Risk bounds as a function of complexity~$k=$. \textbf{Right.} Wall clearance at timesteps within the wall's $x$-range.}
    \label{fig:robot_navigation}
\end{figure}

\subsubsection{Minimal Iris Classification}\label{appendix:QP-iris}

\begin{figure}[t!]
    \centering
\includegraphics[width=\textwidth]{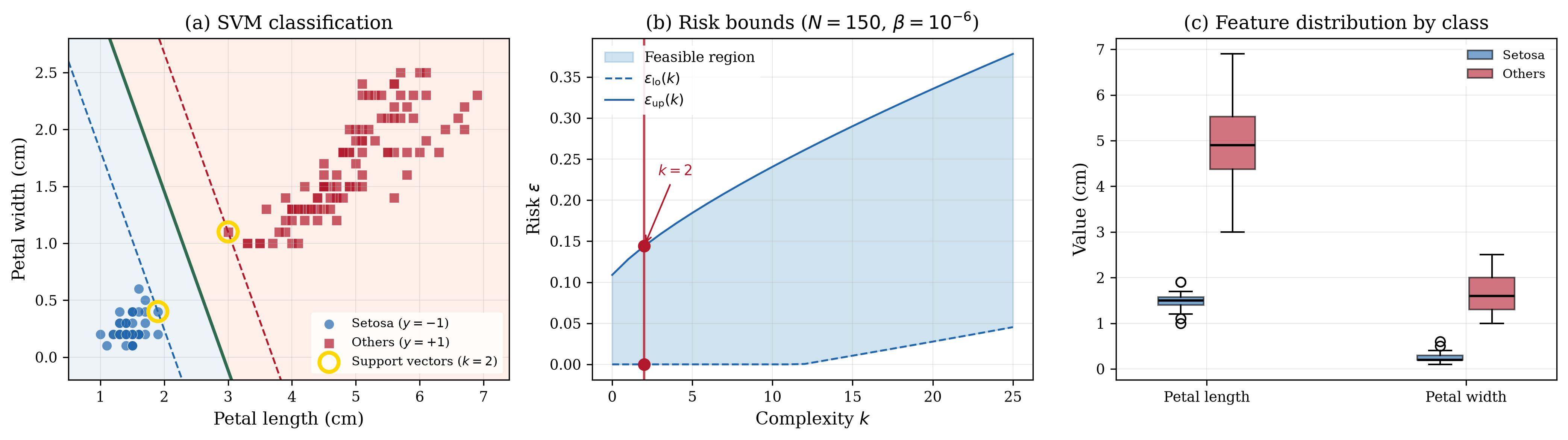}
    \caption{\textbf{Left.} Optimal separating hyperplane with margin boundaries and support vectors. \textbf{Center.} Risk bounds with complexity $k=2$. \textbf{Right.} Petal feature distributions by class.}
    \label{fig:placeholder}
\end{figure}

The small iris classification dataset~\cite{iris_53} is one of the earliest and most widely used benchmarks for evaluating classification algorithms. It includes $150$ iris plants belonging to three species, Iris Setosa, Iris Versicolour, and Iris Virginica,  each described by four measured features: sepal length, sepal width, petal length, and petal width.
    
Using a simple linear classifier, \emph{Iris Setosa} can be reliably distinguished from the other two species. Moreover, only the petal-related measurements are needed for this separation. We therefore consider a minimal binary classification problem based on two features, \emph{petal length} and \emph{petal width}, to determine whether a given iris sample belongs to the \emph{Iris Setosa} class.
    
We sample points from the $2$-dimensional feature space, labeled in blue and orange, and aim to determine a separating hyperplane that maximizes the margin between the two classes (cf. Fig~\ref{fig:placeholder}). A natural approach for this task is the linear \emph{hard-margin support vector machine (SVM)}, formulated for data points $x_i$ with binary labels $y_i$ as the following optimization problem:

     \begin{align*} 
\min_{\textbf{w},b}& \quad \frac{1}{2} \Vert\textbf{w}\Vert^2 \\
    \text{subject to:}& \quad y_i(\langle\textbf{w}, x_i\rangle + b ) \geq 1.
\end{align*}

This can be formulated as a \emph{robust QP}, where the separating hyperplane is given by $\mathbf{w}^\top x + b = 0$. The optimizer returns $\mathbf{w}^\ast = \begin{bmatrix}\mathbf{w}_1^\ast & \mathbf{w}_2^\ast\end{bmatrix}^\top$ and the corresponding intercept $b^\ast$, so that, abusing notation, the scenario solution is $x^\ast_N = \begin{bmatrix}\textbf{w}_1^\ast & \textbf{w}_2^\ast & b^\ast\end{bmatrix}^\top\!\!.$ Each scenario $\delta_i$ consists of the feature–label pair $(\delta_i(1),\delta_i(2),\delta_i(3))$, where $(\delta_i(1),\delta_i(2))$ are the feature coordinates and $\delta_i(3)$ is the binary class label. The associated matrices defining the robust QP in~\eqref{eq:QP-full} are given in~\eqref{new2}. The risk is computed using the dataset with $N=150$, confidence level $\beta = 10^{-6}$, and complexity $k = 2$, yielding risk bounds $\underline{\epsilon} = 0$ and $\overline{\epsilon} = 0.144$, solved in $1.2$ seconds. The resulting optimal solution $x^\ast_{150}$ corresponds to the linear classifier \[3.79 = 1.29x_1 + 0.82x_2.\]

\subsection{Semidefinite Programming Cases}\label{appendix:SDP}

\subsubsection{Robust Covariance Estimation}

We seek the minimum-trace positive semidefinite covariance matrix $\Sigma$ that dominates all subsample covariances drawn from the UCI Wine dataset~\cite{wine_109}. The scenario approach provides probabilistic guarantees that $\Sigma$ will also dominate the covariance of a new random subsample drawn from the same population.

The Wine dataset contains 178 samples with 13 chemical measurements per wine. We select the first 5 features (alcohol, malic acid, ash, alcalinity of ash, magnesium) and standardize each to zero mean and unit variance, yielding the full-sample covariance $\Sigma_{\mathrm{full}} \in\mathbb{R}^{5 \times 5}$.

The decision variable is $x = [\sigma_{11},\, \sigma_{12},\, \ldots,\, \sigma_{55}] \in \mathbb{R}^{15}$, the 15 free entries of the $5 \times 5$ symmetric matrix $\Sigma = \sum_{k=1}^{15} x_k \Pi_k$, where $\Pi_k$ are the standard basis matrices for the symmetric $5 \times 5$ space. Each scenario $\delta_i \in \mathbb{R}^{5}$ is one of the $178$ wine data points ($N = 178$ scenarios, one per wine).

\begin{figure}[t]
\centering
\includegraphics[width=\textwidth]{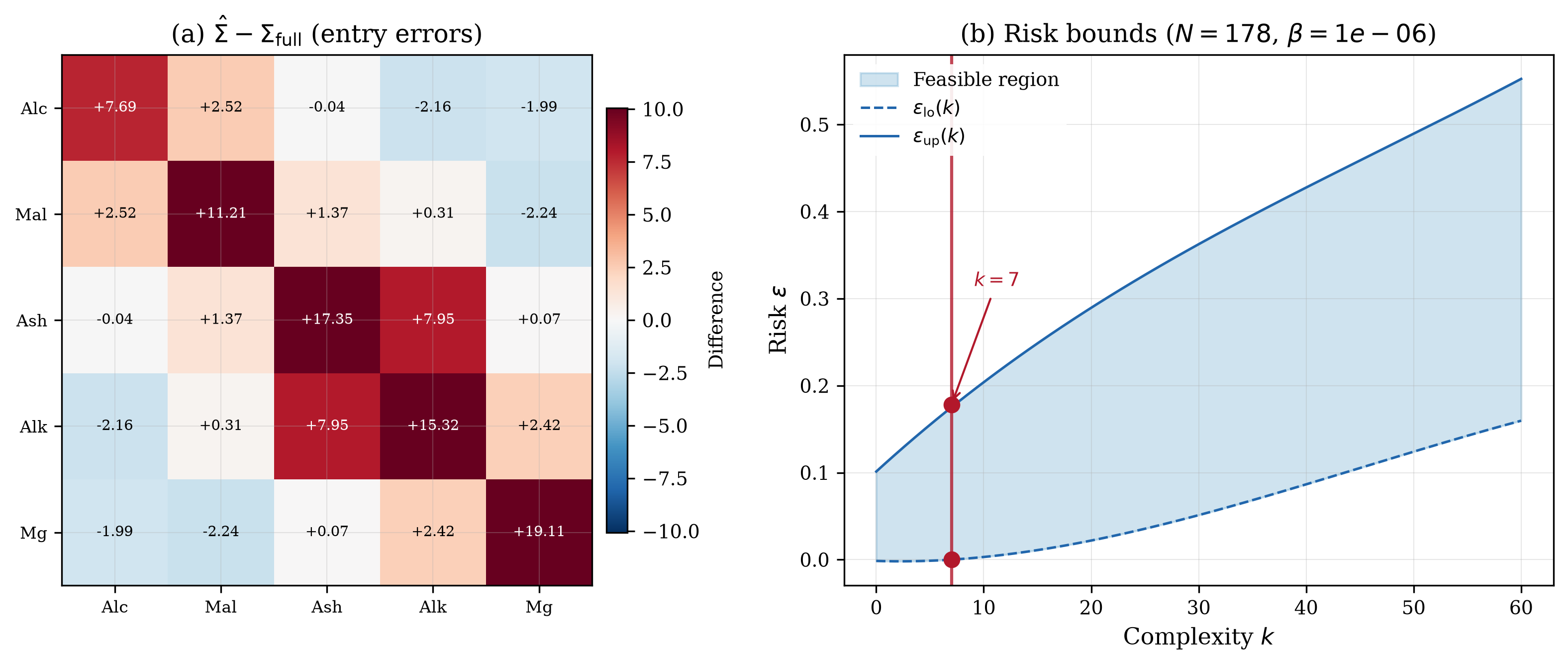}
\caption{\textbf{Left.}~Entry-wise error $\hat{\Sigma} - \Sigma_{\mathrm{full}}$; diagonal inflation reflects the domination requirement. \textbf{Center.}~Risk bounds with $k = 7$ support constraints. \textbf{Right.}~Eigenvalue spectrum comparison: the estimated covariance uniformly dominates the full-sample covariance in every principal direction.}
\label{fig:covariance_wine}
\end{figure}

This is a Semidefinite Program (SDP) in the scenario approach inequality form. The scenario-dependent constraint matrices encode covariance domination as a $5 \times 5$ LMI:
\begin{equation*}
F_0(\delta_i) + \sum_{j=1}^{15} x_j\,F_j(\delta_i) \preceq 0,
\end{equation*}
where $F_0(\delta_i) = S_{\mathrm{sub}}(\delta_i)$ and $F_j = -\Pi_j$, so that $S_{\mathrm{sub}}(\delta_i) - \Sigma \preceq 0$, i.e.\ $\Sigma \succeq S_{\mathrm{sub}}(\delta_i)$ for each scenario. The hard constraint enforces $\Sigma \succ 0$ via $E_0 = 0$ and $E_j = -\Pi_j$. The objective minimizes $\operatorname{trace}(\Sigma)$ using $c = [1,\,0,\,0,\,1,\,\ldots,\,1]^\top$ and $Q = 0$. There is no slack penalty ($\rho = 0$) and no regularization ($\tau = 0$).

Figure~\ref{fig:covariance_wine} presents the solution. The left panel shows the entry-wise error $\hat{\Sigma} - \Sigma_{\mathrm{full}}$ as a heatmap. The diagonal entries are uniformly positive (ranging from $+0.54$ to $+0.93$), reflecting the conservatism inherent in the domination requirement: the estimated covariance must exceed \emph{every} subsample covariance, which inflates the diagonal. Off-diagonal errors are smaller and mixed in sign.

The right panel shows the risk bounds. The complexity is $k = 7$ with no degeneracy, yielding risk bounds $[\varepsilon_{\ell},\, \varepsilon_{u}] = [0.0000,\,0.178]$ at $99.9999\%$ confidence ($\beta = 10^{-6}$). This guarantees that $\hat{\Sigma}$ will dominate the covariance of a new random $50$-wine subsample with probability at least $82.2\%$. \toolname~solved the SDP in $2.1$ seconds.

\subsubsection{Minimum Enclosing Ellipsoid - Breast Cancer Wisconsin}

Given a collection of data points, we seek the tightest ellipsoid (in the trace sense) that contains all sampled observations. This is a fundamental problem in robust statistics and anomaly detection: the scenario approach provides probabilistic guarantees that unseen data points from the same distribution will also be contained.

The Breast Cancer Wisconsin dataset~\cite{breast_cancer_wisconsin_diagnostic_17} contains $569$ samples with $30$ features. We select the first $5$ features (mean radius, mean texture, mean perimeter, mean area, mean smoothness) and standardize each to zero mean and unit variance, with data center $\bar{x} = \operatorname{mean}(\text{data}) \approx \mathbf{0}$.

The decision variable is $x = [p_{11},\,p_{12},\,\ldots,\,p_{55}] \in\mathbb{R}^{15}$, the entries of the $5 \times 5$ symmetric shape matrix $P = \sum_{k=1}^{15} x_k \Pi_k$, defining the ellipsoid
\begin{equation*}
\mathcal{E} = \bigl\{ z \in \mathbb{R}^5 : (z - \bar{x})^\top P\, (z -
\bar{x}) \leq 1 \bigr\}.
\end{equation*}

Each scenario $\delta_i \in \mathbb{R}^5$ is a data point drawn from the dataset ($N = 569$ scenarios).

\begin{figure}[t]
\centering
\includegraphics[width=\textwidth]{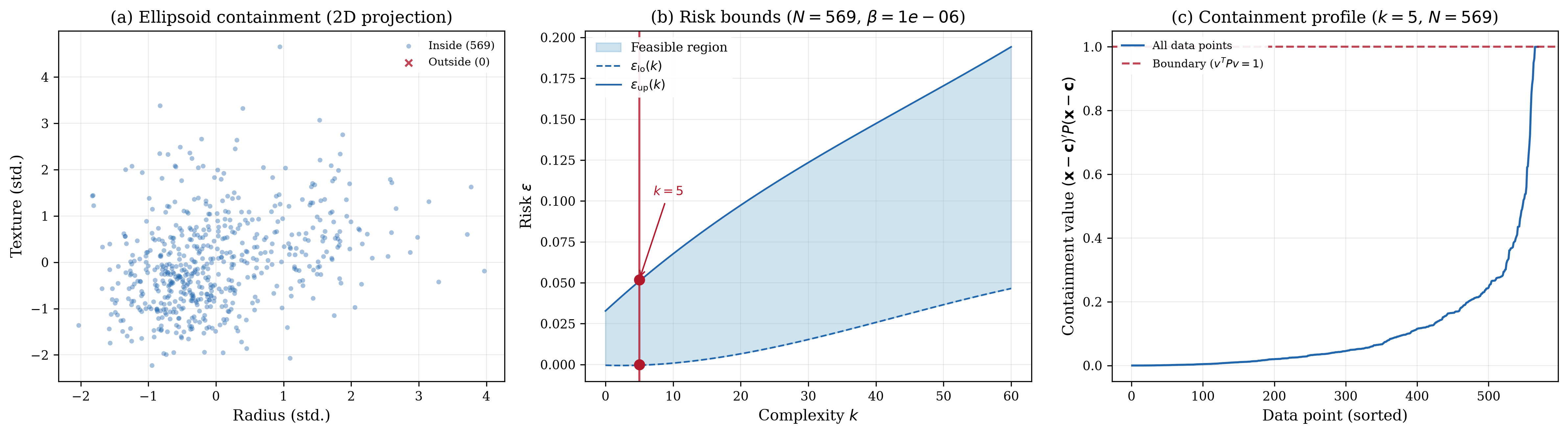}
\caption{
\textbf{Left.}~2D projection showing 569/569 points contained (100\%) inside the region. \textbf{Center.}~Risk bounds with $k = 5$ support constraints.\textbf{Right.}~Containment profile: sorted $(z - \bar{x})^\top P\,(z - \bar{x})$
values for all 569 data points; the boundary at 1.0 separates interior from
exterior points.}
\label{fig:ellipsoid}
\end{figure}

This is a Semidefinite Program (SDP) in the scenario approach inequality form. The scenario-dependent constraint matrices encode point containment as a $1 \times 1$ (scalar) LMI:
\begin{equation*}
F_0(\delta_i) + \sum_{j=1}^{15} x_j\,F_j(\delta_i) \leq 0,
\end{equation*}
where $F_0(\delta_i) = -1$ and $F_j(\delta_i) = (\delta_i - \bar{x})^\top \Pi_j\,(\delta_i - \bar{x})$, so that $(\delta_i - \bar{x})^\top P\, (\delta_i - \bar{x}) - 1 \leq 0$. The hard constraint enforces $P \succ 0$ via $E_0 = 0$ and $E_j = -\Pi_j$. The objective maximizes $\operatorname{trace}(P)$ (i.e.\ minimizes $-\operatorname{trace}(P)$) using $c = [-1,\,0,\,0,\,-1,\,\ldots,\,-1]^\top$ and $Q = 0$, finding the tightest ellipsoid. There is no slack penalty ($\rho = 0$) and no regularization ($\tau = 0$).

Figure~\ref{fig:ellipsoid} presents the solution. The left panel shows the 2D projection of the ellipsoid onto the first two standardized features (mean radius vs.\ mean texture). Since all 569 data points are used as scenarios, the ellipsoid contains all of them (100\% containment).

The center panel shows the risk bounds. The complexity is $k = 5$ with no degeneracy, yielding risk bounds $[\varepsilon_{\ell},\,\varepsilon_{u}] = [0.000,\,0.052]$ at $99.9999\%$ confidence ($\beta = 10^{-6}$), guaranteeing that a new random data point will fall inside the ellipsoid with probability at least $94.8\%$.

The right panel displays the containment profile: for each of the 569 data points (sorted by containment value), the quantity $(z - \bar{x})^\top P\,(z - \bar{x})$ is plotted. Points below the dashed line at 1.0 are contained.
The optimal $P$ has trace~$44.59$ and eigenvalues $\{\approx 0,\;\approx 0,\;\approx 0,\;\approx 0,\;44.59\}$, revealing that the standardized data effectively lies in a 1-dimensional subspace. The first 5 breast cancer features are highly correlated (mean radius, mean perimeter, and mean area are near-linearly dependent), so $P$ concentrates its ``tightness'' along the single direction of genuine data spread. The SDP was solved in $235.6$ seconds.

\subsubsection{LPV Stability}\label{appendix:SDP-lpv}

\begin{figure}[t]
    \centering
    \includegraphics[width=\linewidth]{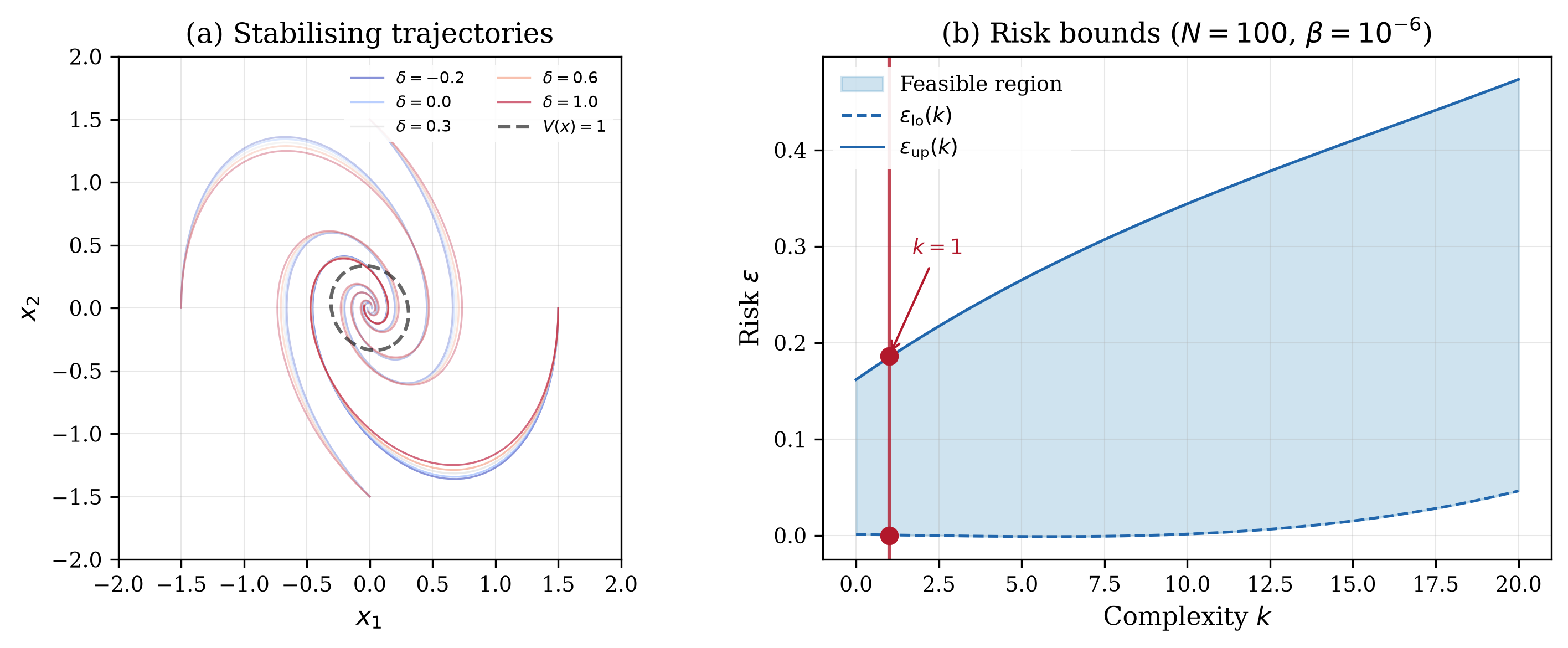}
    \caption{\textbf{Left.} Stabilizing trajectories with for different scenarios $\delta$. \textbf{Right.} Risk bounds for the LPV stability with complexity $k=1$.}
    \label{fig:lpv_stability}
\end{figure}

Consider a linear parameter-varying (LPV) system $\dot{x} = A(\delta)\,x$ with
\begin{equation*}
A(\delta) = \underbrace{\begin{bmatrix} 0 & 1 \\ -2 & -1 \end{bmatrix}}_{A_0}
+ \;\delta\;\underbrace{\begin{bmatrix} 0 & 0 \\ 0.3 & 0.1 \end{bmatrix}}_{A_1},
\qquad \delta \in [-0.22,\; 1.0],
\end{equation*}
where $A_0$ is a stable damped oscillator and $A_1$ captures parameter-varying uncertainty in the restoring and damping terms. The goal is to find a common quadratic Lyapunov function $V(x) = x^\top P\,x$ with $P \succ 0$ certifying stability across all parameter values, i.e.\ $A(\delta)^\top P + P\,A(\delta) \preceq 0$.

This maps to the SDP inequality form~\eqref{eq:SDP-full-ineq} with decision variable $x = [p_{11},\; p_{12},\; p_{22}]^\top \in \mathbb{R}^3$, the upper-triangular entries of the $2\times 2$ symmetric matrix~$P$. Let $\Pi_1 = \left[\begin{smallmatrix}1&0\\0&0\end{smallmatrix}\right]$, $\Pi_2 = \left[\begin{smallmatrix}0&1\\1&0\end{smallmatrix}\right]$, $\Pi_3 = \left[\begin{smallmatrix}0&0\\0&1\end{smallmatrix}\right]$ be the basis matrices for the symmetric $2\times 2$ space, so that $P = \sum_{j=1}^{3} x_j \Pi_j$. The scenario-dependent constraint matrices in~\eqref{eq:SDP-constraint1-ineq} are
\begin{equation*}
F_0(\delta_i) = 0, \qquad
F_j(\delta_i) = A(\delta_i)^\top \Pi_j + \Pi_j\, A(\delta_i),
\quad j = 1,2,3,
\end{equation*}
encoding the Lyapunov inequality $A(\delta_i)^\top P + P\,A(\delta_i) \preceq \zeta_i I$. The hard constraint in~\eqref{eq:SDP-constraint2-ineq} enforces $P \succ 0$ via
\begin{equation*}
E_0 = 0, \qquad E_j = -\Pi_j, \quad j = 1,2,3.
\end{equation*}
The objective uses $c = [-1,\; 0,\; -1]^\top$ and $Q = 0.1\,\mathbb{I}_3$, which maximizes $\mathrm{trace}(P)$ with light quadratic regularization ($\tau = 0$, $\rho = 1$).

\toolname\ solves the SDP with $N = 100$ scenarios. The optimal Lyapunov matrix is
\begin{equation*}
P^* = \begin{bmatrix} 10.67 & 1.30 \\ 1.30 & 9.01 \end{bmatrix},
\end{equation*}
with complexity $k = 1$, no degeneracy, and risk bounds $[0.0, 0.1858]$ at $99.9999\%$ confidence. Only a single scenario constraint supports the solution, corresponding to the smallest sampled value $\delta \approx -0.19$, where the maximum eigenvalue of $A^\top P + PA$ approaches zero. The SDP is solved in $1.7$ seconds.

\subsubsection{Quadratic Stability - Coupled Oscillator}

Consider a $3$-state coupled oscillator with two uncertain parameters:
\begin{equation*}
\dot{x} = \bigl(A_0 + \delta_1 A_1 + \delta_2 A_2\bigr)\,x,\qquad (\delta_1, \delta_2) \in [-1,\;1]^2,
\end{equation*}
where
\begin{equation*}
A_0 = \begin{bmatrix} -0.3 & 1.0 & 0 \\ -2.0 & -1.0 & 0.2 \\ 0.2 & 0 & -1.2 \end{bmatrix}\!,\quad
A_1 = \begin{bmatrix} 0 & 0 & 0 \\ 0.5 & 0.15 & 0 \\ 0 & 0 & 0.3 \end{bmatrix}\!,\quad
A_2 = \begin{bmatrix} 0 & 0 & 0 \\ 0 & 0.3 & 0.15 \\ 0.15 & 0 & 0.45 \end{bmatrix}\!.
\end{equation*}
Here $A_0$ models a well-damped coupled oscillator, $A_1$ perturbs the stiffness of the first mode, and $A_2$ perturbs the inter-mode coupling and damping. This maps to the SDP inequality form~\eqref{eq:SDP-full-ineq} with decision variable \[x = [p_{11}, p_{12}, p_{13}, p_{22}, p_{23}, p_{33}]^\top \in \mathbb{R}^6,\] the six independent entries of the $3 \times 3$ symmetric Lyapunov matrix~$P = \sum_{j=1}^{6} x_j \Pi_j$, where $\Pi_j$ are the standard basis matrices for the symmetric $3\times 3$ space. The scenario constraint matrices in~\eqref{eq:SDP-constraint1-ineq} are
\begin{equation*}
F_0(\delta_i) = 0, \qquad F_j(\delta_i) = A(\delta_i)^\top \Pi_j    + \Pi_j\, A(\delta_i), \quad j = 1,\ldots,6.
\end{equation*}
The hard constraint in~\eqref{eq:SDP-constraint2-ineq} enforces $P \succeq 0.01\,\mathbb{I}_3$ via
\begin{equation*}
E_0 = 0.01\,\mathbb{I}_3, \qquad E_j = -\Pi_j, \quad j = 1,\ldots,6.
\end{equation*}
The objective uses $c = [1, 0, 0, 1, 0, 1]^\top$ and $Q = 0.001\,\mathbb{I}_6$, which minimizes $\mathrm{trace}(P)$ with no regularization ($\tau = 0$, $\rho = 0$).

\begin{figure}
    \centering
    \includegraphics[width=\linewidth]{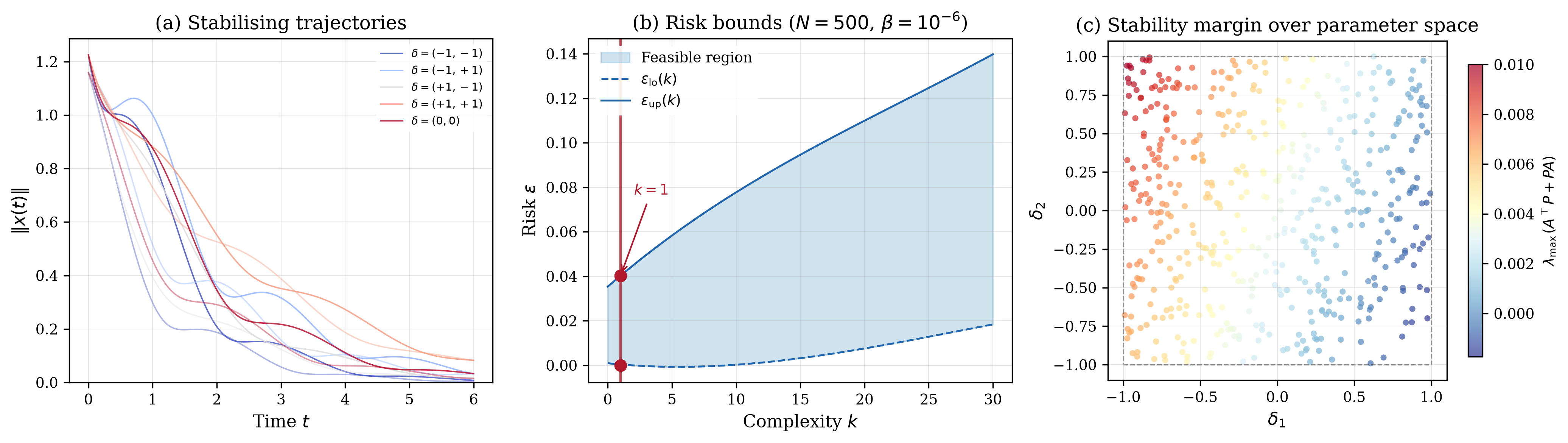}
    \caption{\textbf{Left.} Stabilizing trajectories for example scenarios $\delta$. \textbf{Center.} Risk bounds for the quadratic stability with complexity $k=1$. \textbf{Right.} Stability margins for each sampled scenario, with red less stable and blue more stable.}
    \label{fig:quadratic-stability}
\end{figure}

\toolname\ solves the SDP with $N = 500$ scenarios sampled uniformly over $[-1,1]^2$. The optimal solution is \[P^* \approx \begin{bmatrix} 0.0101 & -0.0005 & 0.0001 \\
-0.0005 & 0.0127 & -0.0003 \\ 0.0001 & -0.0003 & 0.0100 \end{bmatrix}\] with complexity $k = 1$, no degeneracy, and risk bounds $[0, 0.0400]$ at $99.9999\%$ confidence. The Lyapunov matrix is not a scalar multiple of the identity: $P^*_{22} = 0.0127$ is noticeably larger than $P^*_{11}$ and $P^*_{33} \approx 0.010$, reflecting the need for additional Lyapunov weight on the second state to certify stability across the parameter space. One scenario constraint supports the solution ($k=1$), corresponding to a near-worst-case parameter realization close to the $(\delta_1, \delta_2) = (1, 1)$ corner where the stability margin is tightest.

\section{JSON Program Files for the Illustrative Examples}\label{appendix:json}

Instead of entering each matrix through the dialog boxes of the web interface (Section~\ref{sec:gui}), a complete problem can be loaded in one step from a JSON program file through button <10> in Fig.~\ref{fig:web-interface}, either by uploading the file or by pasting its content. Listings~\ref{lst:json-lp}--\ref{lst:json-sdp} give the program files of Examples~\ref{ex:LP}, \ref{ex:QP} and~\ref{ex:SDP1}, taken from the \texttt{benchmarks} folder of the repository. Each listing can be copied and pasted directly into \toolname; after uploading the scenarios through <7>, pressing \emph{Solve} reproduces the results reported in Appendix~\ref{appendix}.

The keys follow the notation of Section~\ref{sec:scenario-convex-optimization}. The key \texttt{type} selects the program class (\texttt{LP}, \texttt{QP} or \texttt{SDP}), and \texttt{mode} selects symbolic or numeric input. The keys \texttt{c}, \texttt{Q}, \texttt{G} and \texttt{h} hold the matrices of the same name, \texttt{A\_d} and \texttt{b\_d} hold $A(\delta)$ and $b(\delta)$, and \texttt{F\_d} and \texttt{E} hold $F_0(\delta),\ldots,F_d(\delta)$ and $E_0,\ldots,E_d$ under the keys \texttt{"0"}, \texttt{"1"}, \ldots. The keys \texttt{rho}, \texttt{tau}, \texttt{x\_ref} and \texttt{p} hold $\rho$, $\tau$, $\bar{x}$ and the norm order $p$, and \texttt{confidence} holds $\beta$. The dimensions $d$, $\mathfrak{m}$ and $\mathfrak{n}$ are given by \texttt{n\_x}, \texttt{rows\_A} and \texttt{rows\_G} for LP and QP, and by \texttt{n\_x}, \texttt{lmi\_size} and \texttt{lmi\_e\_size} for SDP. Matrices are written row by row, and every entry is a string, so that the entries of $A(\delta)$, $b(\delta)$ and $F_j(\delta)$ can be expressions in the scenario components \texttt{delta[0]}, \texttt{delta[1]}, \ldots. The scenarios themselves are not part of the program file.

\begin{lstlisting}[style=json, float=!htbp, label=lst:json-lp, caption={JSON program file for Example~\ref{ex:LP} (smallest enclosing interval, robust LP; case study in Appendix~\ref{appendix:LP-interval}), taken from \texttt{benchmarks/\allowbreak LP\_half\_width\_2d/\allowbreak data/\allowbreak program\_symbolic.json}. Here \texttt{delta[0]} is the sampled point $\delta_i$. The fields \texttt{G} and \texttt{h} (no hard constraints) and \texttt{x\_ref} and \texttt{p} (no regularization) are not used in this example: they can be omitted, as in the repository file, or left empty, as shown. The listing can be copied and pasted directly into \toolname.}]
{
  "type": "LP",
  "A_d": [["-1", "-1"],
          ["1", "-1"]],
  "b_d": [["delta[0]"], ["-delta[0]"]],
  "G": [],
  "h": [],
  "c": [["0"], ["1"]],
  "rho": "0",
  "tau": "0",
  "x_ref": [],
  "p": "",
  "confidence": "1e-06",
  "mode": "symbolic",
  "n_x": 2,
  "rows_A": 2,
  "rows_G": 0
}
\end{lstlisting}

\begin{lstlisting}[style=json, float=!htbp, label=lst:json-qp, caption={JSON program file for Example~\ref{ex:QP} (support vector machine, robust QP; Iris case study in Appendix~\ref{appendix:QP-iris}), taken from \texttt{benchmarks/\allowbreak QP\_Iris\_minimal\_3d/\allowbreak data/\allowbreak program\_symbolic.json}. Here \texttt{delta[0]}, \texttt{delta[1]} and \texttt{delta[2]} are $\delta_i(1)$, $\delta_i(2)$ and $\delta_i(3)$ in~\eqref{new2}. The fields \texttt{G} and \texttt{h} (no hard constraints) and \texttt{x\_ref} and \texttt{p} (no regularization) are not used in this example: they can be omitted, as in the repository file, or left empty, as shown. The listing can be copied and pasted directly into \toolname.}]
{
  "type": "QP",
  "A_d": [["-delta[0]*delta[2]", "-delta[1]*delta[2]", "-delta[2]"]],
  "b_d": [["1"]],
  "G": [],
  "h": [],
  "c": [["0"], ["0"], ["0"]],
  "Q": [["1", "0", "0"],
        ["0", "1", "0"],
        ["0", "0", "0"]],
  "rho": "0",
  "tau": "0",
  "x_ref": [],
  "p": "",
  "confidence": "1e-06",
  "mode": "symbolic",
  "n_x": 3,
  "rows_A": 1,
  "rows_G": 0
}
\end{lstlisting}

\begin{lstlisting}[style=json, float=!htbp, label=lst:json-sdp, caption={JSON program file for Example~\ref{ex:SDP1} (LPV stability, relaxed SDP; case study in Appendix~\ref{appendix:SDP-lpv}), taken from \texttt{benchmarks/\allowbreak SDP\_LPV\_stability\_3d/\allowbreak data/\allowbreak program\_symbolic.json}. The entries \texttt{"0"} to \texttt{"3"} of \texttt{F\_d} and \texttt{E} are $F_0(\delta),\ldots,F_3(\delta)$ and $E_0,\ldots,E_3$, and \texttt{delta[0]} is the scalar parameter $\delta$. The fields \texttt{x\_ref} and \texttt{p} (no regularization, $\tau=0$) are not used in this example: they can be omitted, as in the repository file, or left empty, as shown. The listing can be copied and pasted directly into \toolname.}]
{
  "type": "SDP",
  "c": [["-1"], ["0"], ["-1"]],
  "Q": [["0.1", "0", "0"],
        ["0", "0.1", "0"],
        ["0", "0", "0.1"]],
  "F_d": {
    "0": [["0", "0"],
          ["0", "0"]],
    "1": [["0", "1"],
          ["1", "0"]],
    "2": [["-4+0.6*delta[0]", "-1+0.1*delta[0]"],
          ["-1+0.1*delta[0]", "2"]],
    "3": [["0", "0.3*delta[0]-2"],
          ["0.3*delta[0]-2", "-2+0.2*delta[0]"]]
  },
  "E": {
    "0": [["0", "0"],
          ["0", "0"]],
    "1": [["-1", "0"],
          ["0", "0"]],
    "2": [["0", "-1"],
          ["-1", "0"]],
    "3": [["0", "0"],
          ["0", "-1"]]
  },
  "rho": "1.0",
  "tau": "0",
  "x_ref": [],
  "p": "",
  "confidence": "1e-06",
  "mode": "symbolic",
  "n_x": 3,
  "lmi_size": 2,
  "lmi_e_size": 2
}
\end{lstlisting}

\clearpage

\section{Detailed Function Implementation}\label{appendix:imp}

The callable parameters \lstinline|A_d|, \lstinline|b_d| (LP/QP) and \lstinline|F_d| (SDP) are typically generated by the parsing module (\texttt{parsing.py}), which provides \texttt{generate\_matrix\_function()} and \\\texttt{generate\_tensor\_function()} for converting user-supplied string expressions\\ (e.g. \lstinline|"[[delta[0]+1, 0], [0, delta[1]]]"|) into evaluable Python functions, and their static counterparts \lstinline|generate_matrix()| and \lstinline|generate_tensor()| for constant matrices. All three solvers call the helper function \lstinline|get_support()| (Section~\ref{appendix:imp-get_support}) post-solve to identify support constraints and compute complexity.

\subsection{\lstinline|solve\_lp|}\label{appendix:imp-solve_lp}
\begin{lstlisting}
def solve_lp(deltas, A_d, b_d, G, h, c, tau=0.0, x_ref=np.array([0.0]), rho=0.0, norm_type=2, solver=None, include_slack=None):
return x_out, zeta_out, cost_out, N, complexity, constraints, degeneracy
\end{lstlisting}

\subsubsection*{Parameters:}
\begin{itemize}
  \item \lstinline|deltas|: \lstinline|numpy.ndarray| of shape \((N,K)\). Each row \(\delta_i\) is substituted into \lstinline|A_d| and \lstinline|b_d| to form per-scenario constraints.
  \item \lstinline|A_d|: \lstinline|Callable[[numpy.ndarray], numpy.ndarray]|. Given a delta row returns an \(m\times n\) constraint matrix \(A(\delta)\).
  \item \lstinline|b_d|: \lstinline|Callable[[numpy.ndarray], numpy.ndarray]|. Given a delta row returns an \(m\times 1\) right-hand side vector \(b(\delta)\).
  \item \lstinline|G|: \lstinline|numpy.ndarray| of shape \((p,n)\) (may be an empty array to omit hard constraints).
  \item \lstinline|h|: \lstinline|numpy.ndarray| of shape \((p,1)\) corresponding to \lstinline|G|.
  \item \lstinline|c|: \lstinline|numpy.ndarray| of length \(n\) (or shape \((n,1)\)), objective linear coefficients.
  \item \lstinline|tau|: \lstinline|float|. Weight for the regularization term \(\|x-x_{\mathrm{ref}}\|_{p}\), which is omitted when \(\tau=0\).
  \item \lstinline|x_ref|: \lstinline|numpy.ndarray| broadcastable to same shape as \(x\) (default \lstinline|np.array([0.0])|).
  \item \lstinline|rho|: \lstinline|float|. Weight of the penalty \(\rho\mathbf{1}^\top \zeta\) on the nonnegative slack variables \(\zeta\) of length \(N\) (one per scenario), which are added when \lstinline|include_slack| is true (or, if it is \lstinline|None|, when \(\rho\neq 0\)).
  \item \lstinline|norm_type|: \lstinline|int|, \lstinline|float| or \lstinline|str|. Order \(p\) of the vector norm in the regularization term: any \(p\geq 1\) (default \lstinline|2|), \lstinline|"inf"|, or \lstinline|"fro"| (the same as \(p=2\)).
  \item \lstinline|solver|: \lstinline|Optional[str]|. CVXPY solver\footnote{\url{https://www.cvxpy.org/tutorial/solvers/index.html}} or \lstinline|None| (chooses default solver).
  \item \lstinline|include_slack|: \lstinline|Optional[bool]|. Whether to add the slack variables \(\zeta\) of the relaxation formulation. \lstinline|None| (default) adds them only when \(\rho\neq 0\).
\end{itemize}

\subsubsection*{Return values:}
\begin{itemize}
  \item \lstinline|x_out|: \lstinline|numpy.ndarray| of shape \((n,1)\). Optimizer solution for \(x\).
  \item \lstinline|zeta_out|: \lstinline|numpy.ndarray| of shape \((N,)\). Slack variable values (zeros if no slack variables are used).
  \item \lstinline|cost_out|: \lstinline|float|. Optimal objective value.
  \item \lstinline|N|: \lstinline|int|. Number of rows \(N\) in \lstinline|deltas| used to build robust constraints.
  \item \lstinline|complexity|: \lstinline|int|. Number of scenario constraints in the support list identified post-solve.
  \item \lstinline|constraints|: \lstinline|list| of CVXPY constraint objects: the \(N\) scenario constraints (the hard constraints are not included).
  \item \lstinline|degeneracy|: \lstinline|bool|. Flag indicating whether degeneracy was detected when computing the support list.
\end{itemize}

\subsubsection*{Exceptions and Error Checking:}
\begin{itemize}
  \item Raises \lstinline|ValueError| if \lstinline|deltas| has no rows.
  \item Raises \lstinline|ValueError| if CVXPY problem status is not \lstinline|"optimal"| or \lstinline|"optimal_inaccurate"|.
\end{itemize}

\subsection{ \texttt{solve\_qp}}\label{appendix:imp-solve_qp}
\begin{lstlisting}
def solve_qp(deltas, A_d, b_d, G, h, c, Q, tau=0.0, x_ref=np.array([0.0]), rho=0.0, norm_type=2, solver=None, include_slack=None):
return x_out, zeta_out, cost_out, N, complexity, constraints, degeneracy
\end{lstlisting}

\subsubsection*{Parameters:}
\begin{itemize}
  \item \lstinline|deltas|: \lstinline|numpy.ndarray| of shape \((N,K)\). Each row \(\delta_i\) is substituted into \lstinline|A_d| and \lstinline|b_d| to form per-scenario constraints.
  \item \lstinline|A_d|: \lstinline|Callable[[numpy.ndarray], numpy.ndarray]|. Given a delta row returns an \(m\times n\) constraint matrix \(A(\delta)\).
  \item \lstinline|b_d|: \lstinline|Callable[[numpy.ndarray], numpy.ndarray]|. Given a delta row returns an \(m\times 1\) right-hand side vector \(b(\delta)\).
  \item \lstinline|G|: \lstinline|numpy.ndarray| of shape \((p,n)\) (may be an empty array to omit hard constraints).
  \item \lstinline|h|: \lstinline|numpy.ndarray| of shape \((p,1)\) corresponding to \lstinline|G|.
  \item \lstinline|c|: \lstinline|numpy.ndarray| of length \(n\) (or shape \((n,1)\)), objective linear coefficients.
  \item \lstinline|Q|: \lstinline|numpy.ndarray| of shape \((n,n)\). Symmetric positive semidefinite quadratic cost matrix.
  \item \lstinline|tau|: \lstinline|float|. Weight for the regularization term \(\|x-x_{\mathrm{ref}}\|_{p}\), which is omitted when \(\tau=0\).
  \item \lstinline|x_ref|: \lstinline|numpy.ndarray| broadcastable to same shape as \(x\) (default \lstinline|np.array([0.0])|).
  \item \lstinline|rho|: \lstinline|float|. Weight of the penalty \(\rho\mathbf{1}^\top \zeta\) on the nonnegative slack variables \(\zeta\) of length \(N\) (one per scenario), which are added when \lstinline|include_slack| is true (or, if it is \lstinline|None|, when \(\rho\neq 0\)).
  \item \lstinline|norm_type|: \lstinline|int|, \lstinline|float| or \lstinline|str|. Order \(p\) of the vector norm in the regularization term: any \(p\geq 1\) (default \lstinline|2|), \lstinline|"inf"|, or \lstinline|"fro"| (the same as \(p=2\)).
  \item \lstinline|solver|: \lstinline|Optional[str]|. CVXPY solver or \lstinline|None| (chooses default solver).
  \item \lstinline|include_slack|: \lstinline|Optional[bool]|. Whether to add the slack variables \(\zeta\) of the relaxation formulation. \lstinline|None| (default) adds them only when \(\rho\neq 0\).
\end{itemize}

\subsubsection*{Return values:}
\begin{itemize}
  \item \lstinline|x_out|: \lstinline|numpy.ndarray| of shape \((n,1)\). Optimizer solution for \(x\).
  \item \lstinline|zeta_out|: \lstinline|numpy.ndarray| of shape \((N,)\). Slack variable values (zeros if no slack variables are used).
  \item \lstinline|cost_out|: \lstinline|float|. Optimal objective value.
  \item \lstinline|N|: \lstinline|int|. Number of rows \(N\) in \lstinline|deltas| used to build robust constraints.
  \item \lstinline|complexity|: \lstinline|int|. Number of scenario constraints in the support list identified post-solve.
  \item \lstinline|constraints|: \lstinline|list| of CVXPY constraint objects: the \(N\) scenario constraints (the hard constraints are not included).
  \item \lstinline|degeneracy|: \lstinline|bool|. Flag indicating whether degeneracy was detected when computing the support list.
\end{itemize}

\subsubsection*{Exceptions and Error Checking:}
\begin{itemize}
\item Raises \lstinline|ValueError| if \lstinline|deltas| has no rows.
  \item Raises \lstinline|ValueError| if CVXPY problem status is not \lstinline|"optimal"| or \lstinline|"optimal_inaccurate"|.
  \item Raises \lstinline|AssertionError| if \lstinline|Q| is not positive semidefinite or not symmetric.
\end{itemize}

\subsection{\texttt{solve\_sdp}}\label{appendix:imp-solve_sdp1}

\begin{lstlisting}
def solve_sdp(deltas, F_d, E, c, Q, tau=0.0, x_ref=\texttt{np.array([0.0])}, rho=0.0, norm_type=2, solver=None, include_slack=None):
return x_out, zeta_out, cost_out, N, complexity, constraints, degeneracy
\end{lstlisting}

\subsubsection*{Parameters:}
\begin{itemize}
\item \lstinline|deltas|: \lstinline|numpy.ndarray| of shape \((N,K)\). Each row \(\delta_i\) is substituted into \lstinline|F_d| form per-scenario constraints.
\item \lstinline|F_d|: \lstinline|Callable[[numpy.ndarray], dict]|. Given a delta row returns a dictionary of matrices \(\{k\mapsto F_k(\delta)\}\) with keys like \lstinline|'0','1',...,'d'| denoting the constant term.
\item \lstinline|E|: \lstinline|dict| of \lstinline|numpy.ndarray|. Optional hard constraint matrices keyed as in \lstinline|F_d|.
 \item \lstinline|Q|: \lstinline|numpy.ndarray| of shape \((n,n)\). Symmetric positive semidefinite quadratic cost matrix.
  \item \lstinline|c|: \lstinline|numpy.ndarray| of length \(n\) (or shape \((n,1)\)), objective linear coefficients.
  \item \lstinline|tau|: \lstinline|float|. Weight for the regularization term \(\|x-x_{\mathrm{ref}}\|_{p}\), which is omitted when \(\tau=0\).
  \item \lstinline|x_ref|: \lstinline|numpy.ndarray| broadcastable to same shape as \(x\) (default \lstinline|np.array([0.0])|).
  \item \lstinline|rho|: \lstinline|float|. Weight of the penalty \(\rho\mathbf{1}^\top \zeta\) on the nonnegative slack variables \(\zeta\) of length \(N\) (one per scenario), which are added when \lstinline|include_slack| is true (or, if it is \lstinline|None|, when \(\rho\neq 0\)).
  \item \lstinline|norm_type|: \lstinline|int|, \lstinline|float| or \lstinline|str|. Order \(p\) of the vector norm in the regularization term: any \(p\geq 1\) (default \lstinline|2|), \lstinline|"inf"|, or \lstinline|"fro"| (the same as \(p=2\)).
  \item \lstinline|solver|: \lstinline|Optional[str]|. CVXPY solver or \lstinline|None| (chooses default solver).
  \item \lstinline|include_slack|: \lstinline|Optional[bool]|. Whether to add the slack variables \(\zeta\) of the relaxation formulation. \lstinline|None| (default) adds them only when \(\rho\neq 0\).
\end{itemize}

\subsubsection*{Return values:}
\begin{itemize}
  \item \lstinline|x_out|: \lstinline|numpy.ndarray| of shape \((n,)\). Optimizer solution for \(x\). Note: unlike the LP and QP solvers which return a column vector of shape \((n,1)\), the SDP solver returns a 1-D array.
  \item \lstinline|zeta_out|: \lstinline|numpy.ndarray| of shape \((N,)\). Slack variable values (zeros if no slack variables are used).
  \item \lstinline|cost_out|: \lstinline|float|. Optimal objective value.
  \item \lstinline|N|: \lstinline|int|. Number of rows \(N\) in \lstinline|deltas| used to build robust constraints.
  \item \lstinline|complexity|: \lstinline|int|. Number of scenario constraints in the support list identified post-solve.
  \item \lstinline|constraints|: \lstinline|list| of CVXPY constraint objects: the \(N\) scenario constraints (the hard constraints are not included).
  \item \lstinline|degeneracy|: \lstinline|bool|. Flag indicating whether degeneracy was detected when computing the support list.
\end{itemize}

\subsubsection*{Exceptions and Error Checking:}
\begin{itemize}
\item Raises \lstinline|ValueError| if \lstinline|deltas| has no rows.
  \item Raises \lstinline|ValueError| if CVXPY problem status is not \lstinline|"optimal"| or \lstinline|"optimal_inaccurate"|.
  \item Raises \lstinline|AssertionError| if \lstinline|Q| is not positive semidefinite or not symmetric.
  \item Raises \lstinline|AssertionError| if any \lstinline|F_d(\delta)| matrices are not symmetric.
  \item Raises \lstinline|AssertionError| if any \lstinline|E| matrices are not symmetric.
\end{itemize}

\subsection{\texttt{quantify\_risk}}\label{appendix:imp-quantify_risk}

\begin{lstlisting}
def quantify_risk(k,N,beta):
return epsL, epsU
\end{lstlisting}

\subsubsection*{Parameters:}
\begin{itemize}
\item \lstinline|k|: \lstinline|int|. Complexity, \emph{i.e.} length of support list.
\item \lstinline|N|: \lstinline|int|. Number of scenarios.
\item \lstinline|beta|: \lstinline|float|. User-defined confidence parameter (\(0 < \beta < 1\)).
\end{itemize}

\subsubsection*{Return values:}
\begin{itemize}
  \item \lstinline|epsL|: \lstinline|float|. Lower bound of the risk \(\underline{\varepsilon}\).
  \item \lstinline|epsU|: \lstinline|float|. Upper bound of the risk \(\bar{\varepsilon}\).
\end{itemize}

\subsubsection*{Implementation details:}
\begin{itemize}
  \item Uses the regularized incomplete beta function \lstinline|betainc| from \lstinline|jax.scipy.special|.
  \item Both bounds are computed via binary search with convergence tolerance \lstinline|1e-10|.
  \item Special case: if \(k = N\) the upper bound is set to \lstinline|1.0| immediately (all scenarios are support constraints, so no generalization guarantee is possible).
\end{itemize}

\subsection{\texttt{get\_support}}\label{appendix:imp-get_support}

\begin{lstlisting}
def get_support(constraints, non_risk_constraints, prob, objective, solver=None, threshold=1e-8, sol_tol=1e-6):
return complexity, support, degeneracy
\end{lstlisting}

\subsubsection*{Parameters:}
\begin{itemize}
  \item \lstinline|constraints|: \lstinline|list|. List of CVXPY scenario constraint objects (one per scenario).
  \item \lstinline|non_risk_constraints|: \lstinline|list| (may be empty). Hard constraint(s), always enforced when re-solving and never counted in the complexity.
  \item \lstinline|prob|: CVXPY \lstinline|Problem|. The solved problem instance whose solution and duals are used.
  \item \lstinline|objective|: CVXPY objective expression used to re-solve reduced problems.
  \item \lstinline|solver|: \lstinline|Optional[str]|. Choice of CVXPY solver passed to \lstinline|Problem.solve| when validating candidate supports.
  \item \lstinline|threshold|: \lstinline|float| (default \lstinline|1e-8|). Dual-value threshold for the initial screening of candidate support-list members (violated or active constraints).
  \item \lstinline|sol_tol|: \lstinline|float| (default \lstinline|1e-6|). Relative tolerance for judging whether a re-solve reproduces the reference optimal value; set above the solver's value-reproducibility and below the smallest meaningful support contribution.
\end{itemize}

\subsubsection*{Return values:}
\begin{itemize}
  \item \lstinline|complexity|: \lstinline|int|. Number of items in the support list.
  \item \lstinline|support|: \lstinline|list|. Final support list of CVXPY constraint objects, \emph{i.e.}, violated or active scenario constraints whose removal changes the optimal value.
  \item \lstinline|degeneracy|: \lstinline|bool|. \lstinline|True| if degeneracy was detected during validation/reduction procedures, \emph{i.e.}, when the reduction removes a screened constraint that is active or violated at the optimum, or when the dual screening fails to reproduce the optimum (conservative fallback).
\end{itemize}

\subsubsection*{Implementation details:}
Internally uses the helper function \lstinline|test_support(objective, support, non_risk, ref, solver)| to validate candidate support lists: it re-solves the problem with the candidate constraints plus the always-enforced hard constraints and returns \lstinline|True| if the optimal value matches the reference (cf.\ Section~\ref{sec:implementation}).

\subsubsection*{Exceptions and Error Checking:}
\begin{itemize}
  \item Raises \lstinline|ValueError| if validation fails to produce a consistent support list (suggests trying another solver).
\end{itemize}

\end{document}